\documentclass[10pt]{article}
\usepackage{amsmath}
\usepackage{latexsym}
\usepackage{amssymb}
\usepackage{amsfonts}
\usepackage{amsthm}
\title{ MINIMUM RESOLUTION OF THE\\ MINKOWSKI, SCHWARZSCHILD AND KERR \\DIFFERENTIAL MODULES}
\author{J.-F. Pommaret \\ CERMICS, Ecole des Ponts ParisTech, France \\
 jean-francois.pommaret@wanadoo.fr \\
 ORCID: 0000-0003-0907-2601 }
\date{  }
\begin{document}
\maketitle

\noindent
{\bf ABSTRACT}

Our recent arXiv preprints and published papers on the solution of the Riemann-Lanczos and Weyl-Lanczos problems have brought our attention on the importance of revisiting the algebraic structure of the Bianchi identities in Riemannian geometry. We also discovered in the meantime that, in our first book of 1978, we had already used a new way for studying the various compatibility conditions (CC) of an operator that may not be necessarily formally integrable (FI) in order to construct canonical formally exact differential sequences on the jet level. The purpose of this paper is to prove that the combination of these two facts clearly shows the specific importance of the Spencer operator and the Spencer $\delta$-cohomology, totally absent from mathematical physics today. The results obtained are unavoidable because they only depend on elementary combinatorics and diagram chasing. They also provide for the first time the purely intrinsic interpretation of the respective numbers of successive first, second, third  and higher order generating CC. However, if they of course agree with the linearized Killing operator over the Minkowski metric, they largely disagree with recent publications on the respective numbers of generating CC for the linearized Killing operator over the Schwarzschild and Kerr metrics. Many similar examples are illustrating these new techniques, providing in particular the only symbol existing in the literature which is 2-ayclic witout being of finite type, contrary to the conformal situation.  \\

\vspace{4cm} 

\noindent
{\bf KEY-WORDS}  \\
Formal integrability; Involutivity; Compatibility conditions; Spencer operator; Janet sequence; Spencer sequence; Differential module; Homological algebra; Extension module.\\

\newpage

\noindent
{\bf 1) INTRODUCTION} \\

The present study is mainly local and we only use standard notations of differential geometry. For simplicity, we shall also adopt the same notation for a vector bundle $(E, F, \dots)$ and its set of sections $(\xi, \eta,\zeta, \dots)$. Now, if $X$ is  the ground manifold $X$ with dimension $n$ and local coordinates $(x^1, \dots,x^n)$ and $E$ is a vector bundle over $X$ with local coordinates $(x,y)$, we shall denote by $J_q(E)$ the $q$-jet bundle of $E$ with local coordinates $(x,y_q)$ and sections ${\xi}_q$ transforming like the $q$-derivatives $j_q(\xi)$ of a section $\xi={\xi}_0$ of $E$. If $F$ with section $\eta$ is another vector bundle over $X$ and $\Phi:J_q(E) \rightarrow F$ is an epimorphism with kernel the linear system $R_q \subset J_q(E)$, we shall associate the differential operator ${\cal{D}}=\Phi \circ j_q: E \rightarrow F: \xi \rightarrow \eta$ and set $\Theta = ker({\cal{D}})$. All the operators considered will be locally defined over a differential field $K$ whith $n$ derivations $({\partial}_1,\dots, {\partial}_n)$ and we shall indicate the order of an operator under its arrow. It is well known and we shall provide many explicit examples, that, if we want to solve, at least locally the linear inhomogeneous system $\cal{D}\xi=\eta$, one usually needs {\it compatibility conditions} (CC) of the form ${\cal{D}}_1 \eta=0$ defined by another differential operator ${\cal{D}}_1:F=F_0 \rightarrow F_1:\eta \rightarrow \zeta$ that may be of high order in general but still locally defined over $K$. However, two types of " {\it phenomena} " can arise for exhibiting such CC but, though they can be quite critical in actual practice, we do not know any other reference on the possibility to solve them effectively because most people rely on the work of E. Cartan.\\ 

1) As shown in ([11], Introduction) or ([13]) with the Janet system $({\xi}_{33} - x^2 {\xi}_{11}=0, {\xi}_{22}=0)$ over the differential field $K = \mathbb{Q}(x)$ and in ([22]), it may be possible to find no CC of order one, no CC of order two, one CC of order three, then nothing new but one additional CC of order six and so on with no way to know when to stop. For the fun, when we started computer algebra around $1990$, we had to ask a special permit to the head of our research department for running the computer a full night and were not even able after a day to go any further on. Hence, a first basic problem is to establish a preliminary list of {\it generating} CC and know their maximum order. \\

2) Once the previous problem is solved, we do know a generating ${\cal{D}}_1$ of order $q_1$ and may start anew with it in order to obtain a generating ${\cal{D}}_2$ of order $q_2$ and so on as a way to work out a differential sequence. Contrary to what can be found in the Poincar\'{e} sequence for the exterior derivative where all the successive operators are of order one, things may not be so simple in actual practice and " {\it jumps} " may apear, that is the orders may go up and down in a apparently surprising manner that only the use of " {\it acyclicity} " through the Spencer cohomology can explain. As we shall see with more details in the case of the conformal Killing operator of order $1$, the successive orders are $(1,3,1)$ when $n=3$, $(1,2, 2 ,1)$ when $n=4$, $(1,2,1,2,1)$ when $n=5$ ([27]).  \\

A we have shown in our seven books, the only possibility to escape from these two types of problems is to start with an involutive operator ${\cal{D}}$ and construct in an intrinsic way two {\it canonical} differential sequences, namely the {\it linear Janet sequence} ([8], p 185, 391 for a global definition): 
\[  0 \rightarrow \Theta \rightarrow E \underset q{\stackrel{\cal{D}}{\longrightarrow}} F_0 \underset 1{\stackrel{{\cal{D}}_1}{\longrightarrow}} F_1\underset 1{\stackrel{{\cal{D}}_2}{ \longrightarrow}}  \dots \underset 1 {\stackrel{{\cal{D}}_{n-1}}{\longrightarrow}} F_{n-1} \underset 1 {\stackrel{{\cal{D}}_n}{\longrightarrow }}  F_n \rightarrow 0  \]
and the {\it linear Spencer sequence} ([8], p 185 for a global definition): 
\[   0 \rightarrow \Theta \stackrel{j_q}{\rightarrow} C_0 \underset 1{\stackrel{D_1}{\longrightarrow}} C_1 \underset 1{\stackrel{D_2}{\longrightarrow}} C_2\underset 1{\stackrel{D_3}{ \longrightarrow}}  \dots \underset 1 {\stackrel{D_{n-1}}{\longrightarrow}} C_{n-1} \underset 1 {\stackrel{D_n}{\longrightarrow }}  C_n \rightarrow 0  \]
As in {\it both cases}, the central operator is the Spencer operator but {\it not the exterior derivative}, contrary to what is done in ([1]) and the corresponding references, in particular ([8]), we do not agree on the effectivity of their definition of " {\it involutivity} " (p 1608/1609). In fact, the most important property of theses two sequences is that they are {\it formally exact} on the jet level as follows. Introducing the (composite) $r$-{\it prolongation} by means of the {\it formal derivatives} $d_i$:   
\[  {\rho}_r(\Phi):J_{q+r}(E) \rightarrow  J_r(J_q(E)) \rightarrow J_r(F_0):(x,y_{q+r}) \rightarrow 
(x,z_{\nu}=d_{\nu}\Phi         , 0 \leq \mid \nu \mid \leq r )  \] 
with kernel $R_{q+r}= {\rho}_r(R_q)=J_r (R_q) \cap J_{q+r}(E) \subset J_r(J_q(E))$, we have the long exact sequences: 
\[ 0  \rightarrow R_{q+r} \rightarrow J_{q+r}(E) \rightarrow J_r(F_0)   \] 
\[ 0 \rightarrow R_{q+q_1+r} \rightarrow J_{q+q_1+r}(E) \rightarrow J_{q_1+r}(F_0) \rightarrow J_r(F_1)\]
\[  0 \rightarrow R_{q+q_1+q_2 +r}  \rightarrow   J_{q+q_1+q_2 + r}(E)  \rightarrow    J_{q_1+q_2+r}(F_0) \rightarrow  J_{q_2+r}(F_1) \rightarrow J_r(F_2)   \]
{\it and so on} till the similar ones stopping at $J_r(F_n), \forall r\geq 0$. As shown by the counterexample exhibited in ([18], p 119-126), {\it all these sequences may be absolutely useful till the last one}. We shall also define the symbol $g_q=R_q\cap S_qT^* \otimes E$ and its $r$-prolongations $g_{q+r}={\rho}_r(g_q)$ only depends on 
$g_q$ in a purely algebraic way, that is no differentiation is involved. On the contrary, we shall say that $R_q$ or ${\cal{D}}$ is {\it formally integrable} (FI) if $R_{q+r}$ is a vector bundle $\forall r \geq 0$ {\it and} all the epimorphisms ${\pi}^{q+r+1}_{q+r}: J_{q+r+1}(E) \rightarrow J_{q+r}(E): (x,y_{q+r+1}) \rightarrow (x,y_{q+r})$ are inducing epimorphisms $R_{q+r+1} \rightarrow R_{q+r}$ of constant rank $ \forall r \geq 0$, which is a true purely differential property.  \\

Of course, for people familar with functional analysis, the definition of $\Theta$ could seem strange and uncomplete as it is not clear where to look for solutions. In our opinion (See [12] and review Zbl 1079.93001) it  is mainly for this reason that differential modules or simply $D$-modules have been introduced but we shall explain why such a procedure leads in fact to a (rather) {\it vicious circle} as follows. Working locally for simplicity with $dim(E)=m, dim(F)=p$, we may turn the definition backwards by introducing the non-commutative ring $D=K[d_1,\dots,d_n]=K[d]$ of differential polynomials $(P,Q,\dots)$ with coefficients in $K$. Then, instead of acting on the " {\it left} "  of column vectors of sections by differentiations as in the previous differential setting, {\it we shall use the same operator matrix} still denoted by ${\cal{D}}$ but now acting on the " {\it right} " of row vectors by composition. Introducing the canonical projection onto the residual module $M$, we obtain the exact sequence $D^p \underset q {\stackrel{\cal{D}}{\longrightarrow }} D^m \rightarrow M \rightarrow 0$ of differential modules also called " {\it free resolution} " of $M$ because $D^m$ and $D^p$ are clearly free differential modules. However, as $D$ is filtred by the order of operators, then $I =im({\cal{D}})\subset D^m$ is filtred too and, as we shall clearly see on the motivating examples, the induced filtration of $M=D^m/I$ can only been obtained in {\it any} applications if and only if  $R_q$ or $ {\cal{D}}$ is FI. Accordingly, all the difficulty will be to use the following key theorem (For Spencer cohomology and acyclicity or involutivity, see [8]-[13],[18],[19]):  \\

\noindent
{\bf THEOREM 1.1}: There is a finite {\it Prolongation/Projection} (PP) algorithm providing two integers $r,s\geq 0$ by successive increase of each of them such that the new system $R^{(s)}_{q+r}= {\pi}^{q+r+s}_{q+r}(R_{q+r+s})$ has the same solutions as $R_q$ but is FI with a $2$-acyclic or involutive symbol and first order CC. The order of a generating  
${\cal{D}}_1$ is thus bounded by $ r+s+1$ as we used $r+s$ prolongations.   \\ 

\noindent
{\bf EXAMPLE 1.2}: In the Janet example we have $R_2 \rightarrow R^{(1)}_3 \rightarrow R^{(2)}_4 \rightarrow R^{(2)}_5$ with $8 < 11 < 12 = 12$ and 
$dim_K(M)=12 \Rightarrow rk_D(M)=0 $. The final system is trivially involutive because it is FI with a zero symbol, a fact highly not evident {\it a prori} because it needs $5$ prolongations and the maximum order of the CC is thus equal to $3 + 2 + 1=6$. We obtain therefore a minimum resolution of the form $0 \rightarrow D \underset 4 {\stackrel{{\cal{D}}_2}{\longrightarrow}} D^2 \underset 6 { \stackrel{{\cal{D}}_1}{\longrightarrow}} D^2 \underset 2 {\stackrel{{\cal{D}}}{\longrightarrow }}
D \rightarrow M \rightarrow 0$ (See the introduction of [11] or [13] for details). \\

When a system is FI, we have a {\it projective limit} $R= R_{\infty}\rightarrow \dots \rightarrow R_q \rightarrow \dots\rightarrow R_1 \rightarrow R_0$. \\
As we are dealing with a differential field $K$, there is a bijective correspondence: 
\[  M_q = hom_K(R_q,K)  \hspace{5mm} \Leftrightarrow \hspace{5mm} R_q=hom_K(M_q,K)  \] 
and we obtain the {\it injective limit} $0 \subseteq M_0 \subseteq M_1 \subseteq \dots  \subseteq M_q \subseteq \dots  M_{\infty}=M$ providing the {\it filtration} of $M$. We have in particular $d_iM_q \subseteq M_{q+1}$ and $M=DM_q$ for $q \gg 0$. \\

\noindent
{\bf THEOREM 1.3}: $R=hom_K(M,K)$ is a differential module for the Spencer operator. \\

{\it Proof}: As the ring $D$ is generated by $K$ and $T=\{ a^id_i\mid a^i\in K \}$, we just need to define: 
\[   (af)(m)=a(f(m))=f(am) , \,    (d_i f)(m)={\partial}_i(f(m)) - f(d_im), \,    \forall a \in K, \forall m \in M, \forall d_i \in T, \forall f \in R   \]
and obtain $d_ia= ad_i + {\partial}_ia$ in  the operator sense. Choosing $m\in M$ to be the residue of $d_{\mu}y^k=y^k_{\mu}$ and setting $f(y_q)={\xi}_q:f(y^k_{\mu})={\xi}^k_{\mu}\in K$, we obtain in actual practice {\it exactly} the Spencer operator:   
$ d:R \rightarrow T^* \otimes R:f \rightarrow dx^i \otimes d_if$ with $(d_i f)^k_{\mu}= {\partial}_i{\xi}^k_{\mu} - {\xi}^k_{\mu + 1_i}$ or 
$d{\xi}_{q+1}=j_1({\xi}_q) - {\xi}_{q+1}$ or simply $d=\partial - \delta$ with a slight abuse of language. We notice that a "section"  ${\xi }_q\in R_q$ has in general, particularly for the non-commutative case (See [27] for examples), {\it nothing to do} with a "solution", a concept missing in ([1]-[4]).  \\
\hspace*{12cm}  $\Box$  \\

As we shall see in the motivating examples, once a differential module $M$ or the dual system $R=hom_K(M,K)$ is given, there may be quite different differential sequences or quite different resolutions and the problem will be to choose the one that could be the best in the application considered. During the last world war, many mathematicians discovered that a few concepts, called  {\it extension modules}, were not depending on the sequence used in order to compute them but {\it only} on $M$. A (very) delicate theorem of (differential) homological algebra even proves that no others can exist ([28]). Let us explain in a way as simple as possible these new concepts.   \\

As a preliminary crucial definition, if $P =a^{\mu}d_{\mu} \in D$, we shall define its (formal) {\it adjoint} by the formula $ad(P)= {(-1)}^{\mid \mu \mid}d_{\mu}a^{\mu}$ where we have set $\mid \mu \mid={\mu}_1 + \dots  + {\mu}_n$ whenever $\mu = ({\mu}_1, \dots , {\mu}_n)$ is a multi-index. Such a definition can be extended by linearity in order to define the formal adjoint $ad({\cal{D}})$ to be the transposed operator matrix obtained after taking the adjoint of each element. The main property is that $ad(PQ)=ad(Q)ad(P), \forall P,Q \in D \Rightarrow ad({\cal{D}}_1 \circ{\cal{D}}) = ad({\cal{D}}) \circ ad({\cal{D}}_1)$. \\

\noindent
{\bf EXAMPLE  1.4}: With ${\partial}_{22}\xi={\eta}^2, {\partial}_{12}\xi={\eta}^1$ for $\cal{D}$, we get  ${\partial}_1{\eta}^2-{\partial}_2{\eta}^1=\zeta$ for ${\cal{D}}_1$. Then $ad({\cal{D}}_1)$ is defined by ${\mu}^2=-{\partial}_1\lambda, {\mu}^1={\partial}_2\lambda$ while $ad(\cal{D})$ is defined by $\nu={\partial}_{12}{\mu}^1+{\partial}_{22}{\mu}^2$ but the CC of $ad({\cal{D}}_1)$ are generated by ${\nu}'={\partial}_1{\mu}^1+{\partial}_2{\mu}^2$. In the operator framework, we have the differential sequences:\\  
\[  \begin{array}{rcccl}
 \xi & \stackrel{\cal{D}}{\longrightarrow} & \eta & \stackrel{{\cal{D}}_1}{\longrightarrow} & \zeta   \\
  \nu& \stackrel{ad(\cal{D})}{\longleftarrow} & \mu & \stackrel{ad({\cal{D}}_1)}{\longleftarrow} & \lambda \\
      &  \swarrow &  &  &   \\
 {\nu}' &  &  &  &
  \end{array}  \]
where the upper sequence {\it is} formally exact at $\eta$ but the lower sequence {\it is not} formally exact at $\mu$.  \\
Passing to the module framework, we obtain the sequences:  \\
\[  \begin{array}{rccccl}
 D & \stackrel{{\cal{D}}_1}{\longrightarrow} & D^2 & \stackrel{\cal{D}}{\longrightarrow} & D & \longrightarrow M \longrightarrow  0  \\
  D& \stackrel{ad({\cal{D}}_1)}{\longleftarrow} & D^2 & \stackrel{ad(\cal{D})}{\longleftarrow} & D &  
  \end{array}  \]
where the lower sequence is not exact at $D^2$. The "{\it extension modules} " have been introduced in order to study this kind of " {\it gaps} ". \\

Therefore, we have to prove that the extension modules vanish, that is $ad({\cal{D}})$ generates the CC of $ad({\cal{D}}_1)$ and, conversely, that ${\cal{D}}_1$ generates the CC of ${\cal{D}}$. We also remind the reader that it has not been easy to exhibit the CC of the Maxwell or Morera parametrizations when $n=3$ and that a direct checking for $n=4$ should be strictly impossible ([17]). It has been proved by L. P. Eisenhart in 1926 (Compare to [8]) that the solution space $\Theta$ of the Killing system has $n(n+1)/2$ {\it infinitesimal generators} $\{ {\theta}_{\tau}\}$ linearly independent over the constants {\it if and only if} $\omega $ had constant Riemannian curvature, namely zero in our case. As we have a Lie group of transformations preserving the metric, the three theorems of Sophus Lie assert than $[{\theta}_{\rho},{\theta}_{\sigma}]=c^{\tau}_{\rho \sigma} {\theta}_{\tau}$ where the {\it structure constants} $c$ define a Lie algebra ${\cal{G}}$. We have therefore $\xi \in \Theta \Leftrightarrow \xi = {\lambda}^{\tau}{\theta}_{\tau}$ with ${\lambda}^{\tau}=cst$. Hence, we may replace the Killing system by the system ${\partial}_i{\lambda}^{\tau}=0$, getting therefore the differential sequence:  \\
\[  0 \rightarrow \Theta \rightarrow {\wedge}^0T^*\otimes {\cal{G}} \stackrel{d}{\longrightarrow} {\wedge}^1T^*\otimes {\cal{G}} \stackrel{d}{\longrightarrow} ... \stackrel{d}{\longrightarrow} {\wedge}^nT^* \otimes {\cal{G}} \rightarrow 0  \]
 which is the tensor product of the Poincar\'{e} sequence for the exterior derivative by the Lie algebra ${\cal{G}}$. Finally, as the extension modules do not depend on the resolution used and that most of them do vanish because the Poincar\'{e} sequence is self adjoint (up to sign), that is $ad(d)$ generates the CC of $ad(d)$ at any position, exactly like $d$ generates the CC of $d$ at any position. We invite the reader to compare with the situation of the Maxwell equations in electromagnetisme ([10]). However, we have proved in ([14],[21],[24],[25]) why neither the Janet sequence nor the Poincar\'{e} sequence can be used in physics and must be replaced by another resolution of $\Theta$ called {\it Spencer sequence} ([19]). \\

After this long introduction, the content of the paper will become clear: \\ \\
In section 2 we provide the mathematical tools from homological algebra and differential geometry needed for finding the generating CC of various orders. \\
Then, section 3 will provide motivating examples in order to illustrate these new concepts. \\
They are finally applied to the Killing systems for the S and K metrics in section 4 in such a way that the results obtained, though surprising they are, cannot be avoided because they will only depend on diagram chasing and elementary combinatorics.  \\

\noindent
{\bf 2) MATHEMATICAL TOOLS}. \\

\noindent
{\bf A) HOMOLOGICAL ALGEBRA}\\

We now need a few definitions and results from homological algebra ([12],[13],[28]). In all that follows, $A,B,C,...$ are
modules over a ring or vector spaces over a field and the linear maps are making the diagrams commutative. We introduce the notations $rk=rank$, $nb=number$, $dim=dimension$, $ker=kernel$, $im=image$, $coker=cokernel$. When $\Phi :A\rightarrow B$ is a linear map (homomorphism),
we may consider the so-called ker/coker exact sequence where where $coker(\Phi)=B/im(\Phi) $:  \\
\[0\longrightarrow ker(\Phi) \longrightarrow A \stackrel{\Phi}{\longrightarrow} B \longrightarrow coker(\Phi)  \longrightarrow 0 \]
\noindent

In the case of vector spaces over a field $k$, we successively have $rk(\Phi)=dim(im(\Phi))$, $dim(ker(\Phi))=dim(A)-rk(\Phi)$, 
$dim(coker(\Phi))=dim(B)-rk(\Phi)=nb$ of compatibility conditions, and obtain by substraction:
\[ dim(ker(\Phi))-dim(A)+dim(B)-dim(coker(\Phi))=0 \]
In the case of modules, using localization, we may replace the dimension by the rank and obtain the same relations because of the additive property of
the rank. The following result is essential:\\

\noindent
{\bf SNAKE LEMMA 2.A.1}: When one has the following commutative diagram resulting from the the two central vertical short exact sequences by
exhibiting the three corresponding horizontal ker/coker exact sequences:
\[   \begin{array}{ccccccccccc}
 & & 0 & & 0 & & 0 & & & & \\
 & &\downarrow & & \downarrow & & \downarrow & & & & \\
0&\longrightarrow &K&\longrightarrow &A&\longrightarrow &A'&\longrightarrow &Q&\longrightarrow & 0  \\
 & &\downarrow & &\;\;\;\downarrow \! \Phi& &\;\;\;\;\downarrow \! {\Phi}'& &\downarrow & & \\
0&\longrightarrow &L&\longrightarrow &B & \longrightarrow &B'&\longrightarrow &R& \longrightarrow & 0 \\
 & &\downarrow & &\;\;\;\downarrow \! \Psi & &\;\;\;\; \downarrow \! {\Psi}'& & \downarrow & & \\
0 & \longrightarrow &M& \longrightarrow &C& \longrightarrow &C'& \longrightarrow &S& \longrightarrow & 0 \\
 & & & & \downarrow & & \downarrow & & \downarrow & & \\
 & & & &0& &0& &0& &
\end{array}   \]
then there exists a connecting map $M \longrightarrow Q$ both with a long exact sequence:
\[0 \longrightarrow K \longrightarrow L \longrightarrow M \longrightarrow Q \longrightarrow R \longrightarrow S \longrightarrow 0.\]  \\

\noindent
{\it Proof}: We start constructing the connecting map by using the following succession of elements:
\[   \begin{array}{ccccccc}
 & & a & \cdots & a'& \longrightarrow &  q \\
 & & \vdots & & \downarrow & & \vdots  \\
 & & b & \longrightarrow &b' & \cdots & 0  \\
 & & \downarrow & & \vdots & & \\
m& \longrightarrow &c& \cdots &0& & 
\end{array}   \]
Indeed, starting with $m\in M$, we may identify it with $c\in C$ in the kernel of the next horizontal map. As $\Psi$ is an epimorphism, we may find
$b\in B$ such that $c=\Psi(b)$ and apply the next horizontal map to get $b'\in B'$ in the kernel of ${\Psi}'$ by the commutativity of the lower
square. Accordingly, there is a unique $a'\in A'$ such that $b'={\Phi}'(a')$ and we may finally project $a'$ to $q\in Q$. The map is
well defined because, if we take another lift for $c$ in $B$, it will differ from $b$ by the image under $\Phi$ of a certain $a\in A$ having zero
image in $Q$ by composition. The remaining of the proof is similar and left to the reader as an exercise. The above explicit procedure will not be repeated. \\
\hspace*{12cm} $\Box $   \\

We may now introduce {\it cohomology theory} through the following definition:\\

\noindent
{\bf DEFINITION 2.A.2}: If one has any sequence $A \stackrel{\Phi}{\longrightarrow} B \stackrel{\Psi}{\longrightarrow} C $,
then one may introduce $coboundary=im(\Phi)\subseteq ker(\Psi)=cocycle \subseteq B$ and the cohomology at $B$ is the quotient  $cocycle/coboundary $.\\

\noindent
{\bf THEOREM 2.A.3}: The following commutative diagram where the two central  vertical sequences are long exact sequences and the horizontal lines are
ker/coker exact sequences:
\[    \begin{array}{ccccccccccccc}
 & &0& &0& &0& & & & & & \\
 & &\downarrow & & \downarrow & & \downarrow & & & & & &     \\
0&\longrightarrow &K&\longrightarrow &A&\longrightarrow &A'&\longrightarrow &Q&\longrightarrow &0& & \\
 & &\downarrow & &\;\;\;\downarrow \! \Phi & &\;\;\;\;\downarrow \! \Phi' & &\downarrow & & & & \\
0&\longrightarrow &L&\longrightarrow&B&\longrightarrow &B'&\longrightarrow &R&\longrightarrow &0& & \\
\cdots &\cdots &\downarrow &\cdots &\;\;\;\downarrow \! \Psi &\cdots &\;\;\;\;\downarrow \! \Psi'&\cdots &\downarrow &\cdots &\cdots &\cdots &cut \\
0&\longrightarrow &M&\longrightarrow &C&\longrightarrow &C'&\longrightarrow &S&\longrightarrow &0& & \\
 & &\downarrow & &\;\;\;\downarrow \! \Omega & &\;\;\;\;\downarrow \! \Omega'& &\downarrow & & & & \\
0&\longrightarrow &N&\longrightarrow &D&\longrightarrow &D'&\longrightarrow &T&\longrightarrow &0& & \\
 & & & &\downarrow & &\downarrow & &\downarrow & & & & \\
 & & & &0& &0& &0 & & & & 
\end{array}    \]
induces an isomorphism between the cohomology at $M$ in the left vertical column and the kernel 
of the morphism $Q\rightarrow R$ in the right vertical column.\\

\noindent
{\it Proof}: Let us ``cut'' the preceding diagram into the following two
commutative and exact diagrams by taking into account the relations 
$im(\Psi)=ker(\Omega), im({\Psi}')=ker({\Omega}')$:
\[   \begin{array}{ccccccccccc}
 & &0& &0& &0 & & & \\
 & &\downarrow & &\downarrow & &\downarrow & & & & \\
 0&\longrightarrow &K&\longrightarrow &A&\longrightarrow &A'&\longrightarrow &Q&\longrightarrow &0 \\
 & &\downarrow & &\;\;\;\downarrow \! \Phi & &\;\;\;\;\downarrow \! \Phi' & &\downarrow & & \\
0&\longrightarrow &L&\longrightarrow &B&\longrightarrow &B'&\longrightarrow &R&\longrightarrow &0 \\
 & &\downarrow & &\;\;\;\downarrow \! \Psi & &\;\;\;\;\downarrow \! \Psi' & & & & \\
0&\longrightarrow & cocycle &\longrightarrow &im \, \Psi &\longrightarrow &im \,\Psi' & & & & \\
 & & & &\downarrow & &\downarrow & & & & \\
 & & & &0& &0& & & & 
\end{array}    \]

\[  \begin{array}{ccccccc}
 & &0& &0& &0 \\
 & &\downarrow & &\downarrow & &\downarrow \\
0&\longrightarrow & cocycle &\longrightarrow & ker \, \Omega &\longrightarrow & ker \, \Omega' \\
 & &\downarrow & &\downarrow & &\downarrow \\
0&\longrightarrow &M&\longrightarrow &C&\longrightarrow & C' \\
 & &\downarrow & &\;\;\;\downarrow \! \Omega & &\;\;\;\;\downarrow \! \Omega' \\
0&\longrightarrow &N&\longrightarrow &D&\longrightarrow &D' \\
 & & & &\downarrow & &\downarrow \\
 & & & &0& &0 
\end{array}   \]
Using the snake theorem, we successively obtain:
\[  \begin{array}{cccccc}
\Longrightarrow &\exists &\qquad &0\longrightarrow K \longrightarrow L \stackrel{\Psi}{\longrightarrow}
cocycle \longrightarrow Q \longrightarrow R &\qquad & exact \\
\Longrightarrow &\exists &\qquad & 0 \longrightarrow coboundary \longrightarrow cocycle \longrightarrow ker \,(Q\longrightarrow R) 
\longrightarrow 0 &\qquad & exact \\
\Longrightarrow & &\qquad & cohomology \,\, at  \,\, M \simeq ker \,(Q \longrightarrow R) & &
\end{array}     \]
\hspace*{12cm}   $\Box $ \\

\noindent
{\bf B) DIFFERENTIAL GEOMETRY}  \\

\noindent
Comparing the sequences obtained in the previous examples, we may state:  \\

\noindent
{\bf DEFINITION 2.B.1}: A differential sequence is said to be {\it formally exact} if it is exact on the jet level composition of the prolongations involved. A formally exact sequence is said to be {\it strictly exact} if all the operators/systems involved are FI (See [5],[8],[11],[16] and [19] for more details). A strictly exact sequence is called 
{\it canonical} if all the operators/systems are involutive. Fourty years ago, we did provide the link existing between the only known canonical sequences, namely the Janet and Spencer sequences ([8], See in particular the pages 185 and 391).   \\

With canonical projection ${\Phi}_0=\Phi:J_q(E) \Rightarrow J_q(E)/R_q=F_0$, the various prolongations are described by the following commutative and 
exact "{\it introductory diagram} " often used in the sequel:  \\
\[ \fbox { $ \begin{array}{rcccccccl}
    &  0  &  &  0  &  &  0  &  &   &   \\
     &  \downarrow &  & \downarrow &  &  \downarrow  &  &   &     \\
0 \rightarrow  &  g_{q+r+1}  & \rightarrow & S_{q+r+1}T^*\otimes E & \stackrel{{\sigma}_{r+1}(\Phi)}{\longrightarrow} &  S_{r+1}T^*\otimes F_0 & \rightarrow &   h_{r+1} &  \rightarrow 0 \\
&  \downarrow &  & \downarrow &  &  \downarrow  &  &  \downarrow    &     \\
0 \rightarrow &  R_{q+r+1} & \rightarrow  & J_{q+r+1}(E) & \stackrel{{\rho}_{r+1}(\Phi)}{\longrightarrow}  &  J_{r+1}(F_0) &  \rightarrow &  
Q_{r+1}  &  \rightarrow 0  \\
    &  \downarrow &  & \downarrow &  &  \downarrow  &  & \downarrow   &     \\
0 \rightarrow &  R_{q+r} & \rightarrow  & J_{q+r}(E) & \stackrel{{\rho}_r(\Phi)}{\longrightarrow}  &  J_r(F_0) &  \rightarrow &  Q_r  & 
\rightarrow 0 \\ 
   &  &  & \downarrow &  &  \downarrow  &  & \downarrow  & \\
   &  &  &  0  && 0  && 0  &
\end{array} $ }  \]
Chasing along the diagonal of this diagram while applying the standard "{\it snake}" lemma, we obtain the useful "{\it long exact connecting sequence} " also often used in the sequel:  \\
 \[ \fbox { $ 0  \rightarrow  g_{q+r+1}  \rightarrow R_{q+r+1}  \rightarrow R_{q+r}  \rightarrow h_{r+1} \rightarrow Q_{r+1}  \rightarrow Q_r \rightarrow 0 $ } \]
which is thus connecting in a tricky way FI ({\it lower left}) with CC ({\it upper right}).  \\

 We finally recall the "{\it fundamental diagram I} " that we have presented in many books and papers, relating the (upper) {\it canonical Spencer sequence} to the (lower) {\it canonical Janet sequence}, that only depends on the left commutative square ${\cal{D}}=\Phi \circ j_q$ with $\Phi = {\Phi}_0$ when one has an involutive system $R_q\subseteq J_q(E)$ over $E$ with $dim(X)=n$ and $j_q:E \rightarrow J_q(E)$ is the derivative operator up to order $q$ while the epimorphisms ${\Phi}_1, ...,{\Phi}_n$ are successively induced by $\Phi$:  \\
 \[ \fbox { $ \footnotesize  \begin{array}{rcccccccccccccl}
 &&&&& 0 &&0&&0&&  &  &0&  \\
 &&&&& \downarrow && \downarrow && \downarrow & &  &   & \downarrow &  \\
  & 0& \rightarrow& \Theta &\stackrel{j_q}{\rightarrow}&C_0 &\stackrel{D_1}{\rightarrow}& C_1 &\stackrel{D_2}{\rightarrow} & C_2 &\stackrel{D_3}{\rightarrow}& ... &\stackrel{D_n}{\rightarrow}& C_n &\rightarrow 0 \\
  &&&&& \downarrow & & \downarrow & & \downarrow & &  &&\downarrow &     \\
   & 0 & \rightarrow & E & \stackrel{j_q}{\rightarrow} & C_0(E) & \stackrel{D_1}{\rightarrow} & C_1(E) &\stackrel{D_2}{\rightarrow} & C_2(E) &\stackrel{D_3}{\rightarrow} & ... &\stackrel{D_n}{\rightarrow} & C_n(E) &   \rightarrow 0 \\
   & & & \parallel && \hspace{6mm}\downarrow  {\Phi}_0& & \hspace{6mm}\downarrow  {\Phi}_1& & \hspace{6mm}\downarrow {\Phi}_2 & &  & & \hspace{6mm} \downarrow  {\Phi}_n & \\
   0 \rightarrow & \Theta &\rightarrow & E & \stackrel{\cal{D}}{\rightarrow} & F_0 & \stackrel{{\cal{D}}_1}{\rightarrow} & F_1 & \stackrel{{\cal{D}}_2}{\rightarrow} & F_2 & \stackrel{{\cal{D}}_3}{\rightarrow} & ... &\stackrel{{\cal{D}}_n}{\rightarrow} & F_n & \rightarrow  0 \\
   &&&&& \downarrow & & \downarrow & & \downarrow & &  &  &\downarrow &   \\
   &&&&& 0 && 0 && 0 && &&0 &  
   \end{array} $ }   \]  \\
This result will be used in order to compare the M, S and K metrics when $n=4$ but it is important to notice that this {\it whole} diagram does not depend any longer on the $(a,m)$ parameters of S and K ([20],[23]).  \\

\noindent
{\bf PROPOSITION 2.B.2}: If $R_q\subset J_q(E)$ and $R_{q+1} \subset J_{q+1}(E)$ are two systems of respective orders $q$ and $q+1$, then 
$R_{q+1} \subset {\rho}_1(R_q)$ if and onlty if ${\pi}^{q+1}_q (R_{q+1}) \subset R_q$ {\it and} $dR_{q+1} \subset T^* \otimes R_q$.  \\

\noindent
{\it Proof}: First we notice that {\it necessarily} we must have  ${\pi}^{q+1}_q (R_{q+1}) \subset R_q$ because, as ${\rho}_1(R_q)$ may not project {\it onto} $R_q$, it is nevertheless defined by (maybe) more equations of strict order $q$ than $R_q$. Now, if ${\xi}_{q+1} \in R_{q+1} \subset J_{q+1}(E)$ is such that $d{\xi}_{q+1} \in T^* \otimes R_q$, then ${\xi}_q \in R_q \Rightarrow j_1({\xi}_q) \in J_1(R_q)$. As $J_1(R_q)$ is an affine bundle over $R_q$ modelled on $T^* \otimes R_q$ (or simply $T^* \otimes R_q \subset J_1(R_q)$) and $J_{q+1}(E) \subset J_1(J_q(E))$, we have thus ${\xi}_{q+1} = j_1({\xi}_q) - d{\xi}_{q+1}\in J_1(R_q) \cap J_{q+1}(E)= {\rho}_1(R_q)$. \\The converse way is similar.  \\
\hspace*{12cm}  $\Box$ \\
The next key idea has been discovered in ([8]) as a way to define the so-called Janet bundles and thus for a totally different reason. \\

\noindent
{\bf DEFINITION 2.B.3}: Let us "cut" the preceding introductory diagram  by means of a central vertical line and define $R'_r = im({\rho}_r(\Phi)) \subseteq J_r(F_0)$ with $R'_0=F_0$. Chasing in this diagram, we notice that ${\pi}^{r+1}_r:J_{r+1}(E) \rightarrow J_r(E)$ induces an epimorphism ${\pi}^{r+1}_r: R'_{r+1} \rightarrow R'_r, \forall r\geq 0$. However, a chase in this diagram proves that {\it the kernel of this epimorphism is not} $im({\sigma}_{r+1}(\Phi)$ unless $R_q$ is FI ({\it care}). For this reason, we shall define it to be {\it exactly} 
$g'_{r+1}$.  \\

\noindent
{\bf THEOREM 2.B.4}:  $R'_{r+1} \subseteq {\rho}_1(R'_r)$ and $ dim ({\rho}_1(R'_r)) - dim (R'_{r+1})$ is the number of new generating CC of order $r+1$ .  \\

\noindent
{\it Proof}: First of all, we have the following commutative and exact diagram obtained by applying the Spencer operator to the top long exact sequence:
\[ \begin{array}{rcccccccl}
  0 \rightarrow & R_{q+r+1} & \rightarrow & J_{q+r+1}(E) & \rightarrow & J_{r+1}(F_0) & \rightarrow & Q_{r+1} &\rightarrow 0 \\
             & \hspace{3mm} \downarrow d &   &  \hspace{3mm} \downarrow d &  &  \hspace{3mm} \downarrow d &  &  \hspace{3mm} \downarrow d &   \\
  0 \rightarrow & T^* \otimes R_{q+r}& \rightarrow & T^* \otimes J_{q+r}(E) & \rightarrow & T^* \otimes J_r(F_0) & \rightarrow &  T^* \otimes Q_r & \rightarrow 0 
\end{array}   \]
"Cutting" the diagram in the middle as before while using the last definition, we obtain the induced map $R'_{r+1} \stackrel{d}{\longrightarrow} T^* \otimes R'_r$ and the first inclusion follows from the last proposition. Such a procedure cannot be applied to the top row of the introductory diagram through the use of $\delta$ instead of $d$ because of the comment done on the symbol in the last definition.  \\
Now, using only the definition of the prolongation for the system and its symbol, we have the following commutative and exact diagram:
 \[  \begin{array}{rcccccccl}
    &  0  &  &  0  &  &  0  &  &  0 &   \\
     &  \downarrow &  & \downarrow &  &  \downarrow  &  & \downarrow  &     \\
0 \rightarrow  & {\rho}_1( g'_r)  & \rightarrow & S_{r+1}T^*\otimes F_0 & \stackrel{{\sigma}_1(\Psi)}{\longrightarrow} &  T^*\otimes Q_r & \rightarrow &   Q'_1 & 
 \rightarrow 0 \\
&  \downarrow &  & \downarrow &  &  \downarrow  &  &  \parallel  &     \\
0 \rightarrow & {\rho}_1( R'_r) & \rightarrow  & J_{r+1}(F_0) & \stackrel{{\rho}_1(\Psi)}{\longrightarrow}  &  J_1(Q_r) &  \rightarrow &  Q'_1  &  \rightarrow 0  \\
    &  \downarrow &  & \downarrow &  &  \downarrow  &  & \downarrow   &     \\
0 \rightarrow &  R'_r & \rightarrow  & J_r(F_0) & \stackrel{ \Psi}{\longrightarrow}  &  Q_r &  \rightarrow & 0  &   \\ 
   & \downarrow  &  & \downarrow &  &  \downarrow  &  &  & \\
   & 0  &  &  0  && 0  &&   &
\end{array}   \]
and obtain the following commutative and exact diagram:
\[  \begin{array}{rccccccl}
  & 0 &   & 0  &   &  0 &   \\
  & \downarrow  &  &\downarrow   &  &\downarrow  &  \\
0 \rightarrow & g'_{r+1} & \rightarrow & {\rho}_1(g'_r) & \rightarrow  & A  & \rightarrow 0  \\
                   &  \downarrow &  &  \downarrow  &  &  \parallel &   \\
0 \rightarrow & R'_{r+1} & \rightarrow & {\rho}_1(R'_r) & \rightarrow  & A  & \rightarrow 0  \\
                     & \downarrow  &  &  \downarrow &  & \downarrow &  \\
0 \rightarrow & R'_r & = & R'_r & \rightarrow  & 0  &  \\
                    &  \downarrow &  &  \downarrow &  &  \\
                    &  0  &  &  0  &  &  &  
\end{array}  \]
The computation of $y=dim(A)= dim({\rho}_1(R'_r)) - dim(R'_{r+1})$ only depends on $x= dim(Q'_1)$ and is rather tricky as follows (See the motivating examples):   \\
\[  dim(Q_r)= dim(R_{q+r}) + dim(J_r(F_0)) - dim(J_{q+r}(E)) \]
\[   dim({\rho}_1(R'_r)) = dim(J_{r+1}(F_0)) + x - dim(J_1(Q_r))  \]
\[  dim(R'_{r+1})= dim(J_{q+r+1}(E)) - dim(R_{q+r+1})   \]
As we shall see with the motivating examples and with the S or K metrics, the computation is easier when the system is FI but can be much more difficult when 
the system is not FI.  \\
However, the number of linearly independent CC of order $r+1$ coming from the CC of order $r$ is $dim(J_1(Q_r)) - x$ while the total number of CC of order $r+1$ is:
\[  dim(Q_{r+1}) = dim(R_{q+r+1})+ dim(J_{r+1}(F_0)) - dim(J_{q+r+1}(E)) = dim(J_{r+1}(F_0)) - dim(R'_{r+1})  \] 
The number of new CC of {\it strict} order $r+1$ is equal to $y$ because $dim(J_{r+1}(F_0))$ disappears by difference. For a later use in GR, we point out the fact that, if the given system $R_q \subset J_q(E)$ depends on parameters that {\it must} be contained in the ground differential field $K$ (only $(m)$ for the S metric but $(a,m)$ for the K metric), all the dimensions considered may {\it highly} depend on them even if the underlying procedure is of course the same.  \\
As an alternative proof, we may say that the number of CC of strict order $r+1$ obtained from the CC of order $r$ is equal to 
$dim(S_{r+1}T^* \otimes F_0) - dim ({\rho}_1(g'_r))$ while the total number of CC of order $r+1$ is equal to $dim(S_{r+1}T^* \otimes F_0) - dim (g'_{r+1})$. The number of new CC of strict order $r+1$ is thus also equal to $y= dim (({\rho}_1(g'_r)) - dim (g'_{r+1})$ because $dim(S_{r+1}T^* \otimes F_0)$ also disappears by difference. However, unless $R_q$ is FI, we have in general $g'_r \neq im({\sigma}_r(\Phi))$ and it thus better to use the systems rather than their symbols.  \\
\hspace*{12cm}  $\Box$  \\ 

\noindent
{\bf COROLLARY 2.B.5}: The system $R'_r \subset J_r(F_0)$ becomes FI with a $2$-acyclic or involutive symbol and $R'_{r+1}={\rho}_1(R'_r)  \subset J_{r+1}(F_0)$ when $r$ is large enough.  \\

\noindent
{\it Proof}: According to the last diagram, we have $g'_{r+1}\subseteq {\rho}_1(g'_r)$ and $g'_{r+1}$ is thus defined by more linear equations than ${\rho}_1(g'_r)$. We are facing a purely algebraic problem over commutative polynomial rings and well known noetherian arguments are showing that $g'_{r+1}={\rho}_1(g'_r)$ or, equivalently, $y=0$ when $r$ is large enough. Chasing in the last diagram, we obtain therefore $R'_{r+1} = {\rho}_1(R'_r)$ for $r$ large enough and $R'_r$ is a vector bundle because because $R_{q+r}$ is a vector bundle. If we denote by $M'$ the differential module obtained from the system $R'_r \subset J_r(F_0)$ exactly like we have denoted by $M $ the differential module obtained from the system $R_q \subset J_q(E)$, we have the short exact sequence $0 \rightarrow M' \rightarrow D^m \rightarrow M \rightarrow 0$. Accordingly, $M' \simeq I \subset D^m$ is a torsion-free differential module and there cannot exist any specialization as an epimorphism $M' \rightarrow M" \rightarrow 0 $ with $rk_D(M')=rk_D(M")$ because the kernel should be a torsion differential module and thus should vanish. This comment is strengthening the fact that the knowledge of $M$ and thus of $I$ can only be done through Theorem 1.1. Therefore, if $(r,s)$ are the ones produced by this theorem, then the order of the CC system must be $r+s+1$. We obtain 
$3+2+1=6$ for the Janet system with systems $R'_r$ of successive dimensions $ 2, 8, 20, 39, 66, 102, 147$ and ask the reader to find $dim(R'_7)=202$ (Hint: [11]). \\
\hspace*{12cm}  $\Box$  \\

We are now ready for working out the generating CC ${\cal{D}}_1:F_0 \rightarrow F_1$ and start afresh in a simpler way because this new operator is FI (Compare to [8], Proposition 2.9, p 173). However, contrary to what the reader could imagine, it is precisely at this point that troubles may start and the best example is the conformal Killing operator. Indeed, it is known that the order of  the generating CC for a system of order $q$ which is FI is equal to $s+1$ if  the symbol $g_{q+s}$ becomes $2$-acyclic {\it before} becoming involutive. This fact will be illustrated in a forthcoming motivating example but we recall that the conformal Killing symbol ${\hat{g}}_1\subset T^* \otimes T$ is such that ${\hat{g}}_2$ is $2$-acyclic when $n\geq 4$ while ${\hat{g}}_3=0$, a fact explaining why the Weyl operator is of order $2$ {\it but} the Bianchi-type operator is also of order $2$, a result still neither known nor even acknowledged today ([18],[27]).  \\  \\

\noindent
{\bf 3) MOTIVATING EXAMPLES}. \\

We now provide three motivating examples in order to illustrate both the usefulness and the limit of the previous procedure.  \\

\noindent
{\bf EXAMPLE 3.1}: With $m=1,n=3, K=\mathbb{Q}$, we revisit the nice example of Macaulay ([7]) presented in ([22]), namely the homogeneous second order linear system $R_2\subset J_2(E)$ defined by ${\xi}_{33}=0, {\xi}_{13} - {\xi}_2=0$ which is far from being formally integrable. We let the reader prove the strict inclusions  $R^{(2)}_2 \subset R^{(1)}_2 \subset R_2 \subset J_2(E)$ with successive dimensions $6 < 7 < 8 < 10$. The respective symbols are involutive but only the final system $R^{(2)}_2$ is involutive. It follows that the generating CC of the operator defined by $R_2$ are at most of order $3$ but there is indeed only {\it one} single generating second order CC ([22]). Elementary combinatorics allows to prove the formulas $ dim(g_{r+2})=r+4, \,\, dim(R_{r+2})=4r+8, \,\, \forall r\geq 0  $. We have the short exact sequences:  \\
\[   0 \rightarrow R_2 \rightarrow J_2(E) \rightarrow F_0 \rightarrow 0 ,  \hspace{1cm}  0 \rightarrow 8 \rightarrow 10  \rightarrow  2  \rightarrow 0   \]
\[  0 \rightarrow R_3 \rightarrow J_3(E) \rightarrow J_1(F_0) \rightarrow 0, \hspace{1cm} 0 \rightarrow 12 \rightarrow 20  \rightarrow  8  \rightarrow 0 \]
an the following commutative diagrams:  \\
\[  \begin{array}{rcccccccl}
    &  0  &  &  0  &  &  0  &  &  0 &   \\
     & \downarrow &  & \downarrow &  &  \downarrow  &  & \downarrow  &     \\
  0 \rightarrow  & g_5 & \rightarrow & S_5T^*\otimes E & \rightarrow &  \fbox{$ S_3T^*\otimes F_0 $} & \rightarrow & T^* \otimes F_1 &  \rightarrow 0 \\
&  \downarrow &  & \downarrow &  &  \downarrow  &  &  \parallel    &     \\
0 \rightarrow &  R_5 & \rightarrow  & J_5(E) & \rightarrow  &  J_3(F_0) &  \rightarrow &  J_1(F_1)  &  \rightarrow 0  \\
    &  \downarrow &  & \downarrow &  &  \downarrow  &  & \downarrow   &     \\
0 \rightarrow &  R_4 & \rightarrow  & J_4(E) & \rightarrow  &  J_2(F_0) &  \rightarrow & F_1  &  \rightarrow 0  \\ 
   &  &  & \downarrow &  &  \downarrow  &  &\downarrow   & \\
   &  &  &  0  &  & 0  &  &  0 &
\end{array}   \]
\[  \begin{array}{rcccccccl}
    &  0  &  &  0  &  &  0  &  & 0  &   \\
     &  \downarrow &  & \downarrow &  &  \downarrow  &  &  \downarrow &     \\
0 \rightarrow  &  7  & \rightarrow &21 & \rightarrow & 20 & \rightarrow &  3  &  \rightarrow 0 \\
&  \downarrow &  & \downarrow &  &  \downarrow  &  &  \downarrow    &     \\
0 \rightarrow &  20 & \rightarrow  & 56  & \rightarrow  & 40  &  \rightarrow &  4  &  \rightarrow 0  \\
    &  \downarrow &  & \downarrow &  &  \downarrow  &  & \downarrow   &     \\
0 \rightarrow & 16 & \rightarrow  &  35   & \rightarrow  & 20  &  \rightarrow &  1  & \rightarrow 0 \\ 
   &  &  & \downarrow &  &  \downarrow  &  & \downarrow  & \\
   &  &  &  0  &  & 0  &  & 0  &
\end{array}   \]
First of all, we have $R'_0=F_0$, $R'_1=J_1(F_0)$, $ dim(R'_2)=35 - 16=19$, $dim(R'_3)=56 - 20 = 36$. \\
It follows that we have successively:  \\
$  dim({\rho}_1(R'_0)) - dim(R'_1)= 8 - 8 = 0 \Rightarrow $ $0$ CC of order $1$.  \\
$  dim({\rho}_1(R'_1)) - dim (R'_2)= 20 - 19 = 1 \Rightarrow $ $1$ new CC of order $2$.  \\
$  dim( {\rho}_1(R'_2)) - dim(R'_3)= 36 - 36 = 0 \Rightarrow $ $0$ new CC of order $3$ and so on with:  \\
\[  dim (R'_{r+3})= dim(J_{r+5}(E)) - dim(R_{r+5})= (r+6)(r+7)(r+8)/6 - (4r + 20)       \]
\[    dim ({\rho}_1(R'_{r+2}))= dim (J_{r+3}(F_0)) - dim(J_{r+1}(F_1))= 2(r+4)(r+5)(r+6)/6 - (r+2)(r+3)(r+4)/6  \]
and check that $dim(R'_{r+3})= dim({\rho}_1(R'_{r+2}))= (r+4)(r^2 + 17r + 54)/6 $, $\forall r\geq 0$.  \\
Then, counting the dimensions, it is easy to check that the two prolongation sequences are exact on the jet level but that the upper symbol sequence is not exact at 
$S_3T^* \otimes F_0$ with coboundary space of imension $21 - 7=14$, cocycle space of dimension $20 - 3 =17$ and thus cohomology space of dimension $17 - 14=3$ that is  
$ dim(R_4/R^{(1)}_4)$ as we check that $7 - 20 + 16 - 3 = 0$. The reader may use the snake theorem to find this result directly through a chase not evident at first sight.   \\
We have then $dim(R^{(1)}_{r+2})=dim(R_{r+3}) - dim(g_{r+3})= 3r + 7$ and similarly $ dim(g^{(1)}_{r+2})=r+3 $ leading to $dim(R^{(2)}_{r+2})= dim(R^{(1)}_{r+3}) - dim(g^{(1)}_{r+3})= 2r + 6$ with $dim(g^{(2)}_{r+2})=2, \forall r\geq 0$. This result is of course coherent with the fact that the involutive system with the same solutions as $R_2$ is $R^{(2)}_2$ which is defined by ${\xi}_{33}=0, {\xi}_{23}=0, {\xi}_{22}=0, {\xi}_{13} - {\xi}_2=0$ .  \\

\noindent
{\bf EXAMPLE 3.2}: With $m=1, n=3, q=2, K=\mathbb{Q}$ and the commutative ring $D=K[d_1,d_2,d_3]$ of PD operators with coefficients in $K$, we revisit another example of Macaulay ([7]), namely the homogeneous second order formally integrable linear system $R_2 \subset J_2(E)$ defined in operator form by $P{\xi}\equiv {\xi}_{33}=0, Q{\xi}\equiv{\xi}_{23} - {\xi}_{11}=0, R {\xi}\equiv {\xi}_{22}=0$ and an epimorphism $R_2 \rightarrow J_1(E) \rightarrow 0 $. As for the systems, we have $dim(R_2)= 7, dim(R_{r+3})=8,  \forall r\geq 0$. As for the symbols, we have $dim(g_2)= 3, dim(g_3)=1, g_{r+4}=0, \forall r\geq 0$. This finite type system has the very particular feature that $g_3$ is $2$-acyclic but not $3$-acyclic (thus involutive) with the short exact $\delta$-sequence:  \\
\[  3 \times 1 =3 = 1 \times 3 \hspace{1cm} \Rightarrow \hspace{1cm}  0 \rightarrow  {\wedge}^2 T^* \otimes g_3 \stackrel{\delta}{\longrightarrow} {\wedge}^3 T^* \otimes g_2  \rightarrow 0  \]
and we have the three linearly independent equations:  \\
\[ \left\{\begin{array}{rcccl}
{\xi}_{11,123} & = & {\xi}_{111,23} + {\xi}_{112,31} + {\xi}_{113,12} & = & {\xi}_{111,23}  \\
{\xi}_{12,123} & = & {\xi}_{112,23} + {\xi}_{122,31} + {\xi}_{123,12} & = & {\xi}_{111,12}  \\
{\xi}_{13,123} & = & {\xi}_{113,23} + {\xi}_{123,31} + {\xi}_{133,12} & = & {\xi}_{111,31}
\end{array} \right. \]
Collecting these results, we get the two following commutative and exact diagrams:  \\
\[  \begin{array}{rcccccccl}
    &    &  &  0  &  &  0  &  &  0 &   \\
     &   &  & \downarrow &  &  \downarrow  &  & \downarrow  &     \\
  &  0  & \rightarrow & S_4T^*\otimes E & \rightarrow &  S_2T^*\otimes F_0 & \rightarrow &   F_1 &  \rightarrow 0 \\
&  \downarrow &  & \downarrow &  &  \downarrow  &  &  \parallel  &     \\
0 \rightarrow &  R_4 & \rightarrow  & J_4(E) & \rightarrow  &  J_2(F_0) &  \rightarrow &  F_1  &  \rightarrow 0  \\
    &  \downarrow &  & \downarrow &  &  \downarrow  &  & \downarrow   &     \\
0 \rightarrow &  R_3 & \rightarrow  & J_3(E) & \rightarrow  &  J_1(F_0) &  \rightarrow &  0  &   \\ 
   & \downarrow &  & \downarrow &  &  \downarrow  &  &  & \\
   & 0 &  &  0  && 0  && &
\end{array}   \]

\[  \begin{array}{rcccccccl}
    &   &  &  0  &  &  0  &  &  0 &   \\
     &  &  & \downarrow &  &  \downarrow  &  & \downarrow  &     \\
 & 0 & \rightarrow & S_5T^*\otimes E & \rightarrow &  S_3T^*\otimes F_0 & \rightarrow &  T^*\otimes F_1 &  \rightarrow 0 \\
&  \downarrow &  & \downarrow &  &  \downarrow  &  &  \downarrow    &     \\
0 \rightarrow &  R_5 & \rightarrow  & J_5(E) & \rightarrow  &  J_3(F_0) &  \rightarrow &  J_1(F_1)  &  \rightarrow 0  \\
    &  \downarrow &  & \downarrow &  &  \downarrow  &  & \downarrow   &     \\
0 \rightarrow &  R_4 & \rightarrow  & J_4(E) & \rightarrow  &  J_2(F_0) &  \rightarrow &  F_1  &  \rightarrow 0   \\ 
   & \downarrow &  & \downarrow &  &  \downarrow  &  & \downarrow  & \\
   & 0 &  &  0  && 0  && 0&
\end{array}   \]

We obtain from these diagrams $R'_1=J_1(F_0) \Rightarrow g'_1=T^*\otimes F_0, {\rho}_1(R'_1)= J_2(F_0) \Rightarrow R'_2 \subset {\rho}_1(R'_1)$ with a strict inclusion because $27 < 30$ and we have at least $30 - 27 = 3$ generating second order CC. However, from the second diagram, we obtain $dim({\rho}_1(R'_2))= 60 - 12 = 48 =56 - 8 = dim(R'_3)$ and thus $R'_3={\rho}_1(R'_2)$, a result showing that there are no new generating CC of order $3$. \\

As $dim(E)=1$, we have $S_qT^* \otimes E \simeq S_qT^*$ and the commutative diagram of $\delta$-sequences:   \\
\[  \begin{array}{rcccccccl}
  & 0 &  & 0 &  & 0 &  & &    \\
   & \downarrow &  & \downarrow &  &\downarrow  &  &   &  \\
 0 \rightarrow &  S_6 T^* & \rightarrow & T^* \otimes S_5 T^* & \rightarrow & {\wedge}^2 T^* \otimes S_4T^* & \rightarrow  & {\wedge}^3 T^* \otimes S_3 T^* & \rightarrow 0  \\
   & \parallel &  &  \parallel  &  &  \parallel  &  &  \downarrow &   \\
0 \rightarrow & g'_4 &  \rightarrow &  T^* \otimes g'_3 & \rightarrow &  {\wedge}^2T^* \otimes g'_2 & \rightarrow & {\wedge}^3T^* \otimes g'_1 & \rightarrow 0 \\
        &  \downarrow  &  &  \downarrow  &  &  \downarrow  &  & \downarrow    &  \\
   &  0  &  &  0  &  &  0  &  & 0  &
\end{array}  \]
Using the fact that the upper sequence is known to be exact and $ dim(g'_1)=9 < 10 = dim(S_3T^*)$, an easy chase proves that the lower sequence {\it cannot} be exact and thus $g'_2$ {\it cannot} be $2$-acyclic. \\
The generating CC of ${\cal{D}}_1$ is thus a second order operator ${\cal{D}}_2: F_1 \rightarrow F_2$ where $F_2$ is defined by the long exact prolongation sequence:  \\
\[  0 \rightarrow R_6  \rightarrow J_6(E) \rightarrow J_4(F_0) \rightarrow J_2(F_1) \rightarrow F_2 \rightarrow 0   \]
or by the long exact symbol sequence ({\it by chance } if one refers to the previous example !):  \\
\[      0 \rightarrow S_6T^* \otimes E \rightarrow S_4T^* \otimes F_0 \rightarrow S_2T^* \otimes F_1 \rightarrow F_2 \rightarrow 0 \]
showing that $dim(F_2)= dim(S_6T^*) - 3\, dim( S_4T^* ) + 3 \, dim(S_2T^*) = 28 - 45 + 18 = 1   $ in a coherent way with ([18],[19]).  \\
We have thus obtained the following formally exact differential sequence which is nevertheless {\it not} a Janet sequence because $R_2$ is FI but {\it not} involutive as $g_2$ is finite type with $g_4=0$:  \\
\[  0 \rightarrow \Theta \rightarrow E \underset 2{\stackrel{{\cal{D}}}{\longrightarrow}}  F_0 \underset 2 {\stackrel{{\cal{D}}_1}{\longrightarrow}} F_1   \underset 2 {\stackrel{{\cal{D}}_2}{\longrightarrow}} F_2  \rightarrow 0          \]
\[  0 \rightarrow \Theta \rightarrow  1 \longrightarrow  3 \underset 2 \longrightarrow 3   \longrightarrow 1 \rightarrow 0          \]
\[ 0  \rightarrow D \rightarrow D^3 \rightarrow D^3 \rightarrow D \stackrel{p}{\rightarrow}M \rightarrow 0  \]
Surprisingly, the situation is {\it even quite worst} if we start with $R_3 \subset J_3(E)$ which has nevertheless a $2$-acyclic symbol $g_3$ which is {\it not} $3$-acyclic (thus  involutive because $n=3$ ). Indeed, we know from the second section or by repeating the previous procedure for this new third order operator ${\cal{D}}$ that the generating CC are described by a {\it first order} operator ${\cal{D}}_1$. However, the symbol of this operator is only $1$-acyclic but {\it not} $2$-acyclic (exercise). Hence, one can prove that the new CC are described by a new second order operator ${\cal{D}}_2$ which is involutive ... by chance, giving rise to a Janet sequence with first order operators as follows ${\cal{D}}_3, {\cal{D}}_4, {\cal{D}}_5$ ([18], p 123,124):  \\
\[  0 \rightarrow \Theta \rightarrow 1 \underset 3{\stackrel{{\cal{D}}}{\longrightarrow}} 12 \underset 1 {\stackrel{{\cal{D}}_1}{\longrightarrow}}  21 \underset 2 {\stackrel{{\cal{D}}_2}{\longrightarrow}} 46 \underset 1 {\stackrel{{\cal{D}}_3}{\longrightarrow}} 72 \underset 1 {\stackrel{{\cal{D}}_4}{\longrightarrow }}   48
\underset 1 {\stackrel{{\cal{D}}_5}{\longrightarrow}} 12 \rightarrow 0   \]

One could also finally use the involutive system $ R_4 \subset J_4(E)$ in order to construct the canonical Janet sequence and consider the first order involutive system $R_5 \subset J_1(R_4)$ in order to obtain the canonical Spencer sequence with $C_r ={\wedge}^rT^* \otimes R_4$ and dimensions $(8,24,24,8)$:    \\
\[  0 \rightarrow \Theta \stackrel{j_4}{\longrightarrow} C_0 \underset 1 {\stackrel{D_1}{\longrightarrow}} C_1 \underset 1 {\stackrel{D_2}{\longrightarrow}} C_2 \underset 1 {\stackrel{D_3}{\longrightarrow}} C_3 \rightarrow 0  \] \\
To recapitulate, this example clearly proves that the differential sequences obtained largely depend on whether we use $R_2,R_3$ or $R_4$ but also whether we look for a sequence of Janet or Spencer type.    \\
We invite the reader to treat similarly the example ${\xi}_{33} - {\xi}_{11}=0, {\xi}_{23}=0, {\xi}_{22}- {\xi}_{11}=0$.  \\

\noindent
{\bf EXAMPLE 3.3}: In our opinion, the best striking use of acyclicity is the construction of differential sequences for the Killing and conformal Killing operators which are both defined over the ground differential field $K=\mathbb{Q}$ for the Minkowski metric in dimension $4$ or the Euclidean metric in dimension $5$. We have indeed ([18],[20]):   \\
\[   0 \rightarrow \Theta  \rightarrow  4 \underset 1{\stackrel{{\cal{D}}}{\longrightarrow}} 10 \underset 2 {\stackrel{{\cal{D}}_1}{\longrightarrow}} 20 \underset 1 {\stackrel{{\cal{D}}_2}{\longrightarrow}}  20 \underset 1{\stackrel{{\cal{D}}_3}{\longrightarrow}} 6 \rightarrow 0  \] 
with $E=T, F_0=S_2T^*$ and, successively, the Killing, Riemann and Bianchi operators acting on the left of column vectors. The differential module counterpart over $D=K[d]$ is the resolution of the differential Killing module $M$:  \\
\[     0 \rightarrow D^{6} \underset 1{\stackrel{{\cal{D}}_3}{\longrightarrow}} D^{20} \underset 1{\stackrel{{\cal{D}}_2}{\longrightarrow}} D^{20} 
\underset 2 {\stackrel{{\cal{D}}_1}{\longrightarrow}} D^{10} \underset 1{\stackrel{{\cal{D}}}{\longrightarrow} }D^4 \stackrel{p}{\longrightarrow} M \rightarrow 0   \]
with the same operators as before but acting now on the right of row vectors by composition.   \\
The conformal situation  for $n=4$ is quite unexpected with a second order Bianchi-type operator:Ê \\
\[   0 \rightarrow \Theta  \rightarrow  4 \underset 1{\stackrel{{\cal{D}}}{\longrightarrow}} 9 \underset 2 {\stackrel{{\cal{D}}_1}{\longrightarrow}} 10 \underset 2 {\stackrel{{\cal{D}}_2}{\longrightarrow}}  9 \underset 1{\stackrel{{\cal{D}}_3}{\longrightarrow}} 4 \rightarrow 0  \] 
\[     0 \rightarrow D^{4} \underset 1{\stackrel{{\cal{D}}_3}{\longrightarrow}} D^{9} \underset 2{\stackrel{{\cal{D}}_2}{\longrightarrow}} D^{10} 
\underset 2 {\stackrel{{\cal{D}}_1}{\longrightarrow}} D^{9} \underset 1{\stackrel{{\cal{D}}}{\longrightarrow} }D^4 \stackrel{p}{\longrightarrow} M \rightarrow 0   \]
The conformal situation for $n=5$ is even quite different with the conformal differential sequence:  \\
\[   0 \rightarrow \Theta  \rightarrow  5 \underset 1{\stackrel{{\cal{D}}}{\longrightarrow}} 14 \underset 2 {\stackrel{{\cal{D}}_1}{\longrightarrow}} 35 \underset 1 {\stackrel{{\cal{D}}_2}{\longrightarrow}}  35 \underset 2{\stackrel{{\cal{D}}_3}{\longrightarrow}} 14 \underset 1{\stackrel{{\cal{D}}_4}{\longrightarrow}}5 
\rightarrow 0  \] 
Though these results and "jumps" highly depend on acyclicity, in particular the fact that the conformal symbol ${\hat{g}}_2$ is $2$-acyclic for $n=4$ but $3$-acyclic for $n\geq 5$, and have been confirmed by computer algebra, they are still neither known nor acknowledged ([18],[27]).\\

\noindent
{\bf 4) APPLICATIONS}. \\

Considering the {\it classical Killing} operator ${\cal{D}}:\xi \rightarrow {\cal{L}}(\xi)\omega=\Omega \in S_2T^*=F_0$ where ${\cal{L}}(\xi)$ is the Lie derivative with respect to $\xi$ and $\omega \in S_2T^*$ is a nondegenerate metric with $det(\omega)\neq 0$. Accordingly, it is a lie operator with ${\cal{D}}\xi=0, {\cal{D}}\eta=0 \Rightarrow {\cal{D}}[\xi,\eta]=0$ and we denote simply by $\Theta \subset T$ the set of solutions with $[\Theta,\Theta ]\subset \Theta$. Now, as we have explained many times, the main problem is to describe the CC of ${\cal{D}}\xi=\Omega \in F_0$ in the form ${\cal{D}}_1\Omega=0$ by introducing the so-called {\it Riemann} operator ${\cal{D}}_1:F_0 \rightarrow F_1$. We advise the reader to follow closely the next lines and to imagine why it will not be possible to repeat them for studying the {\it conformal Killing} operator. Introducing the well known Levi-Civita isomorphism $j_1(\omega)=(\omega, {\partial}_x \omega) \simeq (\omega , \gamma)$ by defining the Christoffel symbols 
${\gamma}^k_{ij}=\frac{1}{2}{\omega}^{kr}({\partial}_i{\omega}_{rj} + {\partial}_j{\omega}_{ir}-
{\partial}_r{\omega}_{ij})$ where $({\omega}^{rs})$ is the inverse matrix of $({\omega}_{ij})$ and the {\it formal Lie derivative}, we get the second order system $R_2 \subset J_2(T)$:  \\
\[  \left\{  \begin{array}{lcccl}
{\Omega}_{ij}& \equiv &(L({\xi}_1)\omega)_{ij}  & = & {\omega}_{rj}(x){\xi}^r_i + {\omega}_{ir}(x){\xi}^r_j + {\xi}^r {\partial}_r {\omega}_{ij}(x) =0  \\
{\Gamma}^k_{ij}& \equiv & (L({\xi}_2)\gamma)^k_{ij} & =  & {\xi}^k_{ij} + {\gamma}^k_{rj}(x) {\xi}^r_i+{\gamma}^k_{ir}(x) {\xi}^r_j +{\gamma}^k_{ir}(x) {\xi}^r_j - {\gamma}^r_{ij}(x) {\xi}^k_r + {\xi}^r{\partial}_r {\gamma }^k_{ij}(x)=0
\end{array} \right.  \]
with sections ${\xi}_2:x \rightarrow ({\xi}^k(x), {\xi}^k_i(x), {\xi}^k_{ij}(x))$ transforming like $j_2(\xi): x \rightarrow ({\xi}^k(x), {\partial}_i{\xi}^k(x), {\partial}_{ij}{\xi}^k(x))$. The system $R_1 \subset J_1(T)$ has a symbol $g_1 \simeq {\wedge}^2T^*\subset T^* \otimes T$ depending only on $\omega$ with $dim(g_1)=n(n-1)/2$ and is finite type because its first prolongation is $g_2=0$. It cannot be thus involutive and we need to use one additional prolongation. Indeed, using one of the main results to be found in ([8],[10],[11],[18],[19]), we know that, when $R_1$ is FI, then the CC of ${\cal{D}}$ are of order $s+1$ where $s$ is the number of prolongations needed in order to get a $2$-acyclic symbol, that is $s=1$ in the present situation, a result that should lead to CC of order $2$ if $R_1$ were FI. However, it is known that $R_2$ is FI, thus involutive, {\it if and only if} $\omega$ has constant Riemannian curvature, a result first found by L.P. Eisenhart in 1926 which is only a particular example of the {\it Vessiot structure equations} discovered b E. Vessiot in $1903$ ([29]), though in a quite different setting (See [8],[11],[18] and [19] for an explicit modern proof) and should be compared to ([6]).  \\

We may introduce the (formal) linearization $\Gamma \in S_2T^* \otimes T $ of the Christoffel symbols by linearizing the relations $ {\omega}_{kr} {\gamma}^k_{ij}= \frac{1}{2} ( {\partial}_i{\omega}_{rj} + {\partial}_j {\omega}_{ir} - {\partial}_r {\omega}_{ij})$ in such a way that $\Omega = 0 \Rightarrow \Gamma =0$ with:  \\
\[  {\omega}_{kr} {\Gamma}^k_{ij} = \frac{1}{2} ( d_i{\Omega}_{rj} + d_j {\Omega}_{ir} - d_r {\Omega}_{ij}) - {\gamma}^k_{ij}  {\Omega}_{kr}  \]
We may also introduce the Riemann tensor ${\rho}^k_{l,ij}$ and its (formal) linearization:  \\
\[  R^k_{l,ij} \equiv (L({\xi}_1)\rho)^k_{l,ij} = - {\rho}^s_{l,ij}{\xi}^k_s + {\rho}^k_{s,ij}{\xi}^s_l + {\rho}^k_{l,sj}{\xi}^s_i + {\rho}^k_{l,is}{\xi}^s_j 
+ {\xi}^r{\partial}_r{\rho}^k_{l,ij}  \]
in order to obtain the Ricci tensor ${\rho}_{ij} = {\rho}^r_{i,rj}={\rho}_{ji}$ and its linearization:  \\
\[  R_{ij}= R^r_{i,rj}={\rho}_{rj}{\xi}^r_i + {\rho}_{ir} {\xi}^r_j + {\xi}^r{\partial}_r{\rho}_{ij}=R_{ji} \]
allowing to introduce the Einstein tensor ${\epsilon}_{ij}={\rho}_{ij} - \frac{1}{2}{\omega}_{ij}{\omega}^{rs}{\rho}_{rs}= {\rho}_{ij}- \frac{1}{2} {\omega}_{ij}\,  \rho$ with linearization:  \\
\[  E_{ij} \equiv (L({\xi}_1)\epsilon)_{ij}= R_{ij} - \frac{1}{2}{\omega}_{ij}{\omega}^{rs}R_{rs} - \frac{1}{2}{\rho} \,{\Omega}_{ij}+ \frac{1}{2}  {\omega}_{ij}{\omega}^{ru}{\omega}^{sv}{\rho}_{rs}{\Omega}_{uv}  \]
and we must notice ({\it care}) that the linearization of ${\rho}={\omega}^{rs}\, {\rho}_{rs}$ is $ R = {\omega}^{rs}R_{rs} - {\rho}_{rs}{\omega}^{ru}{\omega}^{sv} {\Omega}_{uv}$.  \\
These formulas become particularly simple when $\omega$ is a solution of Einstein equations in vacuum, that is when ${\epsilon}_{ij}=0 \Leftrightarrow{\rho}_{ij}=0 \Rightarrow \rho = 0$. \\

\noindent
{\bf LEMMA 4.1}: When $n=4$ and the fixed eucldean metric for simplicity, we have the useful formula:  \\
\[   E_{00}= - (R_{12,12} + R_{13,13} + R_{23,23})  \]

\noindent
{\it Proof}: We have $2E_{00}= 2R_{00} - (R_{00} + R_{11} + R_{22} + R_{33} )= R_{00} - (R_{11} + R_{22} + R_{33})$  \\
and $ R_{00}= (R_{10,10} + R_{20,20} + R_{30,30}) $. However, we have also: \\
\[  \begin{array}{rcl}
R_{11} &  =  & R_{01,01} + R_{21,21} + R_{31,31}  \\
R_{22} &  =  & R_{02,02} + R_{12,12} + R_{32,32}  \\
R_{33} &  =  & R_{03,03} + R_{13,13} + R_{23,23}
\end{array}  \]
Summing, we obtain $(R_{11} + R_{22} + R_{33})= (R_{01,01} + R_{02,02} + R_{03,03}) + 2 (R_{12,12} + R_{13,13} + R_{23,23})$.
It follows that $E_{00}= - (R_{12,12} + R_{13,13} + R_{23,23})$ and the three other $E_{ii}$ are obtained by circular permutations of $(0,1,2,3)$. We let the reader treat 
the general situation as an exercise. \\
\hspace*{12cm}  $ \Box $ \\

\noindent
{\bf A) MINKOWSKI METRIC}:  \\

We have considered this situation in many books or papers and refer the reader to our arXiv page or to the recent references ([20],[23]). All the operators are first order between the vector bundles $E=T, F_0=T^* \otimes T/g_1\simeq S_2T^*, F_1= H^2(g_1), F_2=H^3(g_1), F_3=H^4(g_1)$ that are only depending on $g_1$ with dimensions $4,10, 20, 20, 6$ when $n=4$ and Euler-Poincar\'{e} characteristic $rk_D(M) = 4 -10 + 20 - 20 + 6 = 0$. The case of an arbitrary $n$, provided in ([25]), depends on various chases in commutative diagrams that will be exhibited later on for comparing the respective dimensions. This is {\it not} a Janet sequence because $R_1$ is FI but $g_1$ is {\it not} involutive.\\

\noindent
{\bf B) SCHWARZSCHILD METRIC}:  \\

With the standard Boyer-Lindquist local coordinates $(x^0=t, x^1=r, x^2=\theta, x^3= \phi) $ and a constant parameter $m$, we may introduce the field of constants $k=\mathbb{Q}(m)$ and all the systems or differential modules considered in the sequel will be defined over the ground differential field $K=k(t,r,sin(\theta),cos(\theta))$ with differential structure obtained by setting  ${\partial}_2sin(\theta)=cos(\theta), {\partial}_2cos(\theta)= - sin(\theta)$ together with ${sin}^2(\theta) + {\cos}^2(\theta)=1$ instead of using the so-called "rational coordinates" ([23]). With speed of light $c=1$ and $A=1 - \frac{m}{r}$, we shall introduce the diagonal Schwarzschild metric $\omega=(A(r),-1/A(r), -r^2, -r^2{sin}^2(\theta))$ with $det(\omega)= - r^4 {\sin}^2(\theta)$. Following closely the motivating examples already presented, our challenge is to prove that the purely mathematical formal study of the corresponding Killing system $R_1 \subset J_1(T)$ can be achieved as a simple exercise of formal integrability, with no extra physical technical tool, contrary to ([1]-[4]). As the computations will be expicitly done, the numbers of CC obtained will bring serious doubts about the validity of the results obtained in the above references, later confirmed with the K metric.  
\noindent
First of all we obtain easily the following $10$ first order Killing equations $mod(\Omega)$: \\
\[ R_1\subset J_1(T) \,\,\,  \left\{ \begin{array}{rcl}
{\Omega}_{33} &  \rightarrow &   \fbox{${\xi}^3_3$} + \frac{1}{r}{\xi}^1 + cot(\theta) {\xi}^2 =0  \\
{\Omega}_{23  } &  \rightarrow   &   \fbox{${\xi}^2_3$} + {sin}^2(\theta) {\xi}^3_2   =0   \\
{\Omega}_{13  } &  \rightarrow  &   \fbox{$ {\xi}^1_3$}  + A r^2 {sin}^2(\theta) {\xi}^3_1  =0  \\
{\Omega}_{03  } &  \rightarrow   & \fbox{${\xi}^0_3$} - \frac{r^2}{A}{\sin}^2(\theta) {\xi}^3_0    =0   \\
{\Omega}_{22} &  \rightarrow  &  \fbox{${\xi}^2_2$}  + \frac{1}{r} {\xi}^1 =0  \\
{\Omega}_{12  } &  \rightarrow & \fbox{$ {\xi}^1_2$} + A r^2{\xi}^2_1  =0   \\
{\Omega}_{02  } &  \rightarrow  & \fbox{$ {\xi}^0_2 $} - \frac{r^2}{A}{\xi}^2_0     =0  \\
{\Omega}_{11}  &  \rightarrow  &  \fbox{${\xi}^1_1$} -  \frac{m}{2Ar^2}{\xi}^1=0   \\
{\Omega}_{01}  &  \rightarrow  &  \fbox{${\xi}^1_0$} -  A^2 {\xi}^0_1 =0  \\
{\Omega}_{00} &  \rightarrow  &  \fbox{${\xi}^0_0 $}  + \frac{m}{2Ar^2}{\xi}^1=0
\end{array} \right.  \]
where we have framed the leading jets. \\
This is a finite type system because we get ${\Gamma}^k_{ij}\equiv {\xi}^k_{ij} + ... =0$ with {\it only one prolongation} !. \\
The only $9$ non-zero Christoffel symbols on $40$ are:  \\
\[  \fbox { $ \begin{array}{lll}
   {\gamma}^1_{00}= \,\,\frac{m A}{2r^2}, & \,\, {\gamma}^0_{01}= \,\, \frac{m}{2 Ar^2} , & \,\, {\gamma}^1_{11}= - \frac{m}{2 A r^2}  \\           
   {\gamma}^2_{12} = \frac{1}{r} , &  \,\, {\gamma}^3_{13}= \frac{1}{r}, & \,\, {\gamma}^1_{22}= - A r, \\
   {\gamma}^3_{23}= cot(\theta) , & \,\, {\gamma}^1_{33}= - Ar {sin}^2(\theta)° & \,\, {\gamma}^2_{33}=  - sin(\theta)cos(\theta) 
   \end{array} $  }  \]
 We obtain for example: \\
 \[ {\Gamma}^1_{22}\equiv {\xi}^1_{22} + (1- \frac{3m}{2r}){\xi}^1=0, \hspace{5mm} 
  {\Gamma}^1_{33}\equiv {\xi}^1_{33} + sin(\theta)cos(\theta) {\xi}^1_2 + (1- \frac{3m}{2r}){sin}^2(\theta) {\xi}^1=0  \]
 \[ \Rightarrow  \hspace{1cm} {\Gamma}^1_{33} - {sin}^2(\theta){\Gamma}^1_{22}\equiv {\xi}^1_{33} + sin(\theta)cos(\theta) {\xi}^1_2  
 - {sin}^2(\theta) {\xi}^1_{22} = 0  \]
 {\it after only one prolongation} (care). \\
Then, using $r$ as a summation index, we shall see that we have {\it in general} for the linearization of the Riemann and Ricci tensors:  \\
\[  R_{kl,ij}\equiv {\rho}_{rl,ij}{\xi}^r_k +{\rho}_{kr,ij}{\xi}^r_l+{\rho}_{kl,rj}{\xi}^r_i + {\rho}_{kl,ir}{\xi}^r_j + {\xi}^r {\partial}_r 
{\rho}_{kl,ij}\neq 0  \]
\[   R_{ij}\equiv {\rho}_{rj}{\xi}^r_i + {\rho}_{ir}{\xi}^r_j +{\xi}^r{\partial}_r{\rho}_{ij} \neq 0 \]
The only $6$ non-zero components of the Riemann tensor are:  \\
\[ \fbox {  $   \begin{array}{lll}
  {\rho}_{01,01}= + \frac{m}{r^3},& \,\,{\rho}_{02,02}= - \frac{m\,A}{2r}, &\,\,{\rho}_{03,03}= - \frac{m\,A\,sin^2(\theta)}{2r}  \\
  {\rho}_{12,12}= + \frac{m}{2Ar},& \,\,{\rho}_{13,13}= + \frac{m\,sin^2(\theta)}{2Ar},& \,\, {\rho}_{23,23}= - m\, r \, sin^2(\theta)  
\end{array} $ }    \]
but we must not forget hat we have indeed ${\rho}_{ij}=0$ for the $10$ components of the Ricci tensor, in particular ${\rho}_{ii}=0$ for the diagonal components with $i=0,1,2,3$. We have in particuler:  \\
\[  {\rho}_{22}= {\rho}^0_{2,02} + {\rho}^1_{2,12} + {\rho}^3_{2,32}= \frac{1}{A}{\rho}_{02,02} - A {\rho}_{12,12} - \frac{1}{r^2 {sin}^2(\theta)}{\rho}_{23,23}=
- \frac{m}{2r} - \frac{m}{2r} + \frac{m}{r} = 0  \] 
We also obtain $mod(\Omega)$:  \\
\[  R_{01,01}\equiv 2{\rho}_{01,01} ({\xi}^0_0 + {\xi}^1_1) + {\xi}^r{\partial}_r({\rho}_{01,01})={\xi}^1{\partial}_1{\rho}_{01,01}\Rightarrow 
R_{01,01}\equiv - \frac{3m}{r^4}{\xi}^1 = 0   \Rightarrow {\xi}^1=0 \]
and similarly:  \\
\[  R_{01,02} \equiv {\rho}_{01,01}{\xi}^1_2 + {\rho}_{02,02}{\xi}^2_1 +{\xi}^r{\partial}_r{\rho}_{01,02}= \frac{3m}{2r^3}{\xi}^1_2=0 \Rightarrow {\xi}^1_2=0 \]
\[  R_{02,12}\equiv {\rho}_{12,12}{\xi}^1_0+{\rho}_{02,02}{\xi}^0_1+{\xi}^r{\partial}_r{\rho}_{02,12}= \frac{m}{2rA} ({\xi}^1_0 - A^2 {\xi}^0_1)=0 \]
\[ \fbox { $ \begin{array}{rcl}
R_{23,12} & \equiv  & {\rho}_{21,12}{\xi}^1_3 + {\rho}_{23,32}{\xi}^3_1  + {\xi}^r{\partial}_r{\rho}_{23,12}  \\
                & = &  -  \frac{m}{2rA}{\xi}^1_3 +  mrsin^2(\theta){\xi}^3_1  \\
                & = & - \frac{3m}{2rA}{\xi}^1_3=0
\end{array} $ } \]
\[ \Rightarrow \,\,\,  R_{01,03}= \frac{3m}{2r^3}{\xi}^1_3 = - \frac{A}{r^2}R_{23,12}     \]
\[   {\rho}_{01,23}=0 \Rightarrow  R_{01,23}= {\rho}_{01,23} ({\xi}^0_0 + {\xi}^1_1 + {\xi}^2_2 + {\xi}^3_3) + {\xi}^r{\partial}_r{\rho}_{01,23}=0 \]
and so on, in order to avoid using computer algebra. However, the main consequence of this remark is to explain the existence of the $15$ second order CC. Indeed, denoting by "$\sim$" a linear proportional dependence $mod(\Omega)$, we have the successive three cases:  \\
\[ \fbox {  $ \begin{array}{cc} 
  &   \\ 
  (R_{00},R_{11},R_{22},R_{33}) & R_{01,01} \sim R_{02,02}\sim R_{03,03} \sim R_{12,12} \sim R_{13,13} \sim R_{23,23} \rightarrow {\xi}^1=0 \\
    &    \\  
  (R_{12} ) &  R_{01,02} \sim R_{13,23} \rightarrow {\xi}^1_2=0  \\
  (R_{13} ) &  R_{01,03} \sim R_{12,23} \rightarrow {\xi}^1_3=0  \\
  (R_{02} ) &   R_{01,12} \sim R_{03,23} \rightarrow {\xi}^0_2=0  \\
  (R_{03} ) &  R_{01,13} \sim  R_{02,23}  \rightarrow {\xi}^0_3=0  \\
   &   \\
  (R_{01},R_{23}) &   R_{01,23} \rightarrow 0, R_{02,03} \rightarrow 0, R_{02,12} \rightarrow 0, R_{02,13} \rightarrow 0, R_{03,13} \rightarrow  0,  R_{12,13} \rightarrow 0  \\
   &
 \end{array} $  }  \]
as a way to obtain the $5$ equalities to zero on the right and thus a total of $20-5=15$ second order CC obtained by elimination. However, the present partition $15= 5 + 4 + 6$ is quite different from the partition $15=10 + 5$ used by the authors quoted in the Introduction which is obtained by taking into account the vanishing assumption of the $10$ components of the Ricci tensor. As such a result questions once more the mathematical foundations of general relativity, in particular the existence of gravitational waves, we provide a few additional technical comments.\\

The main point is a tricky formula which is not evident at all. Indeed, using the well known properties of the Lie derivative, we have the following geometric objects ({\it not necessarily tensors }) and their linearizations ({\it generally tensors}):  \\
\[  {\omega}_{ij} \rightarrow {\Omega}_{ij} \in S_2T^*, \,\, {\gamma}^k_{ij} \rightarrow {\Gamma}^k_{ij}\in S_2T^* \otimes T, \]
\[   {\rho}_{kl,ij} \rightarrow R_{kl,ij}\in {\wedge}^2T^* \otimes T^* \otimes T, \,\, {\beta}_{kl,ijr} \rightarrow B_{kl,ijr} \in {\wedge}^3T^* \otimes T^* \otimes T \]
Then, using $r$ as a summation index, we shall see that we have {\it in general}:  \\
\[  R_{kl,ij}\equiv {\rho}_{rl,ij}{\xi}^r_k +{\rho}_{kr,ij}{\xi}^r_l+{\rho}_{kl,rj}{\xi}^r_i + {\rho}_{kl,ir}{\xi}^r_j + {\xi}^r {\partial}_r 
{\rho}_{kl,ij}\neq 0  \]
\[  {\rho}_{kl,ij}={\omega}_{kr}{\rho}^r_{l,ij} \,\,\,  \Rightarrow   \,\,\, R_{kl,ij}= {\omega}_{kr}R^r_{l,ij} + {\rho}^r_{l,ij} {\Omega}_{kr}\,\, \Rightarrow \,\, {\omega}^{rs}R_{ri,sj} = R_{ij} + {\omega}^{rs}{\rho}^t_{i,rj}{\Omega}_{st}\] 
\[  {\rho}_{ij}={\rho}^r_{i,rj}\,\,\,  \Rightarrow  \,\,\,  R_{ij} = R^r_{i,rj}\neq {\omega}^{rs} R_{ri,sj}  \]
We prove these results using local coordinates and the formal Lie derivative obtained while replacing $j_1(\xi)$ by ${\xi}_1$ (See [8],[11],[18],[19] for details). First of all, from the tensorial property of the Riemann tensor and the Killing equations ${\Omega}_{us}= {\omega}_{ku}{\xi}^k_s + {\omega}_{ks}{\xi}^k_u + {\xi}^r {\partial}_r{\omega}_{us}$, we have:  \\
\[  R^k_{l,ij}\equiv (L({\xi}_1)\rho)^k_{l,ij}= - {\rho}^s_{l,ij}{\xi}^k_s + {\rho}^k_{s,ij}{\xi}^s_l + {\rho}^k_{l,sj}{\xi}^s_i + {\rho}^k_{l,is}{\xi}^s_j 
+ {\xi}^r{\partial}_r{\rho}^k_{l,ij}  \]
\[  \begin{array}{rcl}
{\omega}_{ku}( - {\rho}^s_{v,ij} {\xi}^k_s + {\xi}^r{\partial}_r{\rho}^s_{v,ij}) & =  & {\rho}^s_{v,ij}{\omega}_{ks}{\xi}^k_u + ({\xi}^r{\partial}_r{\omega}_{us}){\rho^s_{v,ij}  + {\omega}_{ku}{\xi}^r  {\partial}_r{\rho}^k_{v,ij}} -{\rho}^s_{v,ij}{\Omega}_{su} \\
  &  =  &  {\rho}_{sv,ij}{\xi}^s_u + {\xi}^r{\partial}_r  {\rho}_{uv,ij}  -{\rho}^s_{v,ij}{\Omega}_{su}                       
\end{array}  \]
and thus ${\omega}_{ku}R^k_{v,ij}= R_{uv,ij} - {\rho}^s_{v,ij}{\Omega}_{su}$.
\[  {\rho}^1_{1,11}= 0 \,\, \Rightarrow {\rho}_{11}= {\rho}^0_{1,01}+ {\rho}^2_{1,21} + {\rho}^3_{1,31}= \frac{1}{A}{\rho}_{01,01} - \frac{1}{r^2}{\rho}_{12,12} - \frac{1}{r^2sin^2(\theta)} {\rho}_{13,13} =0  \]
We have for example, {\it in this particular case}:   \\
\[ {\rho}^0_{1,01}= \frac{1}{A} {\rho}_{01,01}=\frac{m}{Ar^3} \Rightarrow R^0_{1,01} = ({\cal{L}}(\xi)\rho)^0_{1,01}=2 {\rho}^0_{1,01}{\xi}^1_1+ {\xi}^1 {\partial}_1{\rho}^0_{1,01} = \frac{2m}{Ar^3}{\xi}^1_1+ {\xi}^1{\partial}_1( \frac{m}{Ar^3})  \]
The only use of $R_{01,01}$ is allowing to get ${\xi}_1=0$ in the previous list, but we have also {\it exactly}: \\
\[  {\omega}^{rs}R_{r1,s2}= \frac{1}{A}R_{01,02} - \frac{1}{r^2sin^2(\theta)} R_{31,32}=- \frac{m}{2r^3}{\Omega}_{12}=R_{12}+{\omega}^{11}{\rho}^2_{1,12}{\Omega}_{12} \, \Rightarrow  \, R_{12}=0       \]
The use of  $R_{01,02}$ or $R_{13,23}$ is allowing to get ${\xi}^1_2=0$ in the previous list with:   \\
\[    R^0_{1,02}  =   - \frac{3m}{2r^3}{\xi}_{1,2} + \frac{m}{2r^3}{\Omega}_{12}, \,\,  \,\, 
R^3_{1,32}  =   + \frac{3m}{2r^3} {\xi}_{1,2} -  \frac{m}{r^3}{\Omega}_{12}  \]    
and thus also {\it exactly}:   \\
\[ {\omega}^{rs}{\rho}^t_{1,r2}{\Omega}_{st}= -{\omega}^{11}{\omega}^{22}{\rho}_{12,12}{\Omega}_{12}= - \frac{m}{2r^3}{\Omega}_{12}  \,\,\,  \Rightarrow  \,\,\, R_{12}= R^0_{1,02} + R^3_{1,32}   +  \frac{m}{2r^3} = 0 \].  \\       
It follows that the $4$ central second order CC of the list successively amounts to $R_{12}=0, R_{13}=0,R_{02}=0, R_{03}=0$, a result breaking the intrinsic/coordinate-free interpretation of the $10$ Einstein equations and the situation is even worst for the other components of the Ricci tensor. Indeed, $R_{01}$ and $R_{23}$ only depend on the vanishing of $R_{02,12}, R_{03,13}$ and $R_{02,03}, R_{12,13}$ among the bottom CC of the list, while the diagonal terms
$R_{00}, R_{11}, R_{22}, R_{33}$ only depend, {\it as we just saw}, on the $6$ non zero components of the Riemann tensor. We have thus obtained the totally unusual partition $10= 4 + 4 + 2$ along the successive blocks of the former list with:  \\
\[   \{ R_{ij}\}=\{ R_{00},R_{11},R_{22},R_{33}\}+ \{ R_{12},R_{13},R_{02},R_{03}\} + \{R_{01},R_{23}\}  \]  

Finally, we notice that $R_{01,23}=0,R_{02,31}=0\Rightarrow R_{03,12}=0$ from the identity in ${\wedge}^3 T^* \otimes T^*$: \\
\[    R_{01,23} + R_{02,31} + R_{03,12}=0  \]
and there is no way to have two identical indices in the first jets appearing through the (formal) Lie derivative just described. As for the third order CC, setting 
${\xi}^1_1= \frac{A'}{2A}{\xi}^1\in j_2(\Omega)$, we have at least the first prolongations of the previous second order CC to which we have to add the three new generating ones:     \\
\[ \fbox { $  d_1{\xi}^1 - {\xi}^1_1=0,  \,\, d_2{\xi}^1-{\xi}^1_2=0,\,\,  d_3{\xi}^1-{\xi}^1_3=0   $ } \]
provided by the Spencer operator, leading to the crossed terms $d_i{\xi}^1_j - d_j{\xi}^1_i=0$ for $ i,j=1,2,3$ because the Spencer operator is not FI. \\

Setting now ${\xi}^1 = U, {\xi}^1_2=V_2, {\xi}^1_3=V_3, {\xi}^0_2=W_2, {\xi}^0_3=W_3$ with $(U,V,W) \in j_2(\Omega)$, we have to look for the CC of the system $R^{(1)}_1=R_1$ already presented, then the system $R^{(2)}_1$ with $dim(R^{(2)}_1)= 5$ and finally $R^{(3)}_1$ with $dim(R^{(3)}_1)=4$ which is formally integrable but not involutive because it is of finite type. 
Beside the only zero order equation ${\xi}^1=U$, we have the following $15$ first order ones:  \\ 
  \[  \fbox {  $  \begin{array}{c}
              \\
  {\xi}^0_0= - \frac{A'}{2A} U, \,\,\, {\xi}^0_1= \frac{1}{A^2}d_0U,  \,\, \,{\xi}^0_2=W_2, \,\,\, {\xi} ^0_3= W_3, \\
          \\
   {\xi}^1_0=d_0U, \,\,\, {\xi}^1_1=\frac{A'}{2A}U, \,\, \,{\xi}^1_2=V_2,  \,\,\, {\xi}^1_3= V_3,  \\
         \\
{\xi}^2_0=\frac{A}{r^2} W_2, \,\,\, {\xi}^2_1= - \frac{1}{Ar^2}V_2, \,\,\, {\xi}^3_0 =\frac{A}{r^2{sin}^2(\theta)}W_3, \,\,\,
{\xi}^3_1= - \frac{1}{Ar^2{sin}^2(\theta)}V_3, \\
   \\
   {\xi}^2_2= - \frac{1}{r}U, \,\,\, {\xi}^2_3 + {sin}^2(\theta){\xi}^3_2=0, \,\,\,  {\xi}^3_3 + cot(\theta) {\xi}^2= - \frac{1}{r} U  \\
   {    }
\end{array}  $  }    \]
Among the CC we {\it must} have $d_2V_3 - d_3V_2=0$ which is among the differential consequences of the Spencer operator as we saw but we {\it must} also have 
$d_2W_3 - d_3W_2=0$ and both seem to be new third order CC, together with the CC obtained by eliminating ${\xi}^2$ and ${\xi}^3$ from the three last equations {\it after two prolongations} as in ([23]):  \\
\[ d_3V_3 + sin(\theta)cos(\theta) V_2 + {sin}^2(\theta)d_2V_2 + 2 {sin}^2(\theta) U=0  \] 
However, {\it things are not so simple}, even if we have in mind that $(V,W) \in j_2(\Omega)$, because the central sign in the previous formula is opposite to the sign found {\it after one prolongation} in the formula:  
\[        {\xi}^1_{33}+ sin(\theta)cos(\theta) {\xi}^1_2 - {sin}^2(\theta){\xi}^1_{22}=0   \]
and it is at this moment that we need introduce new differential geometric methods !.\\

First of all, we have:  \\
\[  {\rho}^k_{l,ij}= {\partial}_i{\gamma}^k_{lj} - {\partial}_j{\gamma}^k_{li} + 
{\gamma}^r_{lj} {\gamma}^k_{ri} - {\gamma}^r_{li} {\gamma}^k_{rj}   \]
and thus, because $\Gamma \in S_2T^*\otimes T$ is a tensor::   \\
\[ \begin{array}{rcl}
R^k_{l,ij}& = & d_i{\Gamma}^k_{lj} - d_j{\Gamma}^k_{li} + 
{\gamma}^r_{lj} {\Gamma}^k_{ri} - {\gamma}^r_{li} {\Gamma}^k_{rj}   +
{\gamma}^k_{ri} {\Gamma}^r_{lj} - {\gamma}^k_{rj} {\Gamma}^r_{li}   \\
& = &  (d_i{\Gamma}^k_{lj} - {\gamma}^r_{li} {\Gamma}^k_{rj}   + {\gamma}^k_{ri} {\Gamma}^r_{lj} ) - ( d_j{\Gamma}^k_{li} - {\gamma}^r_{lj} {\Gamma}^k_{ri}   +{\gamma}^k_{rj} {\Gamma}^r_{li})  \\
& = &(d_i{\Gamma}^k_{lj} - {\gamma}^r_{li} {\Gamma}^k_{rj}  - {\gamma}^r_{ji} {\Gamma}^k_{lr} + {\gamma}^k_{ri} {\Gamma}^r_{lj} ) - 
( d_j{\Gamma}^k_{li} - {\gamma}^r_{lj} {\Gamma}^k_{ri} - {\gamma}^r_{ij} {\Gamma}^k_{lr}  + {\gamma}^k_{rj} {\Gamma}^r_{li})  \\
 & = &  {\nabla}_i {\Gamma}^k_{lj} - {\nabla}_j {\Gamma}^k_{li} 
\end{array}   \]
by introducing the covariant derivative $\nabla$. We recall that ${\nabla}_r{\omega}_{ij}=0,\forall r,i,j$ or, equivalently, that $(id, -\gamma): \xi \in T \rightarrow ({\xi}^k, {\xi}^k_i= - {\gamma}^k_{ir}{\xi}^r) \in R_1$ is a $R_1$-connection with ${\omega}_{sj}{\gamma}^s_{ir} + {\omega}_{is} {\gamma}^s_{jr}={\partial}_r{\omega}_{ij}$, a result allowing to move down the index $k$ in the previous formulas (See [18] for more details). \\
We may thus take into account the Bianchi identities implied by the cyclic sums on $(ijr)$   \\
\[  {\beta}_{kl,ijr}\equiv {\nabla}_r {\rho}_{kl,ij} + {\nabla}_i{\rho}_{kl,jr} + {\nabla}_j{\rho}_{kl,ri}=0 \hspace{1cm} \Leftrightarrow \hspace{1cm} 
\beta \equiv \underset {cycl}{\Sigma} ( \partial \rho - \gamma \rho)=0\]
and their respective linearizations $B_{kl,ijr}=0$ as described below. We shall see later on that $\beta$ and $B$ are sections of the vector bundle $F_2$ defined by the short exact sequence:  \\
\[   0 \rightarrow F_2 \rightarrow {\wedge}^3T^* \otimes g_1 \stackrel{\delta}{\longrightarrow} {\wedge}^4T^* \otimes T \rightarrow 0 \] 
with $dim(F_2)= (n(n-1)(n-2)/6)(n(n-1)/2) - (n(n-1)(n-2)(n-3)/24)n= n^2(n^2-1)(n-2)/24$ because $dim(g_1)=n(n-1)/2$ for {\it any} nondegenerate metric, that is $24 - 4=20$ when $n=4$. \\
{\it Such results cannot be even imagined by somebody not aware of the $\delta$-acyclicity} ([10],[11],[18]). \\
We have the linearized cyclic sums of covariant derivatives both with their respective symbolic descriptions, not to be confused with the non-linear corresponding ones:  \\
\[  \begin{array}{rcl}
B_{kl,rij}\equiv  {\nabla}_rR_{kl,ij} + {\nabla}_iR_{kl,jr} + {\nabla}_jR_{kl,ri}= 0 \hspace{1mm}  mod(\Gamma)  & \Leftrightarrow &
\underset{cycl}{\Sigma} ( dR - \gamma R  - \rho \Gamma) =0 \\
    & \Leftrightarrow & B \equiv \underset{cycl}{\Sigma}(\nabla R) =\underset{cycl}{\Sigma}(\rho \Gamma)
    \end{array}  \]
    
In order to recapitulate these new concepts obtained after one, two or three prolongations, we have successively $ \omega \longrightarrow \gamma \longrightarrow \rho \longrightarrow \beta $  and the respective linearizations $\Omega \longrightarrow \Gamma \longrightarrow R \longrightarrow B $.  \\

The 24 Bianchi identities are related by the $4$ linear relations like $B_{01,023} - B_{02,013} + B_{03,012}=0$ when $n=4$ because $B_{00,123}=0$. These relations are existinging between the $24$ components of the Lanczos tensor because $B \in F_2 \subset ker(\delta)$ in the previous short exact sequence ([25]). \\

With more details, we number the $20$ linearly independent Bianchi identities as follows:  \\
\[ \fbox{1}\, B_{01,012},\fbox{2}\, B_{01,013}, \]
\[\fbox{3} \, B_{02,123}, \fbox{4} \, B_{02,012}, \fbox{5}\, B_{02,013}, \fbox{6} \, B_{02,023}, \]
\[   \fbox{7} \, B_{03,123}, \fbox{8} \, B_{03, 012}, \fbox{9} \, B_{03,013}, \fbox{10} \, B_{03,023}, \]
\[\fbox{11} \, B_{12,123}, \fbox{12} \, B_{12,012}, \fbox{13}\, B_{12,013}, \fbox{14} \, B_{12,023}, \]
\[\fbox{15} \, B_{13,123}, \fbox{16} \, B_{13,012}, \fbox{17}\, B_{13,013}, \fbox{18} \, B_{13,023}, \]
\[   \fbox{19} \, B_{23,123}, \fbox{20} \, B_{23, 023}\]
to which we add the $4$ linearly dependent:  \\
\[ \fbox{21} \, B_{01,123}, \fbox{22} \, B_{01,023}, \fbox{23}\, B_{23,012}, \fbox{24} \, B_{23,013}  \]
We successively study a few situations without any, with one or with two vanishing linearized Riemann components, taking into account that the four Einstein equations are described by:  \\
 \fbox{$12$},\fbox{$17$},\fbox{$20$} for the index $0$, \fbox{$4$},\fbox{$9$}, \fbox{$19$} for the index $1$, \fbox{$1$},\fbox{$10$}, \fbox{$15$} for the index $2$, \fbox{$2$},\fbox{$6$},\fbox{$11$} for the index $3$.  \\
 
\noindent
$BIANCHI \, \fbox{1}$\, : 
\fbox{ $ B_{01,012} \equiv {\nabla}_0 R_{01,12} + {\nabla}_1 R_{01,20} + {\nabla}_2 R_{01,01}= - \frac{3m}{2r^3} ({\Gamma}^0_{02} + {\Gamma}^1_{12}) $ } . \\  \\ 
First of all, we have $R_{01,12} = - \frac{3m}{2r^3}{\xi}^0_2, R_{01,02} = \frac{3m}{2r^3} {\xi}^1_2, R_{01,01} = - \frac{3m}{r^4}{\xi}^1$ and obtain:  \\
\[ \begin{array}{rcl}
{\nabla}_0 R_{01,12} & = & - \frac{3m}{2r^3} {\xi}^0_{02} - {\gamma}^0_{01}R_{01,02} \\
                                 & = & - \frac{3m}{2r^3} ({\xi}^0_{02} + \frac{m}{2Ar^2} {\xi}^1_2)   \\
                                 &= &  - \frac{3m}{2r^3} {\Gamma}^0_{02} \\
{\nabla}_1 R_{01,20} & = & ( - \frac{3m}{2r^3} {\xi}^1_{12} +\frac{9m}{2r^4}{\xi}^1_2) - (2 {\gamma}^0_{01} + {\gamma}^1_{11} + {\gamma}^2_{12})R_{01,20}  \\
                                 & = &  - \frac{3m}{2r^3} ({\xi}^1_{12} - \frac{m}{2Ar^2}{\xi}^1_2) + \frac{6m}{r^4} {\xi}^1_2   \\
                                 & = &  - \frac{3m}{2r^3}{\Gamma}^1_{12} + \frac{6m}{r^4}{\xi}^1_2  \\
{\nabla}_2 R_{01,01} & = &  - \frac{3m}{r^4} {\xi}^1_2 - 2 {\gamma}^2_{12} R_{02,01} \\
                                 & = &  - \frac{3m}{r^4} {\xi}^1_2  - \frac{2}{r}\frac{3m}{2r^3} {\xi}^1_2   \\
                                 & = &  - \frac{6m}{r^4} {\xi}^1_2
 \end{array}   \]
and we notice that ${\Gamma}^0_{02} + {\Gamma}^1_{12}={\xi}^0_{02} + {\xi}^1_{12}= d_2 ({\xi}^0_0 + {\xi}^1_1)=0 \Rightarrow B_{01,012}=0  $. As a byproduct, we have $d_0W_2 + \frac{m}{2Ar^2}V_2=0, d_1V_2 - \frac{m}{2Ar^2}V_2=0 \Rightarrow d_1V_2 + d_0W_2=0$. \\

\noindent
$BIANCHI\,  \fbox {3}$ \, : 
\fbox{ $ B_{02,123} \equiv {\nabla}_1 R_{02,23} + {\nabla}_2 R_{02,31} + {\nabla}_3 R_{02,12} = \frac{3mA}{2r} {\Gamma}^0_{13} $ } . \\   \\
First of all, we have $R_{02,23}= \frac{3mA}{2r} {\xi}^0_3, R_{02,31}=0, R_{02,12}= 0, R_{01,31}= \frac{3m}{2r^3} {\xi}^0_3$ and obtain:  \\
\[ \begin{array}{rcl}
{\nabla}_1R_{02,23} & = & d_1 R_{02,23} -{\gamma}^r_{01}R_{r2,23} - {\gamma}^r_{12} R_{0r,23} - {\gamma}^r_{12} R_{02,r3}
-{\gamma}^r_{13}R_{02,2r} \\
  & = & \frac{3mA}{2r}{\xi}^0_{13} + ( \frac{3m^2}{2r^3}  - \frac{3mA}{2r^2} ) {\xi}^0_3 -({\gamma}^0_{01} - \frac{3}{r}) R_{02,23} \\
   & = & \frac{3mA}{2r} {\xi}^0_{13} - (\frac{12m}{2r^2} - \frac{27 m^2}{4 r^3}) {\xi}^0_3  \\
   & = & \frac{3mA}{2r} {\Gamma}^0_{13} - \frac{9mA}{2r^2}{\xi}^0_3   \\
{\nabla}_2 R_{02,31}&= & 0 - {\gamma}^r_{02} R_{r2,31} - {\gamma}^r_{22}R_{0r,31} - 
{\gamma}^r_{23} R_{02,r1} - {\gamma}^r_{12} R_{02,3r} \\
& = & Ar R_{01,31} - cot(\theta) R_{02,31} - \frac{1}{r} R_{02,32} \\
 &= & \frac{3mA}{2r^2}  {\xi}^0_3 + \frac{3mA}{2r^2} {\xi}^0_3 \\
 & = & \frac{3mA}{r^2} {\xi}^0_3   \\ 
{\nabla}_3 R_{02,12} & = & 0 - {\gamma}^r_{03} R_{r2,12} - {\gamma}^r_{23} R_{0r,12} - {\gamma}^r_{13} R_{02,r2}
 - {\gamma}^r_{23} R_{02,1r} \\
 & = & - {\gamma}^3_{23} R_{03,12} - {\gamma}^3_{13} R_{02,32} - {\gamma}^3_{23} R_{02,13}   \\
  & = & \frac{3mA}{2r^2} {\xi}^0_3
  \end{array} \]
and we may use the fact that ${\Gamma}^0_{13}\equiv {\xi}^0_{13} - \frac{1}{Ar}(1- \frac{3m}{2r}){\xi}^0_3 =0 \Rightarrow d_1W_3 - \frac{1}{Ar}(1 - \frac{3m}{2r})W_3=0$. \\

\noindent
$BIANCHI \,  \fbox{4}$ \, : 
\fbox{ $ B_{02,012}\equiv {\nabla}_0 R_{02,12} + {\nabla}_1 R_{02,20} + {\nabla}_2 R_{02,01} = \frac{3m}{2r^3} {\Gamma}^1_{22} $ }.  \\  \\
First of all we have $R_{02,12}=0, R_{02,02}=\frac{3mA}{2r^2}{\xi}^1, R_{02,01}=\frac{3m}{2r^3}{\xi}^1_2,R_{12,12}= - \frac{3m}{2Ar^2} {\xi}^1$ and obtain: \\
\[  \begin{array}{rcl}
{\nabla}_0R_{02,12} & = & d_0 R_{02,12} - {\gamma}^r_{00}R_{r2,12}  -{\gamma}^r_{02} R_{0r,12} -  {\gamma}^r_{01}R_{02,r2} - {\gamma}^r_{02} R_{02,1r} \\
                                &  = &  0  - {\gamma}^1_{00}R_{12,12} - {\gamma}^0_{01} R_{02,02} \\
                                &  = &  (- \frac{mA}{2r^2})(- \frac{3m}{2Ar^2}){\xi}^1 - (\frac{m}{2Ar^2})(\frac{3mA}{2r^2}){\xi}^1  \\
                                &  =  &  0   \\
{\nabla}_1R_{02,20} & =  & d_1(- \frac{3mA}{2r^2}{\xi}^1) - 2{\gamma}^r_{01}R_{r2,20} - 2{\gamma}^r_{12} R_{0r,20}  \\
                               &   =  &   - \frac{3mA}{2r^2}{\xi}^1_1 - {\partial}_1(\frac{3mA}{2r^2}) {\xi}^1 + 2 {\gamma}^0_{01} R_{02,02}  + 
                                               2 {\gamma}^2_{12} R_{02,02}  \\
                               & = & ( - \frac{3mA}{2r^2})(\frac{m}{2Ar^2} ){\xi}^1 - {\partial}_1(\frac{3mA}{2r^2} ){\xi}^1 + \frac{3m^2}{2r^4}{\xi}^1 +
                                                              \frac{3mA}{r^3}{\xi}^1  \\
                               & = & - \frac{3m^2}{4r^4} {\xi}^1 + (\frac{3m}{r^3} - \frac{3m^2}{r^4} + \frac{3m}{r^3} - \frac{3m^2}{r^4}){\xi}^1  \\
                               & = & \frac{6m}{r^3}{\xi}^1 - \frac{27m^2}{4r^4} {\xi}^1  \\
{\nabla}_2 R _{02,01} & = & \frac{3m}{2r^3} {\xi}^1_{22} - {\gamma}^r_{02}R_{r2,01} - {\gamma}^r_{22}R_{0r,01} - {\gamma}^r_{02} R_{02,r1} - 
{\gamma}^r_{12} R_{02,0r}  \\
                               & = & \frac{3m}{2r^3} {\xi}^1_{22} - {\gamma}^1_{22} R_{01,01} - {\gamma}^2_{12} R_{02,02}  \\
                               & = & \frac{3m}{2r^3} {\xi}^1_{22} - \frac{9mA}{2r^3} {\xi}^1  \\
                               & = &  \frac{3m}{2r^3}{\Gamma}^1_{22} - \frac{6m}{r^3} {\xi}^1  + \frac{27m^2}{4r^4} {\xi}^1
\end{array}  \]
This delicate checking proves that $B_{02,012}\equiv \frac{3m}{2r^3}{\Gamma}^1_{22}=0 \Rightarrow d_2V_2 + (1- \frac{3m}{2r}) U=0$  is a differential consequence of $R_{02,12}=0$ . We let the reader prove as an exercise that $B_{03,013}\equiv  \frac{3m}{2r^3} {\Gamma}^1_{33}=0 \Rightarrow 
d_3V_3 + sin(\theta)cos(\theta) V_2 + (1 - \frac{3m}{2r}) {sin}^2(\theta) U=0$ in order to recover $\fbox{$19$}$ by eliminating $U$.  \\

\noindent
$BIANCHI\,  \fbox {6}$ \, : 
\fbox{ $ B_{02,023} \equiv {\nabla}_0 R_{02,23} + {\nabla}_2 R_{02,30} + {\nabla}_3 R_{02,02} = \frac{3mA}{2r} {\Gamma}^0_{03} $ } . \\   \\
First of all, we have $R_{02,23}= \frac{3mA}{2r} {\xi}^0_3, R_{02,30}=0, R_{02,02}= \frac{3mA}{2r^2}{\xi}^1$ and obtain:  \\
\[ \begin{array}{rcl}
{\nabla}_0 R_{02,23} & =  & d_0 R_{02,23} - {\gamma}^r_{00}R_{r2,23} - {\gamma}^r_{02} R_{0r,23} - {\gamma}^r_{02} R_{02,r3} - {\gamma}^r_{03} R_{02,2r} \\
                                 & = & d_0 R_{02,23} - {\gamma}^1_{00} R_{12,23} \\
   & =  & \frac{3mA}{2r} {\xi}^0_{03} + \frac{3m^2}{4r^3} {\xi}^1_3  \\
    & =  & \frac{3mA}{2r} {\Gamma}^0_{03}  \\
{\nabla}_2 R_{02,30} & = & 0 - {\gamma}^1_{22}R_{01,30}  \\
         &  =  &  Ar R_{01,30}   \\
     &  =  &  - \frac{3mA}{2r^2} {\xi}^1_3 \\
{\nabla}_3 R_{02,02} & =  & \frac{3mA}{2r^2} {\xi}^1_3 
\end{array}  \]
where ${\Gamma}^0_{03} \equiv {\xi}^0_{03}+ \frac{m}{2Ar^2} {\xi}^1_3= d_3({\xi}^0_0 + \frac{m}{2Ar^2} {\xi}^1)=0$ and $R_{12,23}= - \frac{3m}{2Ar}{\xi}^1_3$, $R_{01,03}= \frac{3m}{2r^3}{\xi}^1_3$.  \\
Again, only this fnal result proves that $B_{02,023}\equiv \frac{3mA}{2r}{\Gamma}^0_{03}\Rightarrow d_0W_3+ \frac{m}{2Ar^2}V_3=0$ is a differential consequence of $R_{02,03}=0$.  \\

\noindent
$BIANCHI \, \fbox{12}$:
\fbox{ $ B_{12,012} \equiv  {\nabla}_0 R_{12,12} + {\nabla}_1 R_{12,20} + {\nabla}_2 R_{12,01} = - \frac{3m}{2r^3}{\Gamma}^0_{22} $ } . \\  \\
First of all, ${\Gamma}^0_{22}\equiv {\xi}^0_{22} + \frac{r}{A}{\xi}^1_0$, $R_{12,12}= - \frac{3m}{2Ar^2}{\xi}^1, R_{12,20}=0, 
R_{12,01}= - \frac{3m}{2r^3}{\xi}^0_2$ and we obtain:  \\
\[  \begin{array}{rcl}
{\nabla}_0 R_{12,12} & = & d_0 R_{12,12} -2 {\gamma}^r_{01}R_{r2,12} - 2 {\gamma}^r_{02} R_{1r,12}  \\
                                 & = & d_0 R_{12,12} - 2 {\gamma}^0_{01}R_{02,12}  \\
                                & = & - \frac{3m}{2Ar^2}{\xi}^1_0  \\
{\nabla}_1 R_{12,20} &  = &  0 - ({\gamma}^1_{11} + 2{\gamma}^2_{12} +{\gamma}^0_{01})R_{12,20}  \\
                                 & = & - 2 {\gamma}^2_{12} R_{12,20}  \\
                                 &  =  &  0  \\
 {\nabla}_2R_{12,01} & = & - \frac{3m}{2r^3} {\xi}^0_{22} - {\gamma}^2_{12} R_{12,02} \\
                                 & = &  - \frac{3m}{2r^3} {\xi}^0_{22}\\
                                 & = &  - \frac{3m}{2r^3} {\Gamma}^0_{22}  + \frac{3m}{2Ar^2} {\xi}^1_0
                                 \end{array}  \]
a result leading to $d_2W_2 + \frac{r}{A}d_0U=0$. 
Again, {\it only the final sum has an intrinsic mathematical meaning} with $B_{12,012}= - \frac{3m}{2r^3} {\Gamma}^0_{22}$.  \\

\noindent
$BIANCHI \, \, \fbox{22} $:
 \fbox{$ B_{01,023}  \equiv  {\nabla}_0R_{01,23} + {\nabla}_2R_{01,30} + {\nabla}_3R_{01,02}=0 $}  \\  \\ 
First of all, we have $R_{01,23}=0, R_{01,03}= \frac{3m}{2r^3}{\xi}^1_3, R_{01,02}= \frac{3m}{2r^3} {\xi}^1_2, R_{02,03}=0$ and, as ${\Omega}_{23}=0$: \\
\[ {\Gamma}^1_{23}\equiv {\xi}^1_{23}+ {\gamma}^1_{33}{\xi}^3_2 + {\gamma}^1_{22} {\xi}^2_3 - {\gamma}^3_{23} {\xi}^1_3 ={\xi}^1_{23} -Ar({\xi}^2_3 + {sin}^2(\theta){\xi}^3_2) - cot(\theta) {\xi}^1_3={\xi}^1_{23} - cot(\theta) {\xi}^1_3 \]
\[  {\Gamma}^2_{13}\equiv {\xi}^2_{13} + {\gamma}^2_{r3}{\xi}^r_1 + {\gamma}^2_{1r} {\xi}^r_3 - {\gamma}^r_{13} = {\xi}^2_{13} + {\gamma}^2_{33}{\xi}^3_1 +
({\gamma}^2_{12} - {\gamma}^3_{13}){\xi}^2_3 = {\xi}^2_{13} - sin(\theta) cos(\theta) {\xi}^3_1 \]
\[  {\Gamma}^3_{12} \equiv {\xi}^3_{12} + {\gamma}^3_{r2}{\xi}^r_1 + {\gamma}^3_{1r} {\xi}^r_2 - {\gamma}^r_{12}{\xi}^3_r ={\xi}^3_{12} + {\gamma}^3_{23}{\xi}^3_1 + ({\gamma}^3_{13} - {\gamma}^2_{12}) {\xi}^3_2 = {\xi}^3_{12} + cot(\theta){\xi}^3_1 \]

\[   \begin{array}{rcl}
{\nabla}_0 R_{01,23} & = &  d_0 R_{01,23} - {\gamma}^r_{00} R_{r1,23} - {\gamma}^r_{01} R_{0r,23} - {\gamma}^r_{02} R_{01,r3} - {\gamma}^r_{03} R_{01,2r}\\
                                 & = & d_0 R_{01,23}  - {\gamma}^1_{00}R_{01,23} - {\gamma}^0_{01}R_{00,13} \\
                                 & = &  0  \\
{\nabla}_2 R_{01,30} & = &  d_2 R_{01,30} -{\gamma}^2_{12} R_{02,30}  - {\gamma}^3_{23}R_{01,30} \\
                                 & = & - \frac{3m}{2r^3} {\xi}^1_{23} + \frac{3m}{2r^3}cot(\theta) {\xi}^1_3 \\
                                 & = &  - \frac{3m}{2r^3} ( {\xi}^1_{23} - cot(\theta) {\xi}^1_3) \\
                                 & = &  - \frac{3m}{2r^3} {\Gamma}^1_{23} \\
{\nabla}_3 R_{01,02} & = & d_3 R_{01,02} - {\gamma}^3_{13}R_{03,02} - {\gamma}^3_{23}R_{01,03}  \\
                                 & = &  \frac{3m}{2r^3} {\xi}^1_{23} - \frac{3m}{2r^3} cot(\theta) {\xi}^1_3  \\
                                 & = &  \frac{3m}{2r^3} ({\xi}^1_{23} - cot(\theta) {\xi}^1_3) \\
                                 & = &  \frac{3m}{2r^3} {\Gamma}^1_{23}
\end{array}  \]
We could also say that $d_2 R_{01,03} = \frac{3m}{2r^3} (d_2 V_3 - {\xi}^1_{23}) + \frac{3m}{2r^3}{\xi}^1_{23}$ and obtain therefore finally the formula 
$B_{01,023}= 0 \Rightarrow d_2V_3 - d_3V_2=0$ is a differential consequence of $R_{01,23}=0$. \\
We also check in particular:  \\
\[   \begin{array}{rcl}
{\nabla}_0 R_{01,23} & \Rightarrow & {\rho}_{r1,23}{\Gamma}^r_{00} + {\rho}_{0r,23} {\Gamma}^r_{01}+ {\rho}_{01,r3}{\Gamma}^r_{02}+ 
{\rho}_{01,2r}{\Gamma}^r_{03} \\
                               & \Rightarrow  &  0   \\
{\nabla}_2 R_{01,30} & \Rightarrow & {\rho}_{r1,30}{\Gamma}^r_{02} + {\rho}_{0r,30} {\Gamma}^r_{12}+ {\rho}_{01,r0}{\Gamma}^r_{23}+ 
{\rho}_{01,3r}{\Gamma}^r_{02} \\
                               & \Rightarrow  &  {\rho}_{03,30} {\Gamma}^3_{12}  + {\rho}_{01,10} {\Gamma} ^1_{23} \\
                               & \Rightarrow &  \frac{mA}{2r}{sin}^2(\theta) {\Gamma}^3_{12} - \frac{m}{r^3} {\Gamma}^1_{23}  \\
                               & \Rightarrow &   \frac{mA}{2r}{sin}^2(\theta) {\xi}^3_{12} - \frac{m}{2r^3}cot(\theta){\xi}^1_3 - \frac{m}{r^3}{\Gamma}^1_{23} \\
                               & \Rightarrow &  - d_2(\frac{m}{2r^3}{\xi}^1_3) - \frac{mA}{r} sin(\theta) cos(\theta) {\xi}^3_1 + \frac{m}{r^3} cot(\theta) {\xi}^1_3 - \frac{m}{2r^3}cot(\theta) {\xi}^1_3 - 
                                                          \frac{m}{r^3} {\Gamma}^1_{23} \\
                               & \Rightarrow &   - \frac{m}{2r^3} ({\xi}^1_{23}- cot(\theta) {\xi}^1_3) - \frac{m}{2r^3} {\Gamma}^1_{23} \\
                               & \Rightarrow &  - \frac{3m}{2r^3} {\Gamma}^1_{23}  \\
{\nabla}_3 R_{01,02} & \Rightarrow & {\rho}_{r1,02}{\Gamma}^r_{03} + {\rho}_{0r,02} {\Gamma}^r_{13}+ {\rho}_{01,r2}{\Gamma}^r_{03}+ 
{\rho}_{01,0r}{\Gamma}^r_{23} \\                              
                               & \Rightarrow & {\rho}_{02,02}{\Gamma}^2_{13} + {\rho}_{01,01} {\Gamma}^1_{23} \\
                               & \Rightarrow &  - \frac{mA}{2r} {\Gamma}^2_{13} + \frac{m}{r^3} {\Gamma}^1_{23}  \\
                               & \Rightarrow &  - \frac{mA}{2r}({\xi}^2_{13} - {sin}^2(\theta) cot(\theta) {\xi}^3_1) + \frac{m}{r^3}{\Gamma}^1_{23} \\
                               & \Rightarrow &  \frac{m}{2r^3}({\xi}^1_{23} - cot(\theta) {\xi}^1_3) + \frac{m}{r^3} {\Gamma}^1_{23} \\
                               & \Rightarrow &  \frac{3m}{2r^3} {\Gamma}^1_{23}
\end{array}    \]
As a tricky exercise too, we advise the reader to treat similarly the case of  $\fbox{13}+ \fbox{16}$ or $\fbox{21}$ in order to obtain $B_{01,123}= 0 \Rightarrow 
d_2W_3 - d_3W_2=0$ because $R_{01,23}= 0$ {\it and} $R_{03,12}=0$ (care).  \\

\noindent
{\bf REMARK 4.B.1}: Though a few conditions like \fbox{$d_2 V_3 - d_3 V_2=0$} \,\, \fbox{ $22$} or \fbox{ $d_2 W_3 - d_3 W_2=0$}\,\, \fbox{ $21$} {\it look like} to be third order CC for $\Omega$, we have thus proved that they come indeed from the first prolongations of the second order CC. The same comment is also valid for a few other striking CC. Using previous results, we have successively $6$ other relations:  \\
\[  {\xi}^1_{12} + {\xi}^0_{02} = d_2({\xi}^1_1 + {\xi}^0_0)= 0 \Rightarrow  \fbox{$d_1 V_2 + d_0 W_2=0 $}\,\, \fbox{ $1$} \] ,
\[  {\xi}^1_{13} + {\xi}^0_{03} = d_3({\xi}^1_1 + {\xi}^0_0)=0  \Rightarrow  \fbox{$d_1 V_3 + d_0 W_3=0 $}\,\, \fbox{ $2$} \]
\[   {\xi}^1_{33}+ sin(\theta)cos(\theta) {\xi}^1_2 - {sin}^2(\theta){\xi}^1_{22}=0 \Rightarrow \fbox{ $ d_3V_3 + sin(\theta)cos(\theta) V_2 - {sin}^2(\theta) d_2 V_2=0$}\,\,\fbox{$19$}  \]
\[    {\xi}^0_{33} + sin(\theta)cos(\theta) {\xi}^0_2 - {sin}^2(\theta){\xi}^0_{22}=0 \Rightarrow \fbox{ $ d_3 W_3 + sin(\theta)cos(\theta) W_2 - {sin}^2(\theta) d_2 W_2=0 $}  \,\,  \fbox {$20$} \]
because \, \fbox{ $ d_2W_2  + \frac{r}{A}d_0U=0$} \, \fbox{$ 12$}\,\, and \,\, \fbox{$d_3W_3 + sin(\theta)cos(\theta) W_2 + \frac{r}{A}{\sin}^2(\theta) d_0U=0$ } \, \fbox{$ 17$}
\[  {\xi}^1_{02} - A^2{\xi}^0_{12}= d_2({\xi}^1_0 - A^2 {\xi}^0_1)= 0 \Rightarrow  \fbox{ $ d_0 V_2 - A^2 d_1 W_2=0 $ }\,\, \fbox{$24$}  \]
\[  {\xi}^1_{03} - A^2{\xi}^0_{13}= d_3({\xi}^1_0 - A^2 {\xi}^0_1)= 0 \Rightarrow  \fbox{ $ d_0 V_3 - A^2 d_1 W_3=0 $ }   \,\, \fbox{$23$} \]
From the $24 \,\,\, B$, we have thus used $8$ of them and are left with $24 - 8=16$ expressions involving the $4 \times 4 =16 $ different first derivatives of the $4$ functions 
$(V,W)$, namely $B \,\, \fbox{3}$ to $B \,\, \fbox{18}$. Now, we notice that, among these $24 \,\,B$, {\it only} $4$ of them do contain three components $R_{kl,ij}$ that are not vanishing for the S-metric, namely $\fbox{1}$, $\fbox{2}$, $\fbox{19}$ and $\fbox{20}$. They are providing the terms $d_rE_{rr}$ for $r=0,1,2,3$ in the divergence type condition for the linearized Einstein equations implied by the linearized Bianchi identities over the Schwarzschild metric. Accordingly, it does not seem possible to obtain any other third order CC apart from these $4$ divergence conditions.  \\

It remains to apply these results to the successive prolongations of the Killing equations, as we know from the intrinsic study achieved in ([20],[23]) that we have the successive Lie algebroids: \\
\[ R^{(4)}_1 = R^{(3)}_1 \subset R^{(2)}_1 \subset R^{(1)}_1 = R_1 \subset J_1(T) \]
with respective dimensions $ 4 = 4 < 5 < 10 = 10 < 20$ and $R^{(3)}_1$ does not depend any longer on the $S$-parameter $m$. \\
The challenge will be to prove that ... {\it the only knowledge of these numbers is sufficient !}.  \\

In an equivalent way as $g_2=0 \Rightarrow g_{r+2}=0, \forall r\geq 0$, we obtain successively:  
\[  dim(R_1)= dim(J_1(T)) - dim (S_2T^*)=20 - 10=10, \]
\[   dim(R_2)=dim(J_2(T)) - dim(J_1(S_2T^*))=60 - 50=10,\]
\[  dim(R_3)= dim(R_2) - 5=5 \Rightarrow  dim(R_ {r+4})= dim(R_4)=dim(R_3) - 1 = 4  \]
and shall use these results from now on.  \\
First of all, using the {\it introductory diagram} when $q=1, r=$, we may apply the Spencer $\delta$-map to the symbol top ro in order to obtain the left:  \\
\[ \begin{array}{rcccccccl}
   &  &  & 0  & &  0  &  &  &    \\
   &  &  &  \downarrow & & \downarrow &  &  &   \\
   & 0 & \rightarrow &S_3T^*\otimes T& \rightarrow &S_2T^ *\otimes F_0 &\rightarrow & \fbox{ $ h_2 $ }  &  \rightarrow 0 \\
   &  &  &  \downarrow & & \downarrow &  &  &   \\
   & 0& \rightarrow &T^*\otimes S_2T^*\otimes T& \rightarrow &T^*\otimes T^ *\otimes F_0 &\rightarrow & 0 &  \\
   &   &  &  \downarrow & & \downarrow & & &  \\
 0 \rightarrow   &  \fbox{$ {\wedge}^2 T^*\otimes g_1 $ }  & \rightarrow &{\wedge}^2T^*\otimes T^*\otimes T& \rightarrow &{\wedge}^2T^*\otimes F_0 &
 \rightarrow  &  0   \\
  & \downarrow  &  &  \downarrow & & \downarrow & &  & \\
0 \rightarrow &{\wedge}^3 T^*\otimes T & = &{\wedge}^3T^*\otimes  T& \rightarrow & 0 & &  \\
 & \downarrow  &  &  \downarrow & &  & & &  \\
 & 0  &  & 0  & &    &  &   &    
\end{array} \]   \\  
Uing the Spencer $\delta$-cohomology $H^r(g_1)=Z^r(g_1)/B^r(g_1)$ at $ \dots \rightarrow \fbox { ${\wedge}^rT^* \otimes g_1 $ }\rightarrow \dots $, we obtain:  \\

\noindent
{\bf PROPOSITION 4.B.2}: $h_2\simeq H^2(g_1) \Rightarrow n(n+1)/2 \leq dim(Q_2) \leq n^2 (n^2 - 1)/12$ whenever $n\geq 3$.  \\

\noindent
{\it Proof}: As there cannot be any CC of order one and thus $Q_1=0$, we have the long exact connecting sequence  $  0 \rightarrow R_3 \rightarrow R_2 \rightarrow h_2 \rightarrow Q_2  \rightarrow 0 $ and counting the dimensions with $F_0 =S_2T^*$, we have:  
\[ dim(Q_2) \leq dim (h_2) = dim(S_2T^*\otimes S_2T^*) - dim(S_3T^* \otimes T)= n^2(n^2 - 1)/12  \]
This result is confirmed by a circular chase  proving that the left bottom $\delta$-map is an epimorphism and a snake chase in the last diagram providing the short exact sequence: \[   0 \longrightarrow h_2 \rightarrow {\wedge}^2 T^* \otimes g_1 \stackrel{\delta}{\longrightarrow}{\wedge}^3 T^*\otimes T \longrightarrow 0  \hspace{15mm}
 0 \rightarrow  20 \rightarrow  36  \rightarrow  16  \rightarrow 0  \]
Indeed, as $det(\omega)\neq 0$ we may use the metric for providing an isoorphism $T \simeq T^*: ({\xi}^r) \rightarrow ({\xi}_i={\omega}_{ri}{\xi}^r)$ in such a way that $g_1 \simeq {\wedge}^2 T^*$ is defined by ${\xi}_{i,j} + {\xi}_{j,i}=0$ for both the M, S and K metrics. \\
However, introducing the conformal Killing system of infinitesimal Lie equations with symbol ${\hat{g}}_1$ defined by the $(n(n+1)/2) - 1$ linear equations 
${\omega}_{rj}{\xi}^r_i + {\omega}_{ir}{\xi}^r_j - \frac{2}{n}{\omega}_{ij}{\xi}^r_r=0$ that do not depend on any conformal factor, we have the Ò {\it fundamental diagram II} ([9],[11]):  \\

 \[ \begin{array}{rcccccccl}
 & & & & & & & 0 & \\
 & & & & & & & \downarrow & \\
  & & & & & 0& & S_2T^* &  \\
  & & & & & \downarrow & & \downarrow  &  \\
   & & & 0 &\longrightarrow & Z^2(g_1) & \longrightarrow & H^2(g_1)  & \longrightarrow 0  \\
   & & & \downarrow & & \downarrow & &  \downarrow  &  \\
   & 0 &\longrightarrow & T^*\otimes {\hat{g}}_2 & \stackrel{\delta}{\longrightarrow} & Z^2({\hat{g}}_1) & \longrightarrow & H^2({\hat{g}}_1) & \longrightarrow 0  \\
    & & & \downarrow & & \downarrow & & \downarrow     &   \\
 0 \longrightarrow & S_2T^* & \stackrel{\delta}{\longrightarrow}& T^*\otimes T^* &\stackrel{\delta}{\longrightarrow} & {\wedge}^2T^* & \longrightarrow & 0 &   \\
   & & & \downarrow &  & \downarrow & & &  \\
   & & & 0 & & 0 & & &  \\
  & & & & &  & & &     
   \end{array}  \]
showing that we have the splitting sequence $ 0 \rightarrow S_2T^* \rightarrow H^2(g_1) \rightarrow H^2({\hat{g}}_1) \rightarrow 0  $ providing a totally unusual interpretation of the successive Ricci, Riemann and Weyl tensors and the corresponding splitting. However, it must be noticed that the $Weyl$-type operator is of order $3$ when $n=3$ because $n^2(n^2 - 1)/12 - n(n+1)/2= n(n+1)(n+2)(n-3)/12$ but of order $2$ for $n\geq 4$ ([18],[27]). Similar results could be obtained for the $Bianchi$-type operator as we shall see. \\
\hspace*{12cm}$\Box$  \\

Using now the same procedure for the {\it introductory diagram} with $r=2$, we get the diagram:
\[  \begin{array}{rcccccccl}
   &  &  & 0  & &  0  &  & 0  &    \\
   &  &  &  \downarrow & & \downarrow & & \downarrow &   \\
   & 0 & \rightarrow &S_4T^*\otimes T& \rightarrow &S_3T^ *\otimes F_0 &\rightarrow & h_3  &  \rightarrow 0 \\
   &  &  &  \downarrow & & \downarrow & \searrow & \downarrow &   \\
   & 0& \rightarrow &T^*\otimes S_3T^*\otimes T& \rightarrow &T^*\otimes S_2T^ *\otimes F_0 &\rightarrow & \fbox{$ T^*\otimes h_2 $ }& \rightarrow  0  \\
   &   &  &  \downarrow & & \downarrow & & &  \\
   & 0  & \rightarrow &{\wedge}^2T^*\otimes S_2T^*\otimes T& \rightarrow &{\wedge}^2T^*\otimes T^ *\otimes F_0 &\rightarrow  &  0   \\
  &   &  &  \downarrow & & \downarrow & &  & \\
0 \rightarrow & \fbox{$ {\wedge}^3 T^*\otimes g_1 $ } & \rightarrow &{\wedge}^3T^*\otimes T^*\otimes T & \rightarrow &{\wedge}^3T^*\otimes F_0 &
    \rightarrow &   0  & \\
  & \downarrow  &  &  \downarrow & & \downarrow & &  &\\
0 \rightarrow &{\wedge}^4 T^*\otimes T & = &{\wedge}^4T^*\otimes  T& \rightarrow & 0 & &  \\
 & \downarrow  &  &  \downarrow & &  & & &  \\
 & 0  &  & 0  & &    &  &   &    
\end{array}  \]   \\  

\noindent
Using a snake chase {\it and} Theorem 3.2.3, we obtain the short exact sequence:  \\
\[  0 \longrightarrow h_3 \longrightarrow T^* \otimes h_2 \longrightarrow H^3(g_1) \longrightarrow 0, \hspace{15mm} 0 \rightarrow 60 \rightarrow 80 \rightarrow 20 
\rightarrow 0  \]
A chase around the upper south-east arrow on the right is leading to the following corollary where $g'_2 \subset S_2T^*\otimes F_0$ is the symbol of the system $R'_2 \subset J_2(F_0)$ which is the 
image of $J_3(T)$ and $Q'_1$ is the cokernel of the central bottom map:  \\

\noindent
{\bf COROLLARY 4.B.3}: There is a long exact connecting diagram:  
\[ \begin{array}{rcccccccccl}
&  & 0 &  & 0 &  &  &  &  &  &   \\
&  &  \downarrow  &  & \downarrow &  &  &  &  &  &   \\
0 &\longrightarrow & S_4T^* \otimes T& \longrightarrow &S_3T^* \otimes F_0& \longrightarrow &T^* \otimes h_2 & \longrightarrow &H^3(g_1) &\longrightarrow &0 \\
  &   & \downarrow  &  &  \parallel & &  \downarrow  &  & \downarrow &  &  \\
 0 & \longrightarrow  & {\rho}_1(g'_2) & \longrightarrow  &S_3T^* \otimes F_0 & \longrightarrow &T^* \otimes Q_2& \longrightarrow & Q'_1 & \longrightarrow & 0 \\
   &  &  &  & \downarrow  &  & \downarrow  &  & \downarrow &  &  \\
    &  &  &  & 0 &  & 0 &  & 0 &  &
\end{array}  \]
 allowing to use the Bianchi indentities as $ B \in F_2 \simeq H^3(g_1)$ and we have $dim(Q'_1) \leq dim(H^3(g_1)$.  \\
 
\noindent
{\it Proof}: Using the notations of the {\it introductory diagram} and the fact that $Q_1=0$, we have the following two commutative and exact diagrams obtained by choosing $F_1=Q_2 $ for the first, then $F_1=Q_3$ for the second and so on, in a systematic manner as in the motivating examples:  \\
\[  \begin{array}{rccccccccl}
    &  0  &  &  0  &  &  0  &  & 0  &   \\
     & \downarrow   &  & \downarrow &  &  \downarrow  &  & \downarrow  &     \\
 0 \rightarrow  & {\rho}_1(g'_2)  & \rightarrow & S_3T^*\otimes F_0 & \longrightarrow &  T^*\otimes Q_2 & \rightarrow &   Q'_1 & \rightarrow 0 \\
&  \downarrow &  & \downarrow &  &  \downarrow  & \searrow &  \parallel    &     \\
0 \rightarrow & {\rho}_1(R'_2) & \rightarrow  & J_3(F_0) & \longrightarrow  &  J_1(Q_2) &  \rightarrow &  Q'_1 & \rightarrow 0   \\
    &  \downarrow &  & \downarrow &  &  \downarrow  &  & \downarrow   &     \\
0 \rightarrow &  R'_2 & \rightarrow  & J_2(F_0) & \longrightarrow  &  Q_2 &  \rightarrow & 0 &   \\
   & \downarrow &  & \downarrow &  &  \downarrow  &  &   & \\
   & 0 &  &  0  &  & 0  &  &  &
\end{array}   \]
First, we have the short exact sequence  $0 \rightarrow R_2 \rightarrow J_2(T) \rightarrow J_1(F_0) \rightarrow 0$ with $10 - 60 + 50 = 0$ and get $R'_0=F_0$, $R'_1=J_1(F_0)$ and $dim({\rho}_1(R'_0)) - dim(R'_1)=0$, that is no CC of order $1$. \\
Now, using the long exact sequence:  \\
\[   0 \rightarrow R_3 \rightarrow J_3(T) \rightarrow J_2(F_0) \rightarrow Q_2 \rightarrow 0  \] 
\[  R'_2 \subset {\rho}_1(R'_1)={\rho}_1(J_1(F_0))=J_2(F_0) \Rightarrow dim({\rho}_1(R'_1)) - dim (R'_2) = 150 - 135=dim(Q_2)=15  \] 
because $dim(R'_2)= dim(J_3(T)) - dim(R_3)= 140 - 5 = 135   $ and there are $15$ second order CC.   \\
\hspace*{12cm}  $\Box $    \\
Then, with $dim(Q'_1)=x$, we obtain by counting the dimensions:  \\
\[ dim({\rho}_1(R'_2))= dim(J_3(F_0))  + x - dim(J_1(Q_2))=350 + x - 75=x+ 275, \]
\[ dim(R'_3)=dim(J_4(T) - dim(R_4)= dim(J_3(F_0)) - dim(Q_3)= 276      \] 
\[  \Rightarrow \hspace{5mm} y = dim({\rho}_1(R'_2)) - dim(R'_3)= x+275 - 276= x-1  \] 
that is $y\geq 3$ because $x\geq 4$ and thus $x=4 \Rightarrow y=3$ if we only take into account the $4$ divergence condition of the Einstein equations. The situation will be worst for the Kerr metric with $y=6$. \\
After one prolongation, we get:  
\[  \begin{array}{rccccccccl}
    &  0  &  &  0  &  &  0  &  & 0  &   \\
     & \downarrow   &  & \downarrow &  &  \downarrow  &  & \downarrow  &     \\
 0 \rightarrow  & {\rho}_1(g'_3)  & \rightarrow & S_4T^*\otimes F_0 & \longrightarrow &  T^*\otimes Q_3 & \rightarrow &   Q'_1 & \rightarrow 0 \\
&  \downarrow &  & \downarrow &  &  \downarrow  &  &  \parallel    &     \\
0 \rightarrow & {\rho}_1(R'_3) & \rightarrow  & J_4(F_0) & \longrightarrow  &  J_1(Q_3) &  \rightarrow &  Q'_1 & \rightarrow 0   \\
    &  \downarrow &  & \downarrow &  &  \downarrow  &  & \downarrow   &     \\
0 \rightarrow &  R'_3 & \rightarrow  & J_3(F_0) & \longrightarrow  &  Q_3 &  \rightarrow & 0 &   \\
   & \downarrow &  & \downarrow &  &  \downarrow  &  &   & \\
   & 0 &  &  0  &  & 0  &  &  &
\end{array}   \]
From this second diagram we obtain the commutative and exact diagram:  \\
\noindent
\[   \begin{array}{rcccccccl}
    &   0  &  &  0  &  &   &  \\
    & \downarrow &  & \downarrow &   &   &  &   &  \\
    0 \rightarrow & R'_4 & \rightarrow & J_4(F_0) & \rightarrow  & Q_4  & \rightarrow & 0 &  \\
     & \downarrow &  & \parallel &   & \downarrow  &  & \\
0 \rightarrow & {\rho}_1(R'_3) & \rightarrow &  J_4(F_0) & \rightarrow & J_1(Q_3) & \rightarrow & Q'_1 & \rightarrow 0  \\
      &  &   &  \downarrow &  &  &  &  &  \\
       &    &  &  0  &  &  & &  &
       \end{array}  \]
Indeed, setting again $dim(Q'_1)=x$, we obtain now similarly:  \\
\[ dim({\rho}_1(R'_3))= dim(J_4(F_0))  + x - dim(J_1(Q_3))= 700 + x - 370 = x+ 330, \]
\[ dim(R'_4)=dim(J_5(T))- dim(R_5)= dim(J_4(F_0)) - dim(Q_4)= 500      \] 
\[  \Rightarrow  \hspace{5mm}   y = dim({\rho}_1(R'_3)) - dim(R'_4)= x+330 - 500= x-170   \] 
that is $y=0 \Leftrightarrow  x=170 $. We find {\it exactly} $ dim(F_2)=170 $ like in ([20], p 1996) and the condition $y=0$ just means that the CC of order 4 are generated by the CC of order $3$.  \\ 
With one more prolongation, applying again the $\delta$-map to the top symbol sequence, we get the following commutative diagram:  \\
\[  \begin{array}{rccccccl}
  &  0   & &  0  &  &  0  &     \\
  & \downarrow & & \downarrow & & \downarrow &     \\
0 \rightarrow & S_6T^* \otimes T & \rightarrow & S_5T^* \otimes F_0 & \rightarrow & h_5 & \rightarrow 0     \\
     &   \hspace{3mm} \downarrow \delta  &  &  \hspace{3mm} \downarrow \delta &  & \downarrow & \\
0 \rightarrow & T^* \otimes S_5 T^* \otimes T & \rightarrow &  T^* \otimes S_4T ^* \otimes F_0 & \rightarrow & T^* \otimes h_4 & \rightarrow   0  \\
     &   \hspace{3mm} \downarrow \delta &  &  \hspace{3mm} \downarrow \delta &  & \downarrow &      \\
0 \rightarrow & {\wedge}^2T^*\otimes S_4T^* \otimes T & \rightarrow & {\wedge}^2T^*\otimes S_3T^* \otimes F_0 &\rightarrow   & {\wedge}^2 T^* \otimes h_3 & \rightarrow 0       \\
     & \hspace{3mm} \downarrow \delta &  &\hspace{3mm}\downarrow \delta  &  & \downarrow &    \\
0 \rightarrow &  {\wedge}^3T^*\otimes S_3T^* \otimes T & \rightarrow & {\wedge}^3T^* \otimes S_2T^* \otimes F_0 & \rightarrow  & {\wedge}^3 T^*\otimes h_2 &  \rightarrow 0  \\
    &  \hspace{3mm} \downarrow \delta &  & \hspace{3mm} \downarrow \delta  &  &  \downarrow  &  &  \\
 0 \rightarrow & {\wedge}^4 T^* \otimes S_2T^* \otimes T & \rightarrow & {\wedge}^4 T^* \otimes T^* \otimes  F_0 & \rightarrow & 0 & \\
    &  \downarrow &  &  \downarrow  &  &  &  \\
    & 0  &  &  0
 \end{array}   \]
where the right exact vertical column is $ 0 \rightarrow 224 \rightarrow 504 \rightarrow 360 \rightarrow 80  \rightarrow 0 $. 
It just remains to replace in the two upper right epimorphisms $h_5$  by $T^* \otimes Q_4$ and $h_4$ by $Q_4$ along with  the following commutative diagram where we have chosen $F_1=Q_4$:  \\
\noindent
\[   \begin{array}{rcccc}
0 &  &   0  &  &      \\
 & \searrow  & \downarrow &  &   \\
   & & h_5 &  &  \\
   &  & \downarrow & \searrow & \\
0 &  \rightarrow & T^* \otimes h_4 & \rightarrow &  T^* \otimes Q_4  
      \end{array}  \]
in order to obtain the long exaxt sequence $0 \rightarrow S_6T^*\otimes T \rightarrow S_5T ^*  \otimes F_0 \rightarrow T^* \otimes Q_4 $.  \\
Finally, chasing in the following commutative and exact introductory diagram:  \\
\[  \begin{array}{rcccccccccl}
    &    &  &  0  &  &  0  &  & 0  &  & 0 & \\
     &  &  & \downarrow &  &  \downarrow  &  & \downarrow  &  & \downarrow &    \\
  & 0  & \rightarrow & S_6T^*\otimes T & \stackrel{{\sigma}_5(\Phi)}{\longrightarrow} &  S_5T^*\otimes F_0 & \rightarrow &   T^* \otimes Q_4 & \rightarrow & Q'_1 & \rightarrow 0 \\
&  \downarrow &  & \downarrow &  &  \downarrow  &  &  \downarrow    &  & \parallel &    \\
0 \rightarrow &  R_6 & \rightarrow  & J_6(T) & \stackrel{{\rho}_{5}(\Phi)}{\longrightarrow}  &  J_5(F_0) &  \rightarrow &  
J_1(Q_4)  &  \rightarrow & Q'_1 & \rightarrow 0  \\
    &  \downarrow &  & \downarrow &  &  \downarrow  &  & \downarrow   &  & \downarrow &   \\
0 \rightarrow &  R_5 & \rightarrow  & J_5(T) & \stackrel{{\rho}_4(\Phi)}{\longrightarrow}  &  J_4(F_0) &  \rightarrow &  Q_4  & 
\rightarrow &  0   &  \\ 
   & \downarrow &  & \downarrow &  &  \downarrow  &  & \downarrow  & \\
   & 0 &  &  0  && 0  && 0  & &  &
\end{array}   \]
we deduce that $R'_5 ={\rho}_1(R'_4)$ is involutive with $dim (R'_5)= 840 - 4 = 836$ and symbol $g'_5 \simeq S_6 T^* \otimes T$. \\
Unhappily, the reader will check at once that {\it a similar procedure cannot be applied} in order to prove that $R'_4 = {\rho}_1(R'_3)$. Indeed, {\it if we still have a monomorphism} $0 \rightarrow h_4 \rightarrow Q_4$ {\it we do not have a monomorphism} $h_3 \rightarrow Q_3$ {\it because now this map has a kernel of dimension equal to} $dim(R_3/R^{(1)}_3)=5-4=1$ according to the corresponding long exact connecting sequence.  \\

IT IS THUS NOT POSSIBLE TO PROVE THAT THERE ARE ONLY SECOND AND THIRD ORDER GENERATING CC IN A SIMPLE INTRINSIC WAY.   \\

However, like in the first motivating example in which we {\it should} be waiting for third order CC but a direct computation was proving that only second order ones {\it could} be used, we have:\\

\noindent
{\bf THEOREM 4.B.4}: The CC of the first order operator ${\cal{D}}: T \rightarrow F_0$ are generated by a third order operator ${\cal{D}}_1:F_0 \rightarrow F_1=Q_3$ and we have thus $R'_4={\rho}_1(R'_3)$.   \\

\noindent
{\it Proof}: With $F_0=S_2T^*$ and $ F_1=Q_3$ while applying the Spencer operator, we obtain the following commutative diagram in which the two central vertical columns are locally exact ([8],[11]):  \\
\[  \begin{array}{rcccccccl}
      &  0\,  & &   0 \,  &  &   0 \, &   &  &    \\
   &  \downarrow \, & & \downarrow , &   &  \downarrow \, &  & & \\
0 \longrightarrow  &  \Theta & \longrightarrow & T  &  \stackrel{{\cal{D}}}{\longrightarrow} &\fbox{$ F_0 $} & \stackrel{{\cal{D}}_1}{\longrightarrow}   &  F_1 &  \\
   & \, \,\,\, \downarrow j_4  &  & \, \,\,\, \downarrow j_4  &  &   \,\,\,\, \downarrow j_3  & \searrow &  \parallel &  \\
   0 \longrightarrow &  R_4  & \longrightarrow & J_4(T) & \longrightarrow & J_3(F_0) & \longrightarrow  & F_1  & \longrightarrow  0 \\
     &  \,\,\,\, \downarrow d &  &  \, \,\,\, \downarrow d  &  &  \,\,\,\, \downarrow d  &  &  &    \\
     0  \longrightarrow &  \fbox { $T^*\otimes R_3 $ }&  \longrightarrow  &  T^*\otimes J_3(T) & \longrightarrow  & 
     T^*\otimes J_2(F_0) &  \\
     & \, \,\,\,  \downarrow d  &  & \, \,\,\,  \downarrow d &  &  &  &  \\
     0  \longrightarrow & {\wedge}^2T^*\otimes R_2 & \longrightarrow  &{\wedge}^2T^*\otimes  J_2(T) &  &  &
\end{array}  \]
Chasing in this diagram by using the Snake lemma of the second section, we discover that the local exactness at $F_0$ of the top row is equivalent to the local exactness at 
$T^*\otimes R_3$ of the left column.  \\
Now, we have the commutative diagram:  \\
\[  \begin{array}{rccccccc}
&  &  &  0  &  &  0  &  &  0    \\
 &    &  & \downarrow &  &  \downarrow &  &  \downarrow  \\
 0 \longrightarrow &  \Theta  &  \stackrel{j_5}{\longrightarrow} & R_5 & \stackrel{d}{\longrightarrow} & T^*\otimes R_4 &  \stackrel{d}{\longrightarrow } & {\wedge}^2T^*\otimes R_3  \\
   &  \parallel  &  &  \downarrow &  &  \downarrow &  &  \downarrow  \\
 0 \longrightarrow &  \Theta  &  \stackrel{j_4}{\longrightarrow} & R_4 & \stackrel{d}{\longrightarrow} & \fbox{$T^*\otimes R_3 $}&  \stackrel{d}{\longrightarrow } & {\wedge}^2T^*\otimes R_2  \\
  &  &  &  \downarrow &  &  \downarrow &  &  \downarrow    \\
  &  &  &  0  &  \longrightarrow  & T^*\otimes (R_3/R^{(1)}_3) & \stackrel{d}{\longrightarrow} & {\wedge}^2T^*\otimes (R_2/R^{(1)}_2)  \\
  &  &  &  &  & \downarrow & & \downarrow   \\
  &  &  &  &  &    0  &  &  0  
\end{array}  \]
The top row is known to be locally exact as it is isomorphic to a part of the Poincar\'{e} sequence according to the commutative diagram with 
$R_4\simeq R_5\simeq R_6$:  \\
\[  \begin{array}{rccccccc}
&  &  &  0  &  &  0  &  &  0    \\
 &    &  & \downarrow &  &  \downarrow &  &  \downarrow  \\
 0 \longrightarrow &  \Theta  &  \stackrel{j_6}{\longrightarrow} & R_6 & \stackrel{d}{\longrightarrow} & T^*\otimes R_5 &  \stackrel{d}{\longrightarrow } & {\wedge}^2T^*\otimes R_4  \\
   &  \parallel  &  &  \downarrow &  &  \downarrow &  &  \downarrow  \\
 0 \longrightarrow &  \Theta  &  \stackrel{j_5}{\longrightarrow} & R_5 & \stackrel{d}{\longrightarrow} & T^*\otimes R_4 &  \stackrel{d}{\longrightarrow } & {\wedge}^2T^*\otimes R_3  \\
  &  &  &  \downarrow &  &  \downarrow &  &   \\
  &  &  &  0  &   & 0  &  & 
 \end{array}  \]
 The bottom row is purely algebraic as it is induced by the exact sequence obtained by applying the Spencer operator to the long exact connecting sequence and 
 chasing along the south west diagonal:  \\
 \[    0 \rightarrow h_4 \stackrel{\delta}{\rightarrow} T^* \otimes h_3 \stackrel {\delta}{\rightarrow} {\wedge}^2 T^* \otimes h_2 \rightarrow {\wedge}^4 T^* \otimes g_1 \rightarrow 0   \]
 \[        0 \rightarrow 126 \rightarrow 240  \rightarrow  120 \rightarrow 6 \rightarrow 0  \]
Changing the confusing notations used in ([20]), we prove that the bottom Spencer operator is injective. Indeed, we have the following representative parametric jets for the various Lie equations:  \\
\[  dim(R_2)=10 \Rightarrow   \{ {\xi}^0, {\xi}^1, {\xi}^2, {\xi}^3, {\xi}^1_0, {\xi}^0_2, {\xi}^0_3, {\xi}^1_2, {\xi}^1_3, {\xi}^2_3 \}   \]
\[  dim(R_3)=5 \Rightarrow    \{ {\xi}^0, {\xi}^2, {\xi}^3, {\xi}^1_0 , {\xi}^2_3 \}, \hspace{1cm}  dim(R_4)=4 \Rightarrow \{  {\xi}^0, {\xi}^2, {\xi}^3, {\xi}^2_3 \}  \]
\[  dim(R_3/R^{(1)}_3)= 1 \Rightarrow  \{ {\xi}^1_0 \}, \hspace{1cm}dim(R_2/R^{(1)}_2)=5 \Rightarrow \{ {\xi}^1, {\xi}^1_2, {\xi}^1_3, {\xi}^0_2, {\xi}^0_3 \}  \]
We also recall the definition of the Spencer operator  $d:T^* \otimes J_{q+1}(T)\rightarrow {\wedge}^2 T^* \otimes J_q(T)$:  
\[  ({\xi}^k_{\nu,i}) \rightarrow  {\xi}^k_{\mu,ij}= ({\partial}_i{\xi}^k_{\mu,j} - {\partial}_j{\xi}^k_{\mu,i} + {\xi}^k_{\mu + 1_j,i} - {\xi}^k_{\mu + 1_i,j})  \]
Accordingly, we may choose local coordinates $({\xi}^1_{0,i})$ for a representative and a representative of the image by $d$ is for example 
$({\xi}^1_{,0i} = {\xi}^1_{0,i} - {\xi}^1_{i,0})$. Now, as $dim(R_3/R^{(1)}_3)= 4-3=1$, we may introduce the four local coordinates ${\xi}^1_{0,i}$ and ${\xi}^0_{1,i}$ such that ${\xi}^1_{0,i} - A^2 {\xi}^0_{1,i}=0$, $ \forall i=0,1,2,3$. We may also use the $6 \times 5=30$ local coordinates $({\xi}^1_{,ij}, {\xi}^1_{2,ij}, {\xi}^1_{3,ij}, {\xi}^0_{2,ij}, {\xi}^0_{3,ij})$ in order to describe ${\wedge}^2 T^*\otimes (R_2/R^{(1)}_2)$. In the kernel of $d$, we have in particular ${\xi}^1_{,0i}={\xi}^1_{0,i} - {\xi}^1_{i,0}=0 \Rightarrow {\xi}^1_{0,i}={\xi}^1_{i,0}=0, \forall i=1,2,3$ because ${\xi}^1_1= \frac{m}{2Ar^2} {\xi}^1$ in $R_1$ but also ${\xi}^0_{,01}= {\xi}^0_{0,1} - {\xi}^0_{1,0}=0 \Rightarrow {\xi}^0_{1,0}=0  \Rightarrow {\xi}^1_{0,0}=0 $ because $\{{\xi}^0 \}$ is among the parametric jets of $R_3$ and thus ${\xi}^1_{0,i}=0, \forall i=0,1,2,3$. The bottom Spencer operator is thus injective and the bottom sequence is thus exact. A circular chase ends the proof: If $b \in T^* \otimes R_3$ is killed by $d$, then its projection $c\in  T^*\otimes (R_3/R^{(1)}_3) $ is also killed by $d$ and is such that $c=0$. Accordingly, $\exists a \in T^* \otimes R_4$ with image $b$ under the monomorphism $T^* \otimes R_4 \rightarrow T^* \otimes R_3$  and such that $da=0$. We may thus find $e \in R_5$ and $f\in R_4$ because $R_5 \simeq R_4$ with $a=de \Rightarrow  b=df$.  \\
\hspace*{12cm}  $\Box$  \\

Like in the second motivating example, the sequence constructed in the previous theorem may have "{\it jumps} " in the order of the successive operators and we have therefore (Compare to [1]):  \\

\noindent
{\bf COROLLARY 4.B.5}: The symbol of ${\cal{D}}_1$ is {\it not} $2$-acyclic and the CC operator ${\cal{D}}_2$ is thus of order $2$. Accordingly, if one does want a formally exact canonical Janet sequence, {\it the only possibility} is to use the involutive operator ${\cal{D}}_1$ of order $4$ defined by $R'_4={\rho}_1(R'_3)$.  \\

\noindent
{\it Proof}: Recapitulating the results so far obtained, we have successively $ R'_{r+1} \subseteq {\rho}_1(R'_r)$ with:  \\
\[ R'_0=F_0, dim(R'_1)=dim(J_2(T)) - dim (R_2)=60-10=50\Rightarrow R'_1=J_1(F_0), {\rho}_1(R'_1)=J_2(F_0)    \]
\[  dim(R'_2)=dim(J_3(T))-dim(R_3)=140-5=135, dim({\rho}_1(R'_1)) - dim (R'_2)=150 - 135=15   \]
\[ dim(R'_3) = dim(J_4(T)) - dim(R_4)= 280 -4=276, \hspace{10mm}dim({\rho}_1(R'_2)) - dim(R'_3)\geq 3,  \]
\[  dim(R'_4)=dim(J_5(T)) - dim (R_5)=504-4=500, \hspace{10mm} R'_4={\rho}_1(R'_3) ,  \hspace{10mm} \]
\[ dim(R'_5)= dim(J_6(T))- dim(R_6)=840 -4=836, \hspace{10mm} R'_5 = {\rho}_1(R'_4).  \]
the long exact sequence: $ 0 \rightarrow {\rho}_1(R'_2) \rightarrow J_3(F_0) \rightarrow J_1(Q_2) \rightarrow Q'_1 \rightarrow 0 $ with $dim(Q'_1)=x\geq 4$ because of the divergence CC condition for Einstein equations implied by the Bianchi identities.\\
 It also follows that:  \\
\[g'_1 \simeq T^*\otimes F_0 \Rightarrow dim(g'_1)=40, \hspace{10mm} S_3T^* \otimes T \subset g'_2 \Leftrightarrow 80<dim(g'_2)=135 -50=85,\]
\[ S_4T^* \otimes T \subset g'_3 \Leftrightarrow 140 < 141=276 - 135, \hspace{10mm}  S_{5+r}T^* \otimes T \simeq g'_{4+r}, \forall r\geq 0  \]
and we have the basic commutative and exact " {\it defining diagram} " of the system $R_2 \subset J_2(T)$:  \\

\[  \begin{array}{rcccccccl}
 & & & & 0 & & 0 & &   \\
 & & & & \downarrow & &\downarrow & &  \\
 & & 0  & \rightarrow & S_2T^*\otimes T & \rightarrow & T^* \otimes F_0 & \rightarrow &  0    \\
 & &\downarrow & & \downarrow & &\downarrow & &   \\
 0  &  \rightarrow & R_2  & \rightarrow &  J_2(T) &  \rightarrow &  J_1(F_0) & \rightarrow &  0 \\
 & &\downarrow & &  \downarrow  &  &   \downarrow & &  \\
0  & \rightarrow &  R_1 &  \rightarrow &  J_1(T)  &  \rightarrow  &  F_0  & \rightarrow & 0 \\
 & & \downarrow & &  \downarrow & & \downarrow  &  &   \\
 & & 0  & & 0 & & 0 &  & 
\end{array}   \]
allowing to obtain the central vertical short exact sequence $0 \rightarrow g'_1 \rightarrow R'_1 \rightarrow F_9 \rightarrow 0 $.  \\
Now, it is known that a symbol $g_q$ of finite type is involutive if an only if it is vanishing ([8],[11],[12]). Using a similar proof, let us consider the commutative diagram of $\delta$-sequences:   \\
\[  \begin{array}{ccccccl}
     & 0 &  & 0  & & 0  &  \\
  & \downarrow     &   & \downarrow & & \downarrow  & \\
...   \rightarrow & T^* \otimes S_6 T^* \otimes T& \rightarrow  & {\wedge}^3 T^* \otimes S_5 T^*\otimes T  & \rightarrow & {\wedge}^4 T^* \otimes S_4T^* \otimes T & \rightarrow 0    \\
   & \downarrow   &  &  \downarrow &  & \downarrow &    \\
 ... \rightarrow    &T^* \otimes g'_5 & \rightarrow &  {\wedge}^3T^* \otimes g'_4 & \rightarrow & \fbox{${\wedge}^4T^* \otimes g'_3$} & \rightarrow 0  \\
      &  \downarrow  &  & \downarrow  && &    \\
&0  &  &  0  &  & & 
\end{array}  \]
Using the fact that the upper sequence is known to be exact as a $\delta$-sequence and that we have $  dim(S_4T^*\otimes T) = 140 < 141 = dim(g'_3)$, an easy chase proves that the lower sequence {\it cannot} be exact and thus $g'_3$ {\it cannot} be involutive after counting the dimensions. The corollary follows from the fact that $g'_4={\rho}_1(g'_3)\simeq S_5T^* \otimes T$ is indeed $3$-acyclic {\it one step ahead} by chasing and even involutive.  \\
Finally, with vector bundles $A,B$ such that $dim(A)=1,dim(B)=5$, we have the commutative diagram of $\delta$-sequences in which we recall that 
$g'_1\simeq T^*\otimes F_0$:  \\
\[  \begin{array}{cccccccl}
   0 & & 0 &  & 0  & & 0  &  \\
  \downarrow && \downarrow     &   & \downarrow & & \downarrow  & \\
T^* \otimes S_5 T ^* \otimes T  & \rightarrow & {\wedge}^2T^* \otimes S_4 T^* \otimes T& \rightarrow  & {\wedge}^3 T^* \otimes S_3 T^*\otimes T  & \rightarrow & {\wedge}^4 T^* \otimes S_2T^* \otimes T & \rightarrow 0    \\
 \downarrow &  & \downarrow   &  &  \downarrow &  & \downarrow  &    \\
 T^* \otimes g'_4 & \rightarrow    &\fbox{${\wedge}^2T^* \otimes g'_3$ }& \rightarrow &  {\wedge}^3T^* \otimes g'_2 & \rightarrow & {\wedge}^4T^* \otimes g'_1 &
  \rightarrow 0  \\
  \downarrow  &   &  \downarrow  &  & \downarrow  && \downarrow &    \\
0  &\rightarrow  & {\wedge}^2T^* \otimes A  & \rightarrow  &  {\wedge}^3 T^* \otimes B   & \rightarrow  & 0 &  \\
   &   &  \downarrow  &  & \downarrow  & & &    \\
 & & 0  &  &  0  &  & &  
\end{array}  \]
Taking into account that the top row is exact and proceeding as in the last theorem with similar local coordinates, we get:   \\
$  {\xi}^1_{,123}={\xi}^1_{1,23} + {\xi}^1_{2,31} + {\xi}^1_{3,12}=0+0+0=0 $ always. \\
$  {\xi}^1_{,012}={\xi}^1_{0,12} + {\xi}^1_{1,20} + {\xi}^1_{2,01}={\xi}^1_{0,12} + 0 + 0=0 \Rightarrow  {\xi}^1_{0,12}=0 \Rightarrow {\xi}^1_{0,ij}=0, 
\forall i,j=1,2,3$  \\
$  {\xi}^0_{,10i}= {\xi}^0_{1,0i}+ {\xi}^0_{0,i1} + {\xi}^0_{i,01}={\xi}^0_{1,0i}+ 0 + 0=0 \Rightarrow  {\xi}^0_{1,0i}= 0 \Rightarrow {\xi}^1_{0,0i}=0, \forall i=2,3$.\\
We are thus only left with ${\xi}^1_{0,01}$ that may not vanish though ${\xi}^1_{0,01} + {\xi}^1_{0,10} + {\xi}^1_{1,00}=0$ {\it in any case} and the bottom map 
$\delta$ is not injective. \\
Let us prove that $g'_3$ is not $2$-acyclic because the central $\delta$-sequence {\it cannot} be exact at ${\wedge}^2 T^* \otimes g'_3$. Indeed, if it were, let 
$c\in {\wedge}^2 T^* \otimes A$ be killed by $\delta$. Then, we may lift $c$ to $b\in {\wedge}^2T^* \otimes g'_3$ such that $\delta b = f\in {\wedge}^3 T^* \otimes S_3 T^* \otimes T$ and obtain by commutativity $\delta f =0$ because the last vertial downarrow on the right is an isomorphism, thus a monomorphism. As the upper row is an exact sequence, we may thus find $a \in {\wedge}^2 T^* \otimes S_4 T^* \otimes T$ such that $f=\delta a$. Chasing circularly, it follows from the exactness assumtion at 
$ {\wedge}^2 T^* \otimes g'_3$ that we can find $e \in T^* \otimes g'_4 \simeq T^* \otimes S_5 T^* \otimes T$ such that $b= a + \delta e= a'\in {\wedge}^2 T^* \otimes S_4 T^* \otimes T$. It should follow that $c=0$ and a contradiction, that is $g'_3$ cannot be $2$-acyclic. \\

As we know from $([8],[11],[12])$ that the order of the generating CC for ${\cal{D}}_1$ is equal to $s+1$ if one needs $s$ prolongations in such a way that 
${\rho}_s(g'_3)=g'_{3 + s}$ becomes $2$-acyclic. As we already know that $g'_4={\rho}_1(g'_3) \simeq T^* \otimes S_5T^* \otimes T$ is involutive, we get $s=1$ and the generating CC ${\cal{D}}_2$ of  ${\cal{D}}_1$ are of order $2$. We have just a " {\it jump} " in the order and, for the details, refer the reader to the quite delicate Example 3.14 of ([18], p 119-125) in which it is already difficult to discover how many new second order CC should be introduced though the initial system is trivially FI with coefficients in $\mathbb{Q}$. Such a result could not even be imagined while using the methods of ([1]-[4]). \\
\hspace*{12cm}  $\Box $ 

There are "natural" reason for which we do not believe that these results could be useful in physics. Indeed, considering like in the previous reference the long exact sequence of jet bundles allowing to define $F_2$ when $F_1=Q_3$, namely:  \\
\[   0 \rightarrow R_6 \rightarrow J_6(T) \underset 1{ \rightarrow}J_5(F_0) \underset 3 {\rightarrow} J_2(F_1) \underset 2{\rightarrow} F_2 \rightarrow 0  \]
\[   0 \rightarrow  4  \rightarrow  840  \underset 1{ \rightarrow}  1260  \underset 3 {\rightarrow} 1110  \underset 2{\rightarrow} 686  \rightarrow 0 \]
and the large values of these dimensions need no comment for {\it any} application.  \\
\hspace*{12cm} $ \Box $ \\

\noindent
{\bf  C) KERR METRIC}:  \\
We now write the Kerr metric in Boyer-Lindquist coordinates:  \\
\[  \begin{array}{rcl}
ds^2  &  =  & \frac{{\rho}^2 - mr}{{\rho}^2}dt^2 - \frac{{\rho}^2}{\Delta} dr^2  -  {\rho}^2 d{\theta}^2  \\
  &    &  - \frac{2a m r sin^2(\theta)}{{\rho}^2} dtd\phi - (r^2+a^2 + \frac{mr a^2 sin^2(\theta)}{{\rho}^2})sin^2(\theta) d{\phi}^2          
  \end{array}   \]
where we have set $  \Delta= r^2  -mr +a^2 , \,\,  {\rho}^2=r^2 + a^2 cos^2(\theta) $ as usual and we check that we recover the Schwarschild metric when $a=0$. We notice that $t$ or $\phi$ do not appear in the coefficients of the metric. We shall change the coordinate system in order to confirm theses results by using computer algebra and the idea is to use the so-called " {\it rational polynomial} " coefficients as follows: \\
 \[ (x^0=t, \, x^1=r, \, x^2=c=cos(\theta), \, x^3=\phi)  \Rightarrow dx^2= - sin(\theta)d\theta \Rightarrow (dx^2)^2=(1-c^2)d\theta^2 \]
We obtain over the differential field $K= \mathbb{Q}(a,m)(t,r,c,\phi)=\mathbb{Q}(a,m)(x)$: \\
\[  \begin{array}{rcl}
ds^2  &  =  & \frac{{\rho}^2 - mx^1}{{\rho}^2}(dx^0)^2 - \frac{{\rho}^2}{\Delta} (dx^1)^2  -  \frac{{\rho}^2}{1-(x^2)^2} (dx^2)^2  \\
  &    &  - \frac{2a m x^1(1-(x^2)^2) }{{\rho}^2} dx^0dx^3 -  (1-(x^2)^2)((x^1)^2+a^2 + \frac{m a^2 x^1 (1-(x^2)^2)}{{\rho}^2}) (dx^3)^2          
\end{array}   \]
with now $\Delta= (x^1)^2 - m x^1 +a^2=r^2 - mr + a^2$ and $  {\rho}^2=(x^1)^2 +a^2(x^2)^2=r^2 + a^2c^2$. For a later use, it is also possible to set ${\omega}_{33}= - (1-c^2)((r^2 + a^2)^2 - a^2 ((1-c^2)(a^2 - mr + r^2))/ (r^2 + a^2c^2)$ and we have  $det(\omega)= - (r^2 + a^2c^2)^2$. Framing the leading derivatives, we obtain:   \\
\[ R_1\subset J_1(T) \,\,\,  \left\{  \begin{array}{lcl}
{\Omega}_{33} & \equiv & 2( {\omega}_{33}\fbox{${\xi}^3_3$} + {\omega}_{03} {\xi}^0_3 ) + \xi \partial {\omega}_{33}=0 \\
{\Omega}_{23} & \equiv & {\omega}_{33}\fbox{${\xi}^3_2$} + {\omega}_{03}{\xi}^0_2 + {\omega}_{22}{\xi}^2_3 = 0 \\
{\Omega}_{22} & \equiv & 2 {\omega}_{22}\fbox{${\xi}^2_2$} + \xi \partial {\omega}_{22} =0  \\
{\Omega}_{13} & \equiv  & {\omega}_{33}\fbox{${\xi}^3_1$} + {\omega}_{03}{\xi}^0_1 +{\omega}_{11}{\xi}^1_3 =0 \\
{\Omega}_{12} & \equiv  & {\omega}_{22}\fbox{${\xi}^2_1$} +{ \omega}_{11}{\xi}^1_2 =0  \\
{\Omega}_{11} & \equiv  & 2 {\omega}_{11} \fbox{${\xi}^1_1$} + \xi \partial {\omega}_{11} = 0  \\
{\Omega}_{03} & \equiv  & {\omega}_{33}\fbox{${\xi}^3_0$} + {\omega}_{03}({\xi}^0_0 +{\xi}^3_3)  + {\omega}_{00}{\xi}^0_3 + \xi \partial {\omega}_{03} = 0 \\
{\Omega}_{02} & \equiv  & {\omega}_{22}\fbox{${\xi}^2_0$} + {\omega}_{00}{\xi}^0_2  + {\omega}_{03}{\xi}^3_2 = 0 \\
{\Omega}_{01} & \equiv  & {\omega}_{11}\fbox{${\xi}^1_0$} + {\omega}_{00}{\xi}^0_1 + {\omega}_{03}{\xi}^3_1 =0  \\
{\Omega}_{00} & \equiv  & 2 ({\omega}_{00}\fbox{${\xi}^0_0$} + {\omega}_{03} {\xi}^3_0 ) +\xi \partial {\omega}_{00}=0 
\end{array}  \right.  \]

Now, we know that if $R_q\subset J_q(T)$ is a system of infinitesimal Lie equations, then we have the algebroid bracket and its link with the {\it prolongation/projection} (PP) procedure ([8],[11]-[13]):  \\
\[  [R_q,R_q]\subset R_q \Rightarrow [R^{(s)}_{q+r}, R^{(s)}_{q+r}] \subset R^{(s)}_{q+r}, \forall q,r,s \geq 0  \]
As $R^{(1)}_1={\pi}^2_1(R_2)=R_1$, it follows that $R^{(2)}_1={\pi}^3_1(R_3)$ is such that $[R^{(2)}_1,R^{(2)}_1]\subset R^{(2)}_1$ with $dim(R^{(2)}_1)= 20-16=4$ because we have obtained a total of $6$ {\it new different} first order equations. Using the first general diagram of the Introduction, we discover that the operator defining $R_1$ has $10+4=14$ CC of order $2$, a result obtained {\it totally independently of any specific GR technical object} like the {\it Teukolski scalars} or the {\it Killing-Yano tensors} introduced in ([1]-[4],[6]).  \\
Like in the case of the S metric, two prolongations allow to obtain $6$ additional equations (instead of $5$) that we set on the left side in the following list obtained $mod(\j_2(\Omega)$:  \\
We have {\it on sections} ({\it care}) the $16$ (linear) equations $mod(j_2(\Omega))$ of $R^{(2)}_1$ as follows ([23]):\\
\[ R^{(2)}_1 \subset R_1 \subset J_1(T) \, \left\{   \begin{array}{lcl} 
{\xi}^1=0,{\xi}^2=0  &  \Rightarrow & \fbox{ $ {\omega}_{00}{\xi}^0_1 + {\omega}_{03}{\xi}^3_1$} + {\omega}_{11}{\xi}^1_0 =0 ,  \,\, {\xi}^1_1=0, \,\,{\xi}^2_2=0  \\ 
{\xi}^1_2=0  &  \Rightarrow & {\xi}^2_1=0  \\
{\xi}^1_3 + lin({\xi}^1_0,{\xi}^2_0)=0  & \Rightarrow & \fbox{ $ {\omega}_{03}{\xi}^0_1 + {\omega}_{33}{\xi}^3_1$} + {\omega}_{11}{\xi}^1_3 =0   \\
{\xi}^2_3+lin({\xi}^1_0,{\xi}^2_0) =0  & \Rightarrow & \fbox{ $ {\omega}_{00}{\xi}^0_2 +  {\omega}_{03}{\xi}^3_2 $}+ 
{\omega}_{22}{\xi}^2_0 =0 ,  \\
    &  &  \fbox{ $ {\omega}_{03}{\xi}^0_2 +{\omega}_{33}{\xi}^3_2 $}+ {\omega}_{22}{\xi}^2_3=0 \\
{\xi}^0_3=0  & \Rightarrow & {\xi}^3_0=0, \,\, {\xi}^0_0=0, \,\,{\xi}^3_3=0  
\end{array} \right.   \]
The coefficients of the linear equations $lin$ involved depend on the Riemann tensor as in ([23]). Accordingly, we may choose only the $2$ parametric jets $({\xi}^1_0,{\xi}^2_0)$ among $ ({\xi}^1_0, {\xi}^1_3,{\xi}^2_0, {\xi}^2_3)$ to which we must add $({\xi}^0,{\xi}^3)$ {\it in any case} as they are not appearing in the Killing equations. \\
The system is {\it not} involutive because its symbol is finite type but non-zero.   \\ 

Using one more prolongation, all the {\it sections} ({\it care again}) vanish but ${\xi}^0$ and ${\xi}^3$, a result leading to $dim(R^{(3)}_1)=2$ in a coherent way with the only nonzero Killing vectors $\{ {\partial}_t, {\partial}_{\phi} \}$. We have indeed:  \\
\[   \fbox{$  {\xi}^1_0 =0$} \, ,\,\,\,  \fbox{ $  {\xi}^2_0=0$} \,\,  \Rightarrow  \,\,  {\xi}^1_3=0 , \,\, {\xi}^2_3=0 \,\, 
  \Rightarrow  \,\,  {\xi}^0_1=0 , \,\, {\xi}^0_1=0, \,\,  {\xi}^0_2=0 , \,\, {\xi}^3_2=0 \]

Taking therefore into account that the metric only depends on $(x^1=r,x^2=cos(\theta))$ we obtain {\it after three prolongations} the first order system:  \\
\[R^{(3)}_1 \subset R^{(2)}_1\subset  R^{(1)}_1 = R_1 \subset J_1(T) \,\,\,  \left\{  \begin{array}{lcl}
 {\xi}^3_3   & = & 0 \\
 {\xi}^2_3 & = & 0  \\
 {\xi}^1_3 & = & 0  \\
 {\xi}^0_3 & = & 0  \\
{\xi}^3_2  & = & 0 \\
 {\xi}^2_2  & = & 0  \\
 {\xi}^1_2 & = & 0  \\
 {\xi}^0_2 & = & 0 \\
 {\xi}^3_1  & = & 0 \\
 {\xi}^2_1 & = & 0  \\
 {\xi}^1_1 & = & 0  \\
 {\xi}^0_1  & = & 0  \\
 {\xi}^3_0    & = & 0 \\
 {\xi}^2_0  & = & 0 \\
 {\xi}^1_0  & = & 0  \\
 {\xi}^0_0   & = & 0 \\
 {\xi}^2 & = & 0  \\
 {\xi}^1 & = & 0 
  \end{array} \right. \fbox{ $\begin{array}{llll}
  0 & 1 & 2 & 3  \\
  
    0 & 1 & 2 & 3  \\
     0 & 1 & 2 & 3  \\
    0 & 1 & 2 & 3  \\
  0 & 1 & 2 & \bullet  \\
  0 & 1 & 2 & \bullet  \\
    0 & 1 & 2 & \bullet  \\
  0 & 1 & 2 & \bullet  \\
  0 & 1  &  \bullet &  \bullet \\
  0 & 1  &  \bullet &  \bullet \\
  0 & 1  &  \bullet &  \bullet \\
  0 & 1  &  \bullet &  \bullet \\
0 & \bullet & \bullet & \bullet \\
0 & \bullet & \bullet & \bullet \\
0 & \bullet & \bullet & \bullet \\
0 & \bullet & \bullet & \bullet \\
\bullet & \bullet & \bullet & \bullet \\
\bullet & \bullet & \bullet & \bullet 
  \end{array} $ }   \]   \\ 
{\it Surprisingly and contrary to the situation found for the S metric}, we have now an involutive first order system with only solutions $({\xi}^0=cst, {\xi}^1=0, {\xi}^2=0, {\xi}^3=cst)$ and notice that $R^{(3)}_1$ does not depend any longer on the parameters $(m,a)\in K$. The difficulty is to know what second members must be used along the procedure met for all the motivating examples. In particular, we have again identities to zero like $d_0{\xi}^1 - {\xi}^1_0=0, d_0{\xi}^2 - {\xi}^2_0=0$ and thus {\it at least} $6$ third order CC coming from the $6$ following components of the Spencer operator, 
namely:  \\
\[ \fbox {$ d_1{\xi}^1 - {\xi}^1_1=0,\,\, d_2{\xi}^1 - {\xi}^1_2=0, \,\,  d_3 {\xi}^1 - {\xi}^1_3 =0, \,\, d_1{\xi}^2 - {\xi}^2_1=0, \,\, d_2{\xi}^2 - {\xi}^2_2=0, \,\, 
d_3 {\xi}^2 - {\xi}^2_3 = 0 $ } \] 
a result that cannot be even imagined from ([1]-[4]). Of course, proceeding like in the motivating examples, we must substitute in the right members the values obtained from $j_2(\Omega)$ and set for example ${\xi}^1_1= - \frac{1}{2{\omega}_{11}}\xi \partial {\omega}_{11}$ while replacing ${\xi}^1$ and ${\xi}^2$ by the corresponding linear combinations of the Riemann tensor already obtained for the right members of the two zero order equations. \\

We have the fundamental diagram I {\it no longer depending on} $(m,a)$ with fiber dimensions:  \\
\[   \begin{array}{rcccccccccccccl}
 &&&&& 0 &&0&&0&& 0 &  &0&  \\
 &&&&& \downarrow && \downarrow && \downarrow & & \downarrow &   & \downarrow &  \\
  & 0& \rightarrow& \Theta &\stackrel{j_1}{\rightarrow}& 2 &\stackrel{D_1}{\rightarrow}& 8 &\stackrel{D_2}{\rightarrow} & 12 &\stackrel{D_3}{\rightarrow}& 8 &\stackrel{D_4}{\rightarrow}& 2 &\rightarrow 0 \\
  &&&&& \downarrow & & \downarrow & & \downarrow & & \downarrow &&\downarrow &     \\
   & 0 & \rightarrow & 4 & \stackrel{j_1}{\rightarrow} & 20 & \stackrel{D_1}{\rightarrow} & 40 &\stackrel{D_2}{\rightarrow} & 40 &\stackrel{D_3}{\rightarrow} & 20 &\stackrel{D_4}{\rightarrow} & 4 &   \rightarrow 0 \\
   & & & \parallel && \downarrow  & & \downarrow  & & \downarrow  & & \downarrow & & \downarrow  & \\
   0 \rightarrow & \Theta &\rightarrow & 4 & \stackrel{\cal{D}}{\rightarrow} & 18 & \stackrel{{\cal{D}}_1}{\rightarrow} & 32 & \stackrel{{\cal{D}}_2}{\rightarrow} & 28 & \stackrel{{\cal{D}}_3}{\rightarrow} & 12 &\stackrel{{\cal{D}}_4}{\rightarrow} & 2 & \rightarrow  0 \\
   &&&&& \downarrow & & \downarrow & & \downarrow & & \downarrow &  &\downarrow &   \\
   &&&&& 0 && 0 && 0 &&0 &&0 &  
   \end{array}     \]   \\
providing the Euler-Poincar\'{e} characteristic $4 - 18 +32 - 28 + 12 - 2 = 0  $. However, the only intrinsic concepts associated with a differential sequence are the  " {\it extension modules} " that only depend on the Kerr differential module but {\it not} on the differential sequence and it follows that ([16]): \\
  
 {\it THE ONLY IMPORTANT CONCEPT IS THE GROUP INVOLVED, NOT THE SEQUENCE}.  \\  

In an equivalent way as $g_2=0 \Rightarrow g_{r+2}=0, \forall r\geq 0$, we obtain successively:  
\[  dim(R_1)= dim(J_1(T)) - dim (S_2T^*)=20 - 10=10, \]
\[   dim(R_2)=dim(J_2(T)) - dim(J_1(S_2T^*))=60 - 50=10,\]
\[  dim(R_3)= dim(R_2) - 6=4 \Rightarrow  dim(R_ {r+4})= dim(R_4)=dim(R_3) - 2= 4 - 2 = 2  \]
and shall use these results from now on.  \\
According to a cut of the preliminary diagram with now $m=n=4, q =1, K= \mathbb{Q}(m,a)$, we obtain the following commutative and exact diagrams:  
\[   \begin{array}{rcccccccl} 
     &  0   &   & 0  &   & 0 &     &  0 & \\
      &  \downarrow &  & \downarrow  &  & \downarrow &   &  \downarrow &   \\
 0 \rightarrow  &  {\rho}_1({g}'_2)    & \rightarrow  & S_3T^* \otimes F_0 & \rightarrow  & T^* \otimes Q_2 &\rightarrow & {h}'_1   &  \rightarrow 0  \\
      &  \downarrow &  & \downarrow  &  & \downarrow &   &  \downarrow &   \\
 0 \rightarrow     &  {\rho}_1({R}'_2)  & \rightarrow & J_3(F_0)   & \rightarrow &  J_1(Q_2) & \rightarrow  &  {Q}'_1 &\rightarrow 0   \\
&  \downarrow &  & \downarrow  & &  \downarrow &   &  \downarrow &   \\
 0 \rightarrow     & {R}'_2 & \rightarrow & J_2(F_0)   &  \rightarrow  & Q_2  & \rightarrow   &  0 &   \\
  &  \downarrow &  & \downarrow  &  & \downarrow &   &   &   \\
  &  0   &   & 0  &   & 0 &    &   & 
 \end{array}     \]

\[   \begin{array}{rcccccccl} 
     &  0   &   & 0  &   & 0 & & 0      & \\
      &  \downarrow &  & \downarrow  &  & \downarrow &   &  \downarrow &   \\
 0 \rightarrow  &  144+x   &\rightarrow   & 200  &\rightarrow   & 56 &  \rightarrow   &  x  &  \rightarrow  0  \\
      &  \downarrow &  & \downarrow  &  & \downarrow &   &  \parallel &   \\
 0 \rightarrow     &  280 + x & \rightarrow & 350 &  \rightarrow &  70  &  \rightarrow &  x & \rightarrow 0  \\
&  \downarrow &  & \downarrow  &  & \downarrow &   &  \downarrow &   \\
 0 \rightarrow     &  136 & \rightarrow  & 150  &\rightarrow & 14  &  \rightarrow & 0 &  \\
  &  \downarrow &  & \downarrow  &  & \downarrow &   &      \\
  &  0   &   & 0  &   & 0 &     &   &
 \end{array}     \]

Denoting as before by $y$ the number of additional CC of strict order $3$ and by $x$ the number $dim(h'_1)=dim (Q'_1)$, we discover from the above diagram that the sum of the number of second order CC (that is $14$) and the number of differentially independent third order CC obtained by one prolongation of these second order CC is equal to $70 - x$. As now $dim(Q_3)=72$, we obtain therefore $72- y = 70 - x $ and thus $y = x + 2$. However, as $x \geq 4$ because of the $4$ divergence conditions implied on the Einstein tensor by the $20$ Bianchi identities, we {\it must} have $y\geq 6$. As we have already found {\it effectively} only $6$ CC of order $3$, we must have indeed $x= 4$ {\it effectively } and, {\it in any case}, we cannot have $y=4$ as claimed in ([1],[3]).

From the short exact sequence:  \\
\[    0 \rightarrow R_4 \rightarrow J_4(T) \rightarrow J_3(F_0) \rightarrow Q_3 \rightarrow 0     \]
\[    0 \rightarrow 2 \rightarrow 280 \rightarrow  350 \rightarrow  72  \rightarrow 0    \]
we obtain the commutative and exact diagrams:  \\
\[     \begin{array}{rcccccl}
                 & 0 &    &     &  &  &  \\
               &\downarrow &  &  & &  &  \\  
0 \rightarrow &R_4&\longrightarrow &J_4(T) & \longrightarrow &{R}'_3& \rightarrow 0 \\
              &\downarrow &  & \downarrow & & \downarrow  &  \\ 
0 \rightarrow &R_3 & \longrightarrow. & J_3(T) & \longrightarrow & {R}'_2 & \rightarrow 0  \\
              &   &  & \downarrow & & \downarrow  &  \\   
     &  &    &   0  &  & 0 &
     \end{array}  
  \hspace{15mm}   \begin{array}{rcccccl}
                 & 0 &    &     &  &  &  \\
               &\downarrow &  &  & &  &  \\  
0 \rightarrow & 2 &\longrightarrow &  280   & \longrightarrow &  278   & \rightarrow 0 \\
              &\downarrow &  & \downarrow & & \downarrow  &  \\ 
0 \rightarrow & 4 & \longrightarrow & 140  & \longrightarrow & 136 & \rightarrow 0  \\
              &   &  & \downarrow & & \downarrow  &  \\   
     &  &    &   0  &  & 0 & 
\end{array}   \]
As a byproduct, we have the commutative and exact diagrams:  \\
\[    \begin{array}{rcccl}
     &  0   &    &     0  &   \\
      &   \downarrow  &   &  \downarrow &     \\
     0 \rightarrow & {g}'_3& \longrightarrow  &{\rho}_1({g}'_2)&  \\
        &   \downarrow  &   &  \downarrow &     \\
  0 \rightarrow & {R}'_3 & \longrightarrow & {\rho}_1({R}'_2) &    \\
     &  \downarrow  &   & \downarrow  &   \\
0 \rightarrow & {R}'_2 &=&  {R}'_2   & \rightarrow 0 \\ 
   &   \downarrow  &   &  \downarrow &     \\
    &     0     &  &    0
\end{array}  
\hspace{15mm}   \begin{array}{rcccccl}
     &  0   &    &     0  &  &    0  & \\
      &   \downarrow  &   &  \downarrow & &  \downarrow  &   \\
     0 \rightarrow & 142  & \longrightarrow  & 144 + x   & \longrightarrow & y  & \rightarrow 0 \\
        &   \downarrow  &   &  \downarrow &  & \parallel &    \\
  0 \rightarrow & 278   & \longrightarrow & 280 + x  & \longrightarrow & y & \rightarrow 0    \\
     &  \downarrow  &   & \downarrow  &   & \downarrow & \\
0 \rightarrow &  136   &=&  136   & \longrightarrow  &   0   & \\ 
   &   \downarrow  &   &  \downarrow &  & &    \\
    &     0     &  &    0& &
\end{array}  \]
leading thus to the strict inclusions ${g}'_3 \subset {\rho}_1({g}'_2) \Leftrightarrow  {R}'_3 \subset {\rho}_1({R}'_2)$ and to the formula $y=x+2$. \\
We obtain therefore the following most useful diagram with symbolic notations: \\
  \[    \begin{array}{rccccccl}
     & & & & & 0 & &.   \\
     & & & & & \downarrow & &. \\
     &  0   &    &     0  &  & y &  &  \\
      &   \downarrow  &   &  \downarrow &  & \downarrow &  & \\
     0 \rightarrow & {R}'_3 & \longrightarrow  & J_3(F_0) & \longrightarrow & Q_3 & \longrightarrow  & 0 \\ 
        &   \downarrow  &   &  \parallel &&  \downarrow  &  &\\
  0 \rightarrow & {\rho}_1( {R}'_2)& \longrightarrow & J_3(F_0)  &\rightarrow & J_1(Q_2) & \longrightarrow  & {Q}'_1 \rightarrow 0   \\
     &  \downarrow  &   & \downarrow  &  &  \downarrow  & &   \\
      & {\rho}_1({R}'_2)/{R}'_3 & &  0  & &x \\ 
   &   \downarrow  &   &  &  &\downarrow  & &  \\
    &     0     &  &    &    & 0 & &
\end{array}  \]
finally showing that $x=dim(Q'_1) , y = dim ( {\rho}_1({R}'_2)/{R}'_3)=dim({\rho}_1(R'_2)) - dim (R'_3) $, a result leading to the long exact connecting sequence of vector bundles:  \\
\[      0   \rightarrow  R'_3 \longrightarrow  {\rho}_1({R}'_2)  \longrightarrow Q_3 \longrightarrow J_1(Q_2) \longrightarrow Q'_1 \rightarrow 0. \]
in agreement with the main theorem of section $2$. We have the following dimensions:  \\
 \[    \begin{array}{rccccccl}
     & & & & & 0 & &.   \\
     & & & & & \downarrow & &. \\
     &  0   &    &     0  &  & 6 &  &  \\
      &   \downarrow  &   &  \downarrow &  & \downarrow &  & \\
     0 \rightarrow & 278 & \longrightarrow  & 350  & \longrightarrow & 72 & \longrightarrow  & 0 \\ 
        &   \downarrow  &   &  \downarrow &&  \downarrow  &  &\\
  0 \rightarrow & 284 & \longrightarrow & 350 &\rightarrow & 70 & \longrightarrow  & 4 \rightarrow 0   \\
     &  \downarrow  &   & \downarrow  &  &  \downarrow  & &   \\
      & 6 & &  0  & & 4 &  &   \\ 
   &   \downarrow  &   &  &  &\downarrow  & &  \\
    &     0     &  &    &    & 0 & &
\end{array}  \]
Prolonging once while taking into account that $R_5 \simeq R_4$ with common dimension $2$, namely the dimension of the Kerr algebra generated by $\{ {\partial}_t, {\partial}_{\phi}\}$, we obtain the following commutative and exact diagram in which $Q_2$ and $Q_3$ are replaced by $Q_3$ and $Q_4$:   \\
\[   \begin{array}{rcccccl}
     &  0   &    &     0  &  &    0  & \\
      &   \downarrow  &   &  \downarrow & &  \downarrow  &   \\
     0 \rightarrow & g'_4  & \longrightarrow  & S_4T^* \otimes F_0  & \longrightarrow & h_4& \rightarrow 0 \\
        &   \downarrow  &   &  \downarrow &  & \downarrow &    \\
  0 \rightarrow & R'_4   & \longrightarrow & J_4(F_0)  & \longrightarrow & Q_4& \rightarrow 0    \\
     &  \downarrow  &   & \downarrow  &   & \downarrow & \\
0 \rightarrow &  R'_3  & \longrightarrow &  J_3(F_0)    & \longrightarrow  &   Q_3   & \rightarrow 0 \\ 
   &   \downarrow  &   &  \downarrow &  & \downarrow  &    \\
    &     0     &  &    0& &0. &
\end{array}  \]
showing that $g'_4 \simeq S_5T^* \otimes T$ with $dim(R'_4)=504 - 2= 502$ and $dim(Q_4)= 700 - 502 = 198$. It follows that $R'_4={\rho}_1(R'_3) $ is an involutive fourth order system allowing to construct a formally exact Janet sequence following the Killing operator as in ([20]), namely (exercise !!):   \\
\[  0 \rightarrow \Theta \rightarrow 4 \underset 1{\rightarrow}10 \underset 4{\rightarrow} 198 \underset 1{\rightarrow} 568  \underset 1{\rightarrow}  652 \underset 1{\rightarrow}  348 \underset 1{\rightarrow}  72\rightarrow 0    \]
Of course, {\it such a sequence is quite far from being minimum}. However, as the Killing operator for the Kerr metric is not formally integrable as we saw, the corresponding free resolution of the Kerr differential module, namely: \\
\[  0 \rightarrow D^{72} \underset 1{\longrightarrow} D^{348} \underset 1{ \longrightarrow } D^{652} \underset 1{\longrightarrow} D^{568} \underset 1 {\longrightarrow }D^{198} \underset 4 {\longrightarrow} D^{10} \underset 1 {\longrightarrow} D^4 \stackrel{p}{\longrightarrow} M  \rightarrow 0  \]
is {\it not} strictly exact though we have indeed:  \\
\[   rk_D(M)=4 - 10 + 198 - 568 + 652 - 348 + 72 = 0  \]
As the maximum size of the matrices involved is $dim(J_4(198))\times dim(J_3(568))$, that is $13860 \times 19880$, we hope to have convinced the reader that there is no hope for using computer algebra.  \\
As $R'_3 \subset {\rho}_1(R'_2)$ with a strict inclusion, the only posibility to escape from te above difficulty is to use only $R'_3$ and third order CC. However, as we have the strict inclusion $S_4T^* \otimes T \subset g'_3$ with a strict inclusion because $140 < 142$. As for the S metric, we have the crucial theorem:  \\

\noindent
{\bf THEOREM 4.C.1}: The operator ${\cal{D}}_1: F_0 \stackrel{j_3}{\longrightarrow } J_3(F_0) \rightarrow F_1=Q_3$ generates the CC of ${\cal{D}}: T \stackrel{j_1}{\longrightarrow} J_1(T) \rightarrow F_0$. \\

\noindent
{\it Proof}: First of all, as $R'_3$ is {\it strictly} contained into $ {\rho}_1(R'_2)$, we have {\it at least } one third order generating CC but we already know that we have the six $(d_i{\xi}^1 - {\xi}^1_{i}=0, d_i{\xi}^2 - {\xi}^2_{i}=0)   $ for $i=1,2,3$.  \\

Collecting the previous results and applying the Spencer operator, we obtain the following commutative diagram in which the two central columns are known to be locally exact ([8],[11]):  \\
\[  \begin{array}{rcccccccl}
  &  0   & &  0  &  &  0  &  &  &  \\
  & \downarrow & & \downarrow & & \downarrow &  &  &  \\
0 \rightarrow &  \Theta & \rightarrow &  T  & \stackrel{\cal{D}}{\rightarrow} & F_0 &  &  &  \\
     &   \hspace{4mm} \downarrow j_4  &  &  \hspace{4mm} \downarrow j_4 &  & \hspace{4mm} \downarrow j_3 & \hspace{4mm} \searrow {\cal{D}}_1 \\
0 \rightarrow &  R_4 & \rightarrow &  J_4(T) & \rightarrow & J_3(F_0) & \rightarrow   &Q_3  & \rightarrow 0  \\
     &   \hspace{3mm} \downarrow d  &  &  \, \downarrow d &  & \hspace{3mm} \downarrow d  &  &  \downarrow  &      \\
0 \rightarrow &  \fbox{$T^*\otimes R_3$} & \rightarrow & T^* \otimes J_3(T) & \rightarrow &  T^* \otimes  J_2(F_0) & \rightarrow   & T^* \otimes Q_2  & \rightarrow 0  \\
     & \hspace{3mm} \downarrow d  &  & \hspace{3mm} \downarrow d &  & \hspace{3mm} \downarrow d  &  &  \downarrow  &   \\
0 \rightarrow &  {\wedge}^2T^*\otimes R_2 & \rightarrow & {\wedge}^2T^* \otimes J_2(T) & \rightarrow & {\wedge}^2T^* \otimes  J_1(F_0) & \rightarrow   &  0 &     
\end{array}   \]    
Chasing around the right upper commutative triangle, it follows from Theorem .....  that a section $\Omega \in F_0$ with ${\cal{D}}_1 \Omega =0$ is such that there exists $\xi \in T$ with ${\cal{D}}\xi=\Omega$ if and only if the left vertical Spencer sequence is locally exact at $T^*\otimes R_3$.  \\
However, one has isomorphisms $R_4 \simeq R_5 \simeq R_6$ because they have the same dimension equal to $2$ and the map $R_4 \rightarrow R_3$ is a monomorphism because $dim(R_3)= 4$ with parametric jets $({\xi}^0,{\xi}^3, {\xi}^1_0, {\xi}^2_0)$ and $g_4=0$. Accordingly, as the Spencer sequence for the Killing algebroid is locally exact as it is isomorphic to the tensor product of the Poincar\'{e} sequence by a Lie algebra of dimension $2$, it is locally exact. We have therefore the commutative diagram with exact columns:  \\
\[   \begin{array}{rccccccc}
   & 0 & & 0 & & 0 & & 0 \\
   & \downarrow &  &  \downarrow &  &  \downarrow  &  & \downarrow  \\
0 \rightarrow & \Theta & \stackrel{j_5}{\rightarrow} & R_5 & \stackrel{d}{\rightarrow} &  T^* \otimes R_4 &  
\stackrel{d}{\rightarrow} & {\wedge}^2 T^* \otimes R_3 \\
 &  \parallel  &  &  \downarrow &  &  \downarrow &  &  \downarrow  \\
0 \rightarrow & \Theta & \stackrel{j_4}{\rightarrow} & R_4 & \stackrel{d}{\rightarrow} & \fbox{ $ T^* \otimes R_3 $}& \stackrel{d}{\rightarrow} & {\wedge}^2 T^* \otimes R_2  \\
& \downarrow &  &  \downarrow &  &  \downarrow  &  & \downarrow  \\
 &  0   &   &   0  &  \rightarrow & T^* \otimes (R_3/R^{(1)}_3) & \stackrel{d}{\rightarrow} & {\wedge}^2T \otimes (R_2/R^{(1)}_2)   \\
&  &  &  &  & \downarrow &   &  \downarrow  \\
 &  &  &  &  &   0  &  &  0   
\end{array}  \]
in which the upper row is locally exact. Chasing in this diagram, we discover that the central row is locally exact at $T^*\otimes R_3$ if the lower Spencer operator $d$ is injective.\\
Indeed, we have the following representative parametric jets:  \\
\[  dim(R_2)=10 \Rightarrow   \{ {\xi}^0, {\xi}^1, {\xi}^2, {\xi}^3, {\xi}^1_0, {\xi}^2_0, {\xi}^0_3, {\xi}^1_2, {\xi}^1_3, {\xi}^2_3 \}   \]
\[  dim(R_3)=4 \Rightarrow    \{ {\xi}^0, {\xi}^3, {\xi}^1_0, {\xi}^2_0  \}, \hspace{1cm}  dim(R_4)=2 \Rightarrow \{  {\xi}^0, {\xi}^3 \}  \]
\[  dim(R_3/R^{(1)}_3)= 2 \Rightarrow  \{ {\xi}^1_0, {\xi}^2_0 \}, \hspace{1cm}dim(R_2/R^{(1)}_2)=5 \Rightarrow \{ {\xi}^1, {\xi}^2, {\xi}^0_3, {\xi}^1_2, {\xi}^1_3, {\xi}^2_3 \}  \]
Accordingly, we may choose local coordinates $({\xi}^1_{0,i}, {\xi}^2_{0,i})$ for a representative and a representative of the image by $d$ is for example $({\xi}^1_{,0i} = {\xi}^1_{0,i} - {\xi}^1_{i,0}, {\xi}^2_{,0i} = {\xi}^2_{0,i} - {\xi}^2_{i,0})$. In the kernel of $d$, we have 
$( {\xi}^1_{0,i} = {\xi}^1_{i,0},  {\xi}^2_{0,i} = {\xi}^2_{i,0}), \forall i=1,2,3 $ but the situation is more tricky than for the S metric. \\
For $i=1$, we have ${\xi}^1_1=0, {\xi}^1_2=0, {\xi}^2_1=0, {\xi}^2_2=0 \Rightarrow {\xi}^1_{0,i}=0$ and similarly ${\xi}^2_{0,i}=0, \forall i=1,2$.  \\
Then, as $ \{ {\xi}^0, {\xi}^3 \}$ are among the parametric jets of $R_3$, we have ${\xi}^0_{,13}={\xi}^0_{1,3} - {\xi}^0_{3,1}=0 \Rightarrow 
{\xi}^0_{1,3}= {\xi}^0_{3,1}=0$ and similarly ${\xi}^3_{1,3}= {\xi}^3_{3,1}=0$. Using the Lie Equations of $R^{(2)}_1$ we obtain successively: 
\[  {\omega}_{00}{\xi}^0_{1,3} + {\omega}_{03}{\xi}^3_{1,3} + {\omega}_{11}{\xi}^1_{0,3}=0 \Rightarrow {\xi}^1_{0,3}=0,\hspace{1cm}
    {\omega}_{03}{\xi}^0_{1,3} + {\omega}_{33}{\xi}^3_{1,3} + {\omega}_{11}{\xi}^1_{3,3}=0 \Rightarrow {\xi}^1_{3,3}=0   \]
Exchanging $1$ and $2$, we obtain similarly with the two other framed Lie equations the two relations ${\xi}^2_{0,3}= 0, {\xi}^2_{3,3}=0$.
Hence, when $i=3$, it follows that ${\xi}^1_{0,3}={\xi}^1_{3,0}=lin({\xi}^1_{0,0},{\xi}^2_{0,0})=0$ and 
${\xi}^2_{0,3}={\xi}^2_{3,0}=lin({\xi}^1_{0,0},{\xi}^2_{0,0})=0 $, a result leading to ${\xi}^1_{0,0}=0, {\xi}^2_{0,0}=0$ on one side and 
${\xi}^1_{0,3}=0, {\xi}^2_{0,3}=0$ on the other side because ${\xi}^1_0$ and ${\xi}^2_0$ are linearly independent jet coordinates, that is finally 
${\xi}^1_{0,i}=0, {\xi}^2_{0,i}=0, \forall i=0,1,2,3$.  \\ 
\hspace*{12cm}  $\Box$   \\   

Like in the example of the S metric, the sequence constructed in the previous theorem for the K metric have the same "{\it jumps} " in the order of the successive operators. 
Indeed, using the same proof but now with $dim(g'_3)= dim (S_4T^*\otimes T) + 2=142, dim (A)=2, dim(B)=6$, we have:  \\

\noindent
{\bf COROLLARY 4.B.5}: The symbol of ${\cal{D}}_1$ is {\it not} $2$-acyclic and the CC operator ${\cal{D}}_2$ is of order $2$.  \\

Following closely the procedure used for all the motivating examples, we have thus transformed the search for the generating CC of the Killing operator into a {\it purely mathematical problem} of formal integrability and diagram chasing, quite far away from any physical background ([15],[16]).  \\  \\
  
\noindent
{\bf 5) CONCLUSION} \\

To end this paper with a rather personal story, let me come back $60$ years ago when I was preparing the competition for the french "Grandes Ecoles" at the State College Louis le Grand in Paris which is famous for one of his former student Evariste Galois. To give a few statistics, let us say that, for the one I had in mind, 30 000 students were trying, 3000 were selected after the written exam and 300 were only elected after oral exam !. This college was known to have the maximum number of success in France and the teachers   were carefully selected for that purpose, in particular in the best class room where I was. Once, this teacher was writing on the board the text of the problem we had to solve for the next day about what is now called "Desargues theorem". Roughly, if you consider in a plane two triangles $(ABC),(A'B'C')$ that are not flat and such that the $3$ straight lines $AA',BB',CC'$ have a common origin $O$ ({\it Center of perspective}), then the intersection $P$ of $BC$ and $B'C'$, $Q$ of $AC$ and $A'C'$, $R$ of $AB$ and $A'B'$ are on a straight line ({\it Axis of perspective}). Though I knew nothing about this result at that time, I suddenly "saw" the figure as a volume in space and shouted "$P,Q,R$ are on a straight line", even before the teacher had been asking the question in front of the asthonished students. Surprisingly, and I will never forget, the teacher said "Pommaret, 
this is true but how did you find it". When I said "Well, Sir, I have seen in space that the common line is the intersection of the two planes containing the triangles" (the reader may draw the picture for fun), his only comment has been "Better don't do that on the day of the competition". I replied "Sir, a result is important but the way you find it may 
even be more important". As a byproduct he never asked me any question during the full academic year and became a "private ennemy" in my scholar life during $10$ years till he retired.  \\

In a similar way, we point out the fact that during a visit for lecturing at the Albert Einstein Institute (AEI, Berlin/Postdam) in october 23-27, 2017 ([21], arXiv:1802.02430), we discovered that the members of the inviting research team were not interested about the new tools we developed in the many books or papers already quoted, in particular the link existing between the Spencer operator and the bracket of Lie algebroids. We also claim that the few references they quote for defining involutive systems are not the best ones as it happens that we have been regularly lecturing in Aachen during more than fifteen years and we know that the authors involved are only using Janet, Gr\"{o}bner or Pommaret bases for explicit computations but are unable to deal with acyclicity in general. The situation we met previously in the case of the Lie pseudogroup of conformal transformations is a good example. As a byproduct, it became a personal challenge to clarify the CC for the Killing operators over the Schwarzschild and Kerr metrics without using any of their tedious computations. The surprise is that, if we found again the $15$ second order CC for the S metric and the $14$ second order CC for the K metric, we also found explicitly $3$ third order CC for the S metric and $6$ third order CC for the K metric. All the formulas can be written within less than one line provided we use these new methods from differential homological algebra that have never been introduced in GR up to now, mainly because they prove that Einstein equations {\it cannot} be parametrized by a potential like Maxwell equations ... but this is surely another story !.      \\   \\

\noindent
{\bf 6) REFERENCES}  \\

\noindent
[1] Aksteiner, S., Andersson L., Backdahl, T., Khavkine, I., Whiting, B.: Compatibility Complex for Black Hole Spacetimes, Commun. Math. Phys. 384 (2021) 1585-1614.  \\
https://doi.org/10.1007/s00220-021-04078-y      (arXiv:1910.08756) .  \\
\noindent
[2] Aksteiner, S., Backdahl, T: New Identities for Linearized Gravity on the Kerr Spacetime, Phys. Rev. D 99, 044043 (2019), (arXiv:1601.06084) . \\
\noindent
[3] Aksteiner, S., Backdahl, T.: All Local Gauge Invariants for Perturbations of the Kerr Spacetime, Physical Review Letters 121, 051104 (2018), 
(arXiv:1803.05341) . \\
\noindent
[4] Andersson L., Backdahl, T., Blue, P., Ma, S.: Stability for Linearized Gravity on the Kerr Spacetime,Ó (2019), (arXiv:1903.03859) .  \\
\noindent
[5] Goldschmidt, H.: Prolongations of Linear Partial Differential Equations: I Inhomogeneous equations, Ann. Scient. Ec. Norm. Sup., 4, 1 (1968) 617-625.  \\
\noindent
[6] Khavkine, I.: The Calabi Complex and Killing Sheaf Cohomology, J. Geom. Phys., 113 (2017) 131-169. \\
\noindent
[7] Macaulay, F.S.: The Algebraic Theory of Modular Systems, Cambridge Tract 19, Cambridge University Press, 1916.  \\
\noindent
[8] Pommaret, J.-F.: Systems of Partial Differential Equations and Lie Pseudogroups, Gordon and Breach, New York (1978); Russian translation: MIR, Moscow, 1983.\\
\noindent
[9] Pommaret, J.-F.: Differential Galois Theory, Gordon and Breach, New York, 1983. \\
\noindent
[10] Pommaret, J.-F.: Lie Pseudogroups and Mechanics, Gordon and Breach, New York, 1988.\\
\noindent
[11] Pommaret, J.-F.: Partial Differential Equations and Group Theory, Kluwer, Dordrecht, 1994.\\
https://doi.org/10.1007/978-94-017-2539-2    \\
\noindent
[12] Pommaret, J.-F.: Partial Differential Control Theory, Kluwer, 2001  (Zbl 1079.93001).  \\
\noindent
[13] Pommaret, J.-F.: Algebraic Analysis of Control Systems Defined by Partial Differential Equations, in "Advanced Topics in Control Systems Theory", Springer, Lecture Notes in Control and Information Sciences 311, 2005, Chapter 5, pp. 155-223.\\
\noindent
[14] Pommaret, J.-F.: Parametrization of Cosserat Equations, Acta Mechanica, 215 (2010) 43-55.\\
https://doi.org/10.1007/s00707-010-0292-y  \\
\noindent
[15] Pommaret, J.-F.: The Mathematical Foundations of General Relativity Revisited, Journal of Modern Physics, 4 (2013) 223-239. \\
 https://doi.org/10.4236/jmp.2013.48A022   \\
 \noindent
[16] Pommaret, J.-F.: Relative Parametrization of Linear Multidimensional Systems, Multidim. Syst. Sign. Process., 26 (2015) 405-437.  \\
DOI 10.1007/s11045-013-0265-0   \\
\noindent
[17] Pommaret, J.-F.: Airy, Beltrami, Maxwell, Einstein and Lanczos Potentials revisited, Journal of Modern Physics, 7 (2016) 699-728. \\
https://doi.org/10.4236/jmp.2016.77068      (arXiv:1512.05982).   \\
\noindent
[18] Pommaret, J.-F.: Deformation Theory of Algebraic and Geometric Structures, Lambert Academic Publisher (LAP), Saarbrucken, Germany, 2016. A short summary can be found in "Topics in Invariant Theory ", S\'{e}minaire P. Dubreil/M.-P. Malliavin, Springer Lecture Notes in Mathematics, 1478 (1990) 244-254.\\
https://arxiv.org/abs/1207.1964  \\
\noindent
[19] Pommaret, J.-F.: New Mathematical Methods for Physics, Mathematical Physics Books, Nova Science Publishers, New York, 2018, 150 pp. \\
\noindent
[20] Pommaret, J.-F.: Minkowski, Schwarzschild and Kerr Metrics Revisited, Journal of Modern Physics, 9 (2018) 1970-2007.  \\
https://doi.org/10.4236/jmp.2018.910125  (arXiv:1805.11958v2 ).  \\
\noindent
[21] Pommaret, J.-F.: The Mathematical Foundations of  Elasticity and Electromagnetism Revisited, Journal of Modern Physics, 10 2019) 1566-1595, \\
https://doi.org/10.4236/jmp.2019.1013104  (arXiv:1802.02430) .  \\
\noindent
[22] Pommaret, J.-F.: Generating Compatibility Conditions and General Relativity, Journal of Modern Physics, 10, 3 (2019) 371-401.  \\
\noindent
https://doi.org/10.4236/jmp.2019.103025  (arXiv:1811.12186).   \\
\noindent
[23] Pommaret, J.-F.: A Mathematical Comparison of the Schwarzschild and Kerr Metrics, Journal of Modern Physics, Journal of Modern Physics, 11 (2020) 1672-1710.  \\
https://doi.org/10.4236/jmp.2020.1110104   (arXiv:2010.07001).  \\
\noindent
[24] Pommaret, J.-F.: Minimum Parametrization of the Cauchy Stress Operator, Journal of Modern Physics, 12 (2021) 453-482.  \\
https://doi.org/10.4236/jmp.2021.124032  (arXiv:2101.03959).  \\
\noindent
[25] Pommaret, J.-F.: Homological Solution of the Lanczos Problems in Arbitrary Dimensions, Journal of Modern Physics, 12 (2021) 829-858. \\
https://doi.org/10.4236/jmp.2019.1012097  (arXiv:1807.09122). \\
\noindent
[26] Pommaret, J.-F.: Differential Correspondences and Control Theory, Advances in Pure Mathematics, 11 (2021) 835-882.  \\
https://doi.org/10.4236/apm.2021.1111056    (arXiv:2107.08797).   \\
\noindent
[27] Pommaret, J.-F.: The Conformal Group Revisited, Journal of Modern Physics, 12 (2021) 1822-1842.  \\
https://doi.org/10.4236/jmp.2021.1213106  (arXiv:2006.03449 ).  \\
\noindent
[28] Rotman, J.J.: An Introduction to Homological Algebra, (Pure and Applied Mathematics), Academic Press, 1979.  \\
\noindent
[29] Vessiot, E.: Th\'{e}orie des Groupes Continus, Ann. Sc. Ecole Normale Sup., 20 (1903) 411-451 (Can be found in Numdam).  \\

\end{document}